%% file: main.tex
\documentclass[a4paper,onecolumn,11pt,accepted=2026-09-21]{quantumarticle}
\pdfoutput=1
\usepackage[utf8]{inputenc}
\usepackage[english]{babel}
\usepackage[T1]{fontenc}
\usepackage{amssymb}
\usepackage{xcolor}
\usepackage{amsthm}

\usepackage[utf8]{inputenc}
\usepackage[T1]{fontenc}

\makeatletter
\def\@maketitle{%
  \normalfont\rmfamily
  \null
  \let\footnote\thanks
  \noindent
  \begin{minipage}{\textwidth}%
    \iftoggle{@titlepage}{\centering}{}%
    \noindent{\huge\hyphenpenalty=50000 \@printtitle\par}%
  \end{minipage}%
  \vskip 1.5em%
  \iftoggle{@titlepage}{}{\begin{flushleft}}%
  \noindent \@printauthors
  \iftoggle{@titlepage}{}{\end{flushleft}}%
  \vskip 1em%
  \noindent\@printaffiliations
  \vskip 0em%
  \ifdefempty{\@date}{}{\noindent{\footnotesize\color{quantumgray}\@date}}%
  \par
  \vskip 1.5em
  \makeatletter
  \begingroup
  \hypersetup{%
    pdftitle={\detokenize\expandafter{\@title}},
    pdfauthor={\@authorsonly},
    pdfkeywords={\@keywords},
    pdfcreator={LaTeX with hyperref package and class quantumarticle \csname ver@quantumarticle.cls\endcsname},%
  }%
  \endgroup
  \makeatother
}
\makeatother

\newtheorem{definition}{Definition}
\newtheorem{example}{Example}

\input{preample}

\input{commands}

\begin{document}

    \title{GPU-Accelerated Quantum Simulation of Stabilizer Circuits}
	
	\author{Muhammad Osama}
	\email{m.o.mahmoud@liacs.leidenuniv.nl}
	\orcid{0000-0002-5023-5348}
	\author{Dimitrios Thanos}
	\email{d.thanos@liacs.leidenuniv.nl}
	\orcid{0000-0002-0719-1036}
	\author{Alfons Laarman}
	\email{a.w.laarman@liacs.leidenuniv.nl}
	\orcid{0000-0002-2433-4174}
	\affiliation{%
		Leiden Institute of Advanced Computer Science (LIACS), Leiden University,
		Leiden,
		The Netherlands
	}
	
	\maketitle
	
\begin{abstract}

We introduce new parallel algorithms for efficiently simulating stabilizer (Clifford) circuits on GPUs, with a focus on data-parallel tableau evolution and scalable handling of projective measurements. Our approach reformulates key bottlenecks in stabilizer simulation---such as Gaussian elimination and measurement updates---into GPU-tailored primitives that eliminate sequential dependencies and maximize memory coalescing. We implement these techniques in \ourTool, a GPU-accelerated stabilizer simulator designed for large qubit counts and many-shot sampling. Across a broad benchmark suite reaching 180,000 qubits and depth 1,000 (roughly 130M gates), \ourTool shows substantial runtime improvements, with up to 105$\times$ speedup, and over 80\% energy reduction on demanding instances. Moreover, \ourTool consistently outperforms \stim, a state-of-the-art CPU-optimized stabilizer simulator, as well as \qiskitAer (CPU/GPU), \qibo, \cirq, and \pennylane. Finally, \ourTool exhibits a significant advantage in many-shot sampling on large workloads. These results demonstrate that our parallel algorithms can significantly advance the scalability of stabilizer-circuit simulation, particularly for workloads involving extensive measurements and sampling.

	\end{abstract}
	
	\input{acmcode}

	\keywords{Quantum Simulation, Stabilizer Formalism, Clifford Gates, GPUs, Parallel Computing, Measurements}

	\maketitle
	
	\input{introduction}

	\input{preliminaries}
	
	\input{challenges}
	
	\input{simulation}

    \input{sampling}
	
	\input{experiments}

	\input{contributions}

	\bibliographystyle{quantum}
	\bibliography{literature}

\end{document}

%% file: preample.tex
\usepackage{bold-extra}              
\usepackage{textcomp}                
\usepackage{xspace}                  

\usepackage[normalem]{ulem}

\usepackage[dvipsnames]{xcolor}      
\usepackage[textsize=footnotesize]{todonotes} 
\makeatletter
\define@key{todonotes}{fixed}[]{%
    \setkeys{todonotes}{author=fixed,color=green!20}}%
\makeatother   

\usepackage{mathtools, bm}   			
\usepackage{dsfont}                           	
\usepackage{amssymb}
\usepackage{stmaryrd}
\newenvironment{smallmat}                
{\left[\begin{smallmatrix}}
	{\end{smallmatrix}\right]}
\usepackage{graphicx} 					
\usepackage{adjustbox} 				   
\usepackage{wrapfig}                  	 
\usepackage{stfloats}         			  
{\setlength{\columnsep}{0pt}\wrapfloat{figure}}
{\endwrapfloat}

\usepackage{caption}
\usepackage{subcaption}

\usepackage{tikz}
\usetikzlibrary{fit,calc}
\usetikzlibrary{decorations.pathreplacing}
\usepackage{quantikz}                

\usepackage[
  colorlinks=true,
  linkcolor=red!80!black,
  citecolor=green!80!black,
  urlcolor=blue!80!black
]{hyperref}

\usepackage{tabularx}                 
\usepackage{rotating}                 
\usepackage{multirow}                 
\usepackage{tablefootnote}            
\usepackage{siunitx}                  
\usepackage{colortbl}                 
\usepackage{paralist}                 

\usepackage{pifont}                   

\let\llncssubparagraph\subparagraph  
\let\subparagraph\paragraph
\usepackage{titlesec}                 
\let\subparagraph\llncssubparagraph

\usepackage[misc,geometry]{ifsym}     

\usepackage[ruled,linesnumbered,noline,noend]{algorithm2e}
\usepackage{algpseudocode, setspace}
\newcommand\midsmall{\fontsize{7.6pt}{8.5pt}\selectfont}

\usepackage{xspace} 
\newcommand\defmath[2]{\newcommand#1{\ensuremath{#2}\xspace}}

\newcommand\concept[1]{\textit{#1}}

\SetKwFor{ForAllPar}{forall}{do in parallel}{endfor}
\SetKwFor{Forparblockx}{for all $j_x \in [$}{$]$ do in parallel}{end}
\SetKwFor{Forparx}{for all $\tidx \in [$}{$]$ do in parallel}{end}
\SetKwFor{Forpary}{for all $\tidy \in [$}{$]$ do in parallel}{end}

\SetKwProg{Kernel}{\textit{device kernel}}{:}{}
\SetKwProg{Devfunction}{\textit{device function}}{:}{}
\SetKwProg{Function}{\textit{function}}{:}{}

\let\oldnl\nl
\newcommand{\nonl}{\renewcommand{\nl}{\let\nl\oldnl}}

\providecommand{\set}[1]{\ensuremath{\left\lbrace #1 \right\rbrace}}

\providecommand{\gen}[1]{\ensuremath{\left\langle #1 \right\rangle}}

\usepackage{cleveref}   
\crefname{figure}{figure}{figures}
\Crefname{figure}{Figure}{Figures}

%% file: commands.tex
\newcommand{\ttscale}{s*[1.1]}
\DeclareFontFamily{OT1}{cmtt}{\hyphenchar\font -1 }
\DeclareFontShape{OT1}{cmtt}{m}{n}{
        <-9>   \ttscale cmtt8
        <9-10> \ttscale cmtt9
        <10-12>\ttscale cmtt10
        <12->  \ttscale cmtt12
      }{}
\DeclareFontShape{OT1}{cmtt}{m}{it}{
        <->    \ttscale cmitt10
      }{}
\DeclareFontShape{OT1}{cmtt}{m}{sl}{
        <->    \ttscale cmsltt10
      }{}
\DeclareFontShape{OT1}{cmtt}{m}{sc}{
        <->    \ttscale cmtcsc10
      }{}
\DeclareFontShape{OT1}{cmtt}{m}{ui}
       {<->ssub*cmtt/m/it}{}
\DeclareFontShape{OT1}{cmtt}{bx}{n}
       {<->ssub*cmtt/m/n}{}
\DeclareFontShape{OT1}{cmtt}{bx}{it}
       {<->ssub*cmtt/m/it}{}
\DeclareFontShape{OT1}{cmtt}{bx}{ui}
       {<->ssub*cmtt/m/it}{}

\newcommand{\code}[1]{\texttt{#1}}
\newcommand{\var}[1]{\textit{#1}}
\newcommand{\routine}[1]{{\relsize{0.5}{\textsc{#1}}}}
\newcommand{\tool}[1]{\textsc{#1}}
\newcommand{\myparagraph}[1]{{\par\vspace{0.5\baselineskip}\noindent\textit{#1}~}}

\newcommand{\fontFormat}[1]{\textit{#1}}
\newcommand{\I}{\fontFormat{I}\xspace}
\newcommand{\X}{\fontFormat{X}\xspace}
\newcommand{\Xtilde}{\Tilde{\X}\xspace}
\newcommand{\Y}{\fontFormat{Y}\xspace}
\newcommand{\Z}{\fontFormat{Z}\xspace}
\newcommand{\Ztilde}{\Tilde{\Z}\xspace}
\newcommand{\Signs}{\fontFormat{S}\xspace}
\newcommand{\Had}{\fontFormat{H}\xspace}
\newcommand{\Phase}{\fontFormat{P}\xspace}
\newcommand{\CX}{\fontFormat{CX}\xspace}
\newcommand{\M}{\fontFormat{M}\xspace}
\newcommand{\Pauli}{\ensuremath{\mathcal{P}}\xspace}
\newcommand{\qstate}{\ensuremath{\psi}\xspace}
\newcommand{\numQubits}{\ensuremath{n}\xspace}
\newcommand{\depth}{\ensuremath{d}\xspace}
\newcommand{\circuit}{\ensuremath{\mathcal{C}}\xspace}

\newcommand{\INV}{\ensuremath{\,T}\xspace}
\newcommand{\destab}{\ensuremath{\mathrm{D}}\xspace}
\newcommand{\stab}{\ensuremath{\mathrm{S}}\xspace}
\newcommand{\XD}{\ensuremath{\X_{\destab}}\xspace}
\newcommand{\ZD}{\ensuremath{\Z_{\destab}}\xspace}
\newcommand{\XS}{\ensuremath{\X_{\stab}}\xspace}
\newcommand{\ZS}{\ensuremath{\Z_{\stab}}\xspace}
\newcommand{\SignsS}{\ensuremath{\Signs_{\stab}}\xspace}
\newcommand{\SignsD}{\ensuremath{\Signs_{\destab}}\xspace}

\newcommand{\tableau}{\ensuremath{\mathcal{T}}\xspace}
\newcommand{\tableauBlock}{\ensuremath{\tableau}\xspace}
\newcommand{\tableauTilde}{\ensuremath{\Tilde{\tableauBlock}}\xspace}

\newcommand{\numWords}{\ensuremath{k}\xspace}

\newcommand{\numShots}{\ensuremath{f}\xspace}

\newcommand{\newGate}{\ensuremath{g}\xspace}

\newcommand{\circuitWindow}{\ensuremath{\mathcal{W}}\xspace}
\newcommand{\circuitScheduled}{\ensuremath{\mathcal{R}}\xspace}

\newcommand{\qubit}{\ensuremath{q}\xspace}
\newcommand{\td}[1]{\textbf{\var{#1}}\xspace}
\newcommand{\tx}{\ensuremath{\var{t}_{\td{x}}}\xspace}

\newcommand{\blockidx}{\ensuremath{b}\xspace}
\newcommand{\blockDim}{\ensuremath{B_x}\xspace}
\newcommand{\yblockDim}{\ensuremath{B_y}\xspace}

\newcommand{\tid}{\var{tid}\xspace}
\newcommand{\tidz}{\ensuremath{\tid_{\td{z}}}\xspace}
\newcommand{\tidy}{\ensuremath{\tid_{\td{y}}}\xspace}
\newcommand{\tidx}{\ensuremath{\tid_{\td{x}}}\xspace}
\newcommand{\initialState}{\code{initstate}\xspace}

\newcommand{\pivots}{\code{pivots}\xspace}
\newcommand{\allNondet}{\ensuremath{\mathcal{Q}}\xspace}
\newcommand{\pivot}{\ensuremath{p}\xspace}
\newcommand{\target}{\ensuremath{t}\xspace}
\newcommand{\observable}{\ensuremath{q}\xspace}

\newcommand{\transposed}{\ensuremath{\tableauBlock^{\INV}}\xspace}
\newcommand{\transposedX}{\ensuremath{\X^{\INV}}\xspace}
\newcommand{\transposedZ}{\ensuremath{\Z^{\INV}}\xspace}
\newcommand{\wordSize}{\ensuremath{w}\xspace}
\newcommand{\columnLen}{\ensuremath{2\numWords}\xspace}
\newcommand{\rowLen}{\ensuremath{2\numQubits}\xspace}
\newcommand{\wordidx}{\ensuremath{r}\xspace}
\newcommand{\qWord}{\ensuremath{\lfloor\qubit/\wordSize\rfloor}\xspace}
\newcommand{\tabMatrix}{\ensuremath{M}\xspace}
\newcommand{\tileidx}{\var{tileIdx}\xspace}
\newcommand{\lookuptab}{\code{MASKS}\xspace}
\newcommand{\mask}{\var{mask}\xspace}
\newcommand{\offset}{\var{offset}\xspace}
\newcommand{\Control}{\var{c}\xspace}
\newcommand{\Target}{\var{t}\xspace}
\newcommand{\blocksum}{\var{bsum}\xspace}
\newcommand{\localprefix}{\var{prefix}\xspace}
\newcommand{\sharedz}{\ensuremath{\shared_z}\xspace}
\newcommand{\sharedx}{\ensuremath{\shared_x}\xspace}
\newcommand{\prefixes}{\code{prefixes}\xspace}
\newcommand{\blocksums}{\code{blocksums}\xspace}
\newcommand{\x}{\ensuremath{x}\xspace}
\newcommand{\z}{\ensuremath{z}\xspace}
\newcommand{\xc}{\ensuremath{\x_c}\xspace}
\newcommand{\zc}{\ensuremath{\z_c}\xspace}
\newcommand{\xt}{\ensuremath{\x_t}\xspace}
\newcommand{\zt}{\ensuremath{\z_t}\xspace}
\newcommand{\isYPauli}{\var{isYPauli}\xspace}

\newcommand{\measured}{\code{measured}\xspace}
\newcommand{\sign}{\var{s}\xspace}
\newcommand{\shared}{\code{sm}\xspace}
\newcommand{\maxThreads}{\ensuremath{N}\xspace}

\newcommand{\speedup}{\ensuremath{\scriptstyle\mathbf{\times}}\xspace}
\newcommand{\record}{\code{record}\xspace}
\newcommand{\bitstring}{\ensuremath{x}\xspace}

\colorlet{gpu}{violet!90!black}
\newcommand{\sharedMem}{\textbf{\text{shared}}\xspace}
\newcommand{\constMem}{\textbf{\text{constant}}\xspace}

\newcommand{\li}[1]{line~\ref{#1}}
\newcommand{\lis}[2]{lines~\ref{#1}\code{-}\ref{#2}}
\newcommand{\bigO}{\ensuremath{\mathcal{O}}\xspace}

\newcommand{\scheduleParallelGates}{\routine{ScheduleWindows}\xspace}
\newcommand{\initTableau}{\routine{InitializeTableau}\xspace}
\newcommand{\measureWindow}
{\routine{Measure}\ensuremath{}\xspace}
\newcommand{\hasMeasurements}
{\routine{HasMeasurement}\xspace}
\newcommand{\simulateWindow}{\routine{ApplyGates}\xspace}
\newcommand{\copyWindowToDevice}{\routine{CopyWindowToDevice}\xspace}

\newcommand{\allocate}{\routine{AllocateTableau}\xspace}
\newcommand{\Swap}{\routine{Swap}\xspace}
\newcommand{\syncThreads}{\routine{SynchronizeBlock}\xspace}
\newcommand{\atomicXOR}{\routine{AtomicXOR}\xspace}
\newcommand{\collapseSigns}{\routine{CollapseSigns}\xspace}

\newcommand{\transpose}{\routine{InlineTranspose}\xspace}

\newcommand{\GE}{\routine{GaussianEliminationByCX}\xspace}
\newcommand{\swapAntiComm}{\routine{SwapAntiCommuting}\xspace}
\newcommand{\injectInvP}{\routine{PInvGate}\xspace}

\newcommand{\injectHad}{\routine{HGate}\xspace}
\newcommand{\injectCX}{\routine{CXGate}\xspace}
\newcommand{\injectX}{\routine{XGate}\xspace}
\newcommand{\findAllNondetQubits}{\routine{MaxNondeterministic}\xspace}
\newcommand{\compactPivots}{\routine{FindAndCompactPivots}\xspace}
\newcommand{\copyFirstPivot}{\routine{CopyFirstPivotToHost}\xspace}
\newcommand{\paraGE}{\routine{Para}\GE{}\xspace}
\newcommand{\paraSwap}{\routine{Para}\swapAntiComm{}\xspace}
\newcommand{\paraX}{\routine{Para}\routine{XGate}\xspace}
\newcommand{\shuffleTiles}{\routine{ShuffleTiles}\xspace}
\newcommand{\swapTiles}{\routine{SwapTiles}\xspace}
\newcommand{\shiftLeft}{\routine{ShiftLeft}\xspace}
\newcommand{\shiftRight}{\routine{ShiftRight}\xspace}
\newcommand{\prefixPassOne}{\routine{PrefixXORPass1}\xspace}
\newcommand{\prefixPassTwo}{\routine{PrefixXORPass2}\xspace}
\newcommand{\prefixPassThree}{\routine{PrefixXORPass3}\xspace}
\newcommand{\prefixBlock}{\routine{ExclusivePrefix}\xspace}
\newcommand{\injectCXByPrefix}{\routine{InjectCXByPrefix}\xspace}
\newcommand{\scatter}{\routine{Scatter}\xspace}
\newcommand{\compactCub}{\routine{DeviceSelect::If}\xspace}
\newcommand{\getBit}{\routine{GetBit}\xspace}
\newcommand{\syncGrid}{\routine{SynchronizeGrid}\xspace}
\newcommand{\random}{\routine{Random}\xspace}

\newcommand{\copySigns}{\routine{CopyStabilizerSigns}\xspace}
\newcommand{\copyListToHost}{\routine{CopyListToHost}\xspace}
\newcommand{\randomize}{\routine{Randomize}\xspace}
\newcommand{\randomizeWord}{\routine{RandomWord}\xspace}
\newcommand{\measureSample}{\routine{MeasureSample}\xspace}

\newcommand{\stim}{\tool{Stim}\xspace}
\newcommand{\qiskit}{\tool{Qiskit}\xspace}
\newcommand{\qiskitAer}{\tool{Qiskit-Aer}\xspace}
\newcommand{\pennylane}{\tool{PennyLane}\xspace}
\newcommand{\cirq}{\tool{Cirq}\xspace}
\newcommand{\ourTool}{\tool{QuaSARQ}\xspace}

\newcommand{\qibo}{\tool{Qibo}\xspace}
\newcommand{\bqsim}{\tool{BQSim}\xspace}

\newcommand{\quiZX}{\tool{QuiZX}\xspace}

\newcommand\ketbra[1]{\ensuremath{|#1\rangle\langle #1|}}

\defmath{\defn}{\,\triangleq\,}

%% file: acmcode.tex
%

%% file: introduction.tex
\section{Introduction}\label{sec:intro}

Classical simulation of quantum circuits plays an instrumental role in the development of quantum computing. Simulators support algorithm design, verification, and optimization, enabling tasks such as equivalence checking \cite{EC}, circuit compilation and synthesis \cite{MCPaper,Synthesis2023}, and state tomography \cite{statetomography10,AaronsonShadowTomogr20}. They also operate within hybrid quantum-classical workflows, where they are frequently used for pre-optimization routines \cite{PhysRevA.111.062413}, adaptive measurement protocols in measurement-based quantum computing \cite{jozsa2005introductionmeasurementbasedquantum}, and variational algorithms \cite{PhysRevX.6.021043}. 

Within this landscape, \emph{stabilizer} (\emph{Clifford}) circuits form a foundational class. Their rich algebraic structure allows efficient classical simulation via the \emph{stabilizer-tableau formalism} \cite{aaronson2008improved}, yielding an $\bigO(n^2)$ update cost and enabling large-scale applications in quantum error correction, communication, and simulation. 
Recent GPU-accelerated simulators such as IBM’s \qiskitAer \cite{qiskit2024} have demonstrated that GPUs can deliver substantial speedups. Meanwhile, {\stim} -- a state-of-the-art, CPU-optimized stabilizer simulator \cite{stim} -- introduced a series of algorithmic innovations for measurement and sampling. However, these methods have not yet been fully adapted to exploit modern GPU architectures. {\stim}’s paper itself reports a GPU experiment that did not achieve the desired results.\footnote{See ``GPU Experiment Failure'' in \cite[p.~14]{stim}.} 

In prior work, we showed that GPUs can dramatically accelerate certain stabilizer-circuit tasks, most notably with a GPU-based equivalence checker~\cite{paraEC}. Yet equivalence checking does not require full support for projective measurements, nor does it handle extensive sampling workloads. Measurements are pivotal to a wide range of quantum protocols, including error correction \cite{surfaceCode12,gameOfSurface19}, randomized benchmarking \cite{randbench18}, and hybrid quantum-classical algorithms \cite{VQA21}. Efficiently supporting them at scale on GPUs remains a significant challenge. In particular, naive implementations of measurement-induced tableau updates, such as Gaussian elimination \cite{aaronson2008improved, stim}, introduce sequential dependencies and irregular memory access patterns that limit parallelism.

\paragraph{Contributions.}

In this work, we address these challenges by introducing new parallel algorithms for tableau evolution and measurement operations on GPUs, designed to align with GPU memory hierarchies and minimize control divergence.

Our main contributions are as follows:

\begin{itemize}

\item[$\star$] A host-side parallel-gate and measurement scheduler to maximize GPU parallelism. Mutually independent gates are arranged into maximal parallel windows, enabling large batches of Clifford gates to be applied simultaneously (\cref{sec:parallelGates}).

\item[$\star$] A GPU-oriented data-structure with a dual tableau layout. A column-major layout enables fast gate application, while an in-place word/bit transpose provides a row-major layout optimized for measurement operations, ensuring coalesced memory access throughout tableau evolution (\cref{sec:tableauLayout}). 

\item[$\star$] A tree‑based tableau evolution method for applying Clifford gates. We reduce atomic contention and improve occupancy by aggregating sign updates in shared memory via a binary-tree reduction scheme (\cref{sec:applyGates}).

\item[$\star$] Dense stream compaction \cite{streamCompact} of tableau pivots (indices of anti-commuting Pauli strings). By packing pivots into contiguous memory, we eliminate warp divergence and ensure coalesced memory access during Gaussian elimination (\cref{sec:compact}).

\item[$\star$] A three‑pass prefix-XOR formulation, Gaussian elimination. We transform CX injections \cite{stim} into fully parallel prefix operations, eliminating sequential dependencies in the measurement  process (\cref{sec:GEPrefix}). 

\end{itemize}

We have implemented these algorithms in \ourTool, a GPU-accelerated stabilizer simulator designed for large-scale Clifford circuits and many-shot sampling. \ourTool completes 177 heavy-suite circuits within 72 hours, compared to {\stim}’s 125 circuits in 132 hours, reduces energy consumption by more than 80\% on demanding instances, and supports thousands of projective measurements at 180,000 qubits. Across a broad benchmark suite, \ourTool consistently outperforms \stim, \qiskitAer (CPU/GPU), \qibo, \cirq, and \pennylane. In addition, for many-shot sampling, {\ourTool} uses a GPU-based sampler which shows a clear advantage over \stim on our 1,024-shot benchmarks.
These results show that GPU-oriented algorithms can significantly advance the scalability of stabilizer simulation, particularly for workloads involving extensive measurements and sampling. The remainder of this paper describes our parallel algorithms, implementation design, and benchmarking in detail.

\paragraph{Related Work.}

Classical simulation of quantum circuits spans a large body of research, ranging from generic state-vector simulators to specialized algorithms for Clifford circuits, tensor-network contractions, and hybrid approaches. GPU-based simulators have gained increasing attention as quantum workloads grow in scale, but most existing systems do not fully exploit the algebraic structure of stabilizer circuits.

\emph{State-vector simulators.}
General-purpose simulators such as \cirq \cite{Cirq}, \pennylane \cite{pennylane}, and \qiskitAer \cite{qiskit2024}, and \qibo \cite{qibo} provide GPU backends capable of accelerating dense state-vector operations. More recently, NVIDIA’s cuQuantum SDK \cite{cuQuantum} has introduced high-performance GPU primitives for simulation and tensor-network contraction, enabling speedups for circuits that fit within GPU memory.  These simulators support arbitrary quantum circuits but exhibit exponential memory scaling, limiting simulations to approximately 30 qubits on high-end hardware. Although their GPU kernels are optimized for linear-algebra operations, they do not exploit tableau structure. As a result, they are typically far slower than specialized stabilizer simulators, even for modest qubit counts.

\emph{Decision Binary Diagram (BDD)-based simulators.}
An alternative line of work uses graph-based or decision-diagram representations to compress quantum states. Tools such as \bqsim \cite{BQSim} and earlier BDD-based simulators exploit structural regularities and repeated sub-patterns to avoid explicit state-vector storage. These techniques can achieve dramatic compression on circuits with high redundancy, but their performance can degrade significantly when applied to dense Clifford circuits. 

\emph{Weighted Model Counting and Symbolic approaches.}
Another line of work is simulation through symbolic encodings. Weighted Model Counting (WMC) methods, such as Quokka \cite{Quokka}, translate Clifford+T circuits into conjunctive normal form and use SAT or \#SAT/WMC solvers to answer decision or probability queries. However, they suffer from rapid formula growth as circuit size increases. ZX-calculus simulators, such as \quiZX \cite{Kissinger_2022}, apply rewrite rules to represent and simplify circuits heuristically. This approach is powerful for circuits with abundant algebraic structure but can become unpredictable for large or less structured Clifford circuits.

\emph{Stabilizer-tableau simulators.}
Classical simulation of Clifford circuits traces back to the Gottesman–Knill theorem \cite{GottesmanKnillTheorem}, with practical tableau-based implementations developed in CHP \cite{aaronson2008improved} and later improved in \qiskit. The most recent advance is \stim \cite{stim}, a high-performance CPU-based stabilizer simulator that introduced algorithmic optimizations for measurements, Gaussian elimination, and fast sampling. \stim represents the current state of the art in CPU efficiency. However, its techniques are optimized for vectorized instruction sets and branch prediction on CPUs and have not been adapted to fully exploit GPU parallelism. The single GPU experiment reported in the \stim paper achieved very limited speedup due to irregular memory access patterns and sequential dependencies in measurement and tableau updates.

\emph{GPU-based stabilizer simulation.}
Despite the natural parallelism in tableau operations, few stabilizer simulators target GPUs. The GPU backend of \qiskitAer supports Clifford gates but does not restructure measurement operations to avoid sequential operations, leading to limited scalability. Earlier GPU work has focused on specific subtasks, such as our own GPU equivalence checker \cite{paraEC}, that do not require full support for projective measurements or high-volume sampling. To our knowledge, there is no existing GPU simulator that addresses the key challenges of tableau evolution, projective measurements, sampling fully in parallel.

In a different direction, PACOX \cite{pacox} investigates FPGA acceleration of bit-level Pauli-string computations.

%% file: preliminaries.tex
\section{Preliminaries}\label{sec:prelim}

\subsection{Quantum States and Gates}\label{sec:basic}

A qubit is described by a linear combination of the \textit{basis states} $\ket{0}$ and $\ket{1}$ that can be identified with the 2-vectors $[1\quad0]^T$ and $[0\quad1]^T$, respectively. These states are analogous to the classical bits 0 and 1, but in the quantum case, a single qubit can take on the value of the linear combination $\ket{\qstate}= \alpha_0 \ket{0} + \alpha_1 \ket{1} $
where $\alpha_0, \alpha_1$ are complex numbers and $\ket{\qstate}$, like any quantum state, must be normalized to be a unit vector (i.e., $|\alpha_0|^2 + |\alpha_1|^2 = 1$). The linear combination of basis states is also known as \textit{superposition}.

The operation for combining $\numQubits$ single-qubit states $\ket{\qstate}_1,\cdots,\ket{\qstate}_n$ into an $\numQubits$-qubit state $\ket{\qstate}_1\otimes \cdots \otimes \ket{\qstate}_n$  is known as the \textit{tensor product}. It is denoted by $\otimes$ and is typically known in linear algebra as the Kronecker product. The $\numQubits$-qubit states have the form $\sum_{\ell\in \{ 0,1 \}^{\numQubits} } \alpha_{\ell} \ket{\ell}$ where $\ket{\ell}$ is the conventional way of denoting a string of multiple tensored single-qubit states, i.e., we suppress the $\otimes$ symbol and say $\ket{1011}$ instead of $\ket{1}\otimes \ket{0} \otimes \ket{1} \otimes \ket{1}$.
If a quantum state cannot be factored into a tensor product of its individual subsystems, then we say it is an \textit{entangled} state. For example, the state $\frac{1}{\sqrt{2}}(\ket{00}+\ket{11})$ is entangled, while the state $\ket{1001}$ is not.

In general, a quantum state $\ket{\qstate}$ of $\numQubits$ qubits can be written as a vector of $2^\numQubits$ complex numbers: $\ket{\qstate}=[\alpha_0,\alpha_1,\cdots,\alpha_{2^\numQubits}]^T$. For every state, there is an \textit{adjoint} defined as $(\ket{\qstate}^*)^T$ and denoted by $\bra{\qstate}$. The ``$*$'' operator is a complex conjugation, meaning that we replace every entry with its complex conjugate, then transpose the vector (i.e., transforming it into a row vector). The normalization requirement, can be now reformulated into $\bra{\qstate}\cdot \ket{\qstate}=1$. This product is also known as \textit{inner product} and can be performed between distinct states.

Quantum states can be manipulated by two types of operations: quantum gates and quantum measurements. Quantum gates are \emph{unitary} matrices, representing reversible linear transformations denoted as $U\in\mathbb{C}^{2^\numQubits\times 2^\numQubits}$. These gates preserve both the norm and the inner product of quantum states. A key property of unitary gates is that $U^{-1} = U^{\dagger}$, where $U^{\dagger}=(U^*)^T$ represent the conjugate transpose of $U$. The effect of a gate $U$ acting on a state $\ket{\qstate}$, can be calculated using matrix-vector multiplication $U\cdot \ket{\qstate}$.


An important gate set is the Clifford gate set (\CX is also called \fontFormat{CNOT}):
\begin{align*}
	\Had &= \frac{1}{\sqrt{2}}
	\begin{bmatrix*}[r]
		1 & 1 \\
		1 & -1
	\end{bmatrix*} &
	\Phase &=
	\begin{bmatrix*}[r]
		1 & 0 \\
		0 & \phantom-i
	\end{bmatrix*}&
	\CX &= 
	\begin{smallmat}
		1 & 0 & 0 & 0 \\
		0 & 1 & 0 & 0 \\
		0 & 0 & 0 & 1 \\
		0 & 0 & 1 & 0
	\end{smallmat}.
\end{align*}
It is (quantum) non-universal and classically simulatable as explained in the following sections. Extending the Clifford gate set with a $T = \begin{smallmat}
    1 & 0 \\ 0 & \omega 
\end{smallmat}$ gate, where $\omega = \sqrt i = \frac{1+i}{\sqrt 2}$, yields a universal gate set~\cite{KitaevCode}.

To classically simulate circuits consisting of Clifford gates, we will use the following Pauli gate set to represent symmetries of the encountered states.
\begin{align*}
	\I &=
	\begin{bmatrix*}[r]
		1 & 0 \\
		0 & 1
	\end{bmatrix*}&
	\X &=
	\begin{bmatrix*}[r]
		0 & \phantom-1 \\
		1 & 0
	\end{bmatrix*}&
	\Z &=
	\begin{bmatrix*}[r]
		1 & 0 \\
		0 & -1
	\end{bmatrix*}&
	\Y &=
	\begin{bmatrix*}[r]
		0 & -i \\
		i & 0
	\end{bmatrix*}
\end{align*}

\subsection{Circuit Simulation}\label{sec:circuitsim}

Classical simulation of quantum circuits is typically categorized into two distinct tasks: \emph{strong} and \emph{weak} simulation. In this work, we focus on weak simulation, which is essential for systems involving extensive measurements.

\emph{Weak simulation}, also refereed to as \emph{sampling}, more closely mimics the behavior of quantum hardware. Rather than calculating exact probabilities as in strong simulation, the goal is to produce a random bitstring $\bitstring$ derived from the probability distribution defined by the state $\ket{\qstate}$.

\begin{definition}[Weak Simulation]
Given a quantum circuit \circuit and an initial state $\ket{0}^{\otimes \numQubits}$, weak simulation produces an output $\bitstring \in \{0,1\}^{\numQubits}$ such that the probability of obtaining \bitstring is exactly $p(\bitstring) = |\gen{\bitstring | \circuit | 0}^{\otimes\numQubits}|^2$.
\end{definition}

In our simulator \ourTool, we implement weak simulation through two different GPU optimized algorithms:

\begin{itemize}
    \item[$\bullet$] Single-shot simulation: For scenarios where the state tableau must be updated following a projective measurement, we design and implement a parallel algorithm for performing Gaussian Elimination (GE).
    \item[$\bullet$] Many-shot sampling: for applications requiring a high volume of samples or shots, our simulator implements \emph{Pauli frames} \cite{PhysRevA.103.042604,Knill05,stim} to avoid the overhead of GE being repeatedly performed for every sample. Pauli frames allow the simulator to generate multiple samples efficiently by tracking Pauli errors and updates without full tableau restructuring for every shot.
\end{itemize}

Next, we discuss simulation algorithms for the specific case of stabilizer circuits, a non-universal class of quantum circuits that are efficiently simulatable by classical algorithms.





\subsection{Stabilizer Circuits}\label{sec:stabilizerCircuits}
%

Stabilizer circuits, also known as \textit{Clifford circuits}, are quantum circuits composed of Clifford gates. The quantum states generated by applying these circuits to the computational basis \(|0\rangle^{\otimes \numQubits}\) are called \emph{stabilizer states}. 
Interestingly, stabilizer circuits can be simulated efficiently by a classical algorithm \cite{gottesman1997stabilizer,aaronson2008improved}.

The \concept{Pauli group} on $\numQubits$ qubits is defined as the set of operators of the form $\Pauli_n  \defn i^k P_1 \otimes \cdots \otimes P_n$ with $P_i\in \{ \I, \X, \Y, \Z \,\}$ and $k\in\set{0,1,2,3}$, i.e., the phase of \concept{Pauli string} is one of $\set{\pm 1, \pm i}$.
For simplicity, the tensor product is often suppressed, and $\numQubits$-Pauli strings are written in shorthand. For example, $\X \otimes \I \otimes \Y \otimes \X \otimes \Z$ can be compactly written as \fontFormat{XIYXZ}.
The \emph{Clifford group} on $\numQubits$ qubits consists of the unitary operators $V$ that normalize the Pauli group, that is, $\mathcal C_n =\{ V\in \mathbb{C}^{2\numQubits\times 2\numQubits} \mid \forall P \in \Pauli_n \colon VP V^{\dagger}\in\Pauli_n \}$.
The Clifford group is generated by the Clifford gate set \set{H, P, CX} \cite{NielsenChuang2010}. 
Each stabilizer state, generated by a stabilizer circuit, can be uniquely encoded by a subgroup of the Clifford group, requiring exactly $\numQubits$ independent generators, where each generator is an $\numQubits$-Pauli string.

Stabilizer states are called this way because they are \textit{stabilized} by certain operators. A unitary operator $U$ stabilizes a state iff $U \ket{\qstate}=\ket{\qstate}$. 
For example, the Pauli string $-IZ$ stabilizes the state $\ket{0}\otimes \ket{1}= \ket{01}$. A state can be stabilized by multiple Pauli strings; for instance, $\ket{01}$ is also stabilized by $II$, $ZI$ and $-ZZ$. These Pauli strings generate the maximal commutative subgroup $\mathcal{H}_{\max{}}=\{\fontFormat{II}, -\fontFormat{IZ}, \fontFormat{ZI}, -\fontFormat{ZZ}\,\}$, which is the stabilizer group of $\ket{01}$. Each maximal commutative subgroup of the Pauli group stabilizes exactly one stabilizer state \cite{NielsenChuang2010}. 

Although the stabilizer group for an $\numQubits$-qubit stabilizer state has $2^\numQubits$ elements, only $\numQubits$ Pauli strings are needed to generate it. For example, $\mathcal{H}_{\max{}}$ can be generated by the two generators in multiple ways, expressed formally by: $\mathcal{H}_{\max{}} = \gen{\, -\fontFormat{IZ}, -\fontFormat{ZZ}\, }  = \gen{\, -\fontFormat{IZ}, \fontFormat{ZI}\, }$. This property allows stabilizer states to be efficiently and succinctly represented by a data structure such as the tableau formalism.

%
%
%

The density matrix of the state $\ket{\tableauBlock}$ represented by a tableau \tableauBlock consisting of generators $\mathrm{G}_1, \dots, \mathrm{G}_n$, can be established directly as:
\begin{align}
\label{eq:tabstate} 
\ketbra{\tableauBlock} \quad=\quad \prod_{i\in [n]} \frac{I^{\otimes n} + \mathrm{G}_i}2 \quad=\quad
\frac1{2^n} \sum_{P \in \gen{\mathrm{G}_1, \dots, \mathrm{G}_n}} P.
\end{align}

\subsection{Tableau Formalism}
\label{sec:tableauFormalism}
Throughout this paper, we refer to the conventional forward-evolving stabilizer tableau as the \emph{state tableau}.
We use the term \emph{tableau evolution} to denote the application of gate-level update rules to the stabilizer tableau as the circuit executes. 

A stabilizer state's generators can be described by a binary table known as a \emph{stabilizer tableau} \cite{gottesman1997stabilizer}. 
The tableau encodes a Pauli operator $P \in \set{I, X, Y, Z}$ with two bits $x,z$
such that $P = Z^z X^x$ up to global phase (i.e., $Y = i ZX \equiv ZX$).
The ($n$-qubit) generators are then encoded with bits $x_1,\dots,x_n, z_1,\dots,z_n, s$, where $s$ is a bit recording the phase as $(-1)^s$.\footnote{Since all stabilizers in the group mutually commute, the phase $\pm i$ does not occur, because $(\pm iP)^2 = - I^{\otimes n}$ (which only commutes with itself).}
The $n$ generators then form the tableau as follows.


\begin{equation}\label{tableauFormat}
	\tableauBlock =   
	\left(
	\begin{array}{ccc|ccc|c}
		\x_{11} & \cdots & \x_{1n} & \z_{11} & \cdots & \z_{1\numQubits} & \sign_{1} \\
		\vdots & \ddots & \vdots & \vdots & \ddots & \vdots & \vdots \\
		\x_{\numQubits1} & \cdots & x_{\numQubits\numQubits} & \z_{\numQubits1} & \cdots & \z_{\numQubits \numQubits} & \sign_{\numQubits} \\
	\end{array}
	\right)
\end{equation}


\noindent
For example, the tableau below encodes the generators $\{ \fontFormat{ZII}, \fontFormat{-IZI}, \fontFormat{-IXY}\,\}$.

\newcommand{\braceY}{-0.4}
\newcommand{\raiseLabel}{10pt}
\newcommand{\labelY}{-0.7cm}
\begin{equation}\label{tableauExample}
	\left(
		\begin{array}{ccc|ccc|c}
			0 & 0 & 0 & 1 & 0 & 0 & 0 \\
			1 & 0 & 0 & 1 & 1 & 0 & 1 \\
			0 & 1 & 1 & 0 & 0 & 1 & 1 \\
		\end{array}
	\right)
	\begin{tikzpicture}[overlay, remember picture]
		\node at (0, \braceY) (X) {};
		\node at (0, \braceY) (Z) {};
		\node at (0, \braceY) (S) {};
		\draw[decorate,decoration={brace,amplitude=4pt,mirror,raise=\raiseLabel},yshift=0pt]
		(-4.2, \braceY) -- (-2.7, \braceY) node [black,midway,yshift=\labelY] {\scriptsize $\X$ part};
		\draw[decorate,decoration={brace,amplitude=4pt,mirror,raise=\raiseLabel},yshift=0pt]
		(-2.5, \braceY) -- (-1, \braceY) node [black,midway,yshift=\labelY] {\scriptsize $\Z$ part};
		\draw[decorate,decoration={brace,amplitude=4pt,mirror,raise=\raiseLabel},yshift=0pt]
		(-0.8, \braceY) -- (-0.45, \braceY) node [black,midway,yshift=\labelY] {\scriptsize Signs};
	\end{tikzpicture}
	\cong
	\left\langle
	\begin{matrix*}[c]
		&\phantom-\Z\,\I\,\I& \\
		&-\Y\Z\I& \\
		&-\I\X\Y& 
	\end{matrix*}
	\right\rangle
\end{equation}\\

To update $\tableauBlock$ after applying a Clifford gate on a single qubit $q$ or on a pair $(c,t)$, the following rules are applied to every row index $i\in\{1,\dots,\numQubits\}$ \cite{gottesman1997stabilizer}:
\begin{equation}\label{update_rules}
\begin{aligned}
&\Had_q:\;\quad
\bigl(x_{iq},\ z_{iq},\ s_i\bigr)
   \gets 
   \bigl(
     \z_{iq},
    &&x_{iq},
    &&s_i \oplus (\x_{iq}\wedge \z_{iq})
    \bigr),\\[4pt]
&\Phase_q:\;\quad
\bigl(\x_{iq},\ \z_{iq},\ s_i\bigr)
   \gets \bigl(
      \x_{iq}, 
    &&\z_{iq}\oplus \x_{iq},
    &&s_i \oplus (\x_{iq}\wedge \z_{iq})
    \bigr),\\[4pt]
&\CX_{c,t}:\;
\bigl(\x_{it},\ \z_{ic},\ s_i\bigr)
    \gets \bigl(
      \x_{it}\oplus \x_{ic}, 
    &&\z_{ic}\oplus \z_{it}, 
    &&s_i \oplus (\x_{ic}\wedge \z_{it}\wedge \neg(\x_{it}\oplus \z_{ic}))
    \bigr).
\end{aligned}
\end{equation}

\Cref{update_rules} is expressed in the standard state-tableau formulation, where Clifford gates are indexed by qubits and applied column-wise across all generators. \stim uses the operator-tableau representation, which, while very similar to the state-tableau, goes beyond encoding states. The operator-tableau is the state-tableau extended by the \emph{destabilizer} generators \cite{aaronson2008improved}. This doubles the number of rows and yields a structure isomorphic to the Pauli group modulo global phase. 

The column update rules remain identical to those of the state tableau, except that they apply to twice as many generators. This isomorphism originates in \cite{gottesman1997stabilizer}, and is made explicit in later works such as \cite{aaronson2008improved}. Both representations are symplectic encodings, whose structure is rooted in symplectic geometry \cite{deGosson2006}. 



The following example shows how gates can be applied in parallel to $\tableauBlock$ when acting on independent qubits, hence affecting independent columns.

\begin{example}\label{ex_ind_gates}
	Given a 3-qubit system initialized to the $\ket{000}$ state, assume that the gates $\Had_1$ and $\Phase_3$ are applied in order, to the qubits 1 and 3 respectively. The affected columns, according to \cref{update_rules}, are updated and highlighted by colors:

	\begin{equation}\label{tabl_1}
        \left[
		{\arraycolsep=2.4pt\def\arraystretch{1}
			\begin{array}{ccc|ccc|c}
				0 & 0 & 0 & 1 & 0 & 0 & 0\\
				0 & 0 & 0 & 0 & 1 & 0 & 0\\
				0 & 0 & 0 & 0 & 0 & 1 & 0\\
			\end{array}
		}
		\right]
		\overset{\Had_1 }{\Longrightarrow}
		\left[
		{\arraycolsep=2.4pt\def\arraystretch{1}
			\begin{array}{>{\columncolor{red!20}}ccc|>{\columncolor{red!20}}ccc|>{\columncolor{gray!25}}c}
				1 & 0 & 0 & 0 & 0 & 0 & 0\\
				0 & 0 & 0 & 0 & 1 & 0 & 0\\
				0 & 0 & 0 & 0 & 0 & 1 & 0\\
			\end{array}
		}
		\right]
		\overset{\Phase_3 }{\Longrightarrow}
		\left[
		{\arraycolsep=2.4pt\def\arraystretch{1}
			\begin{array}{ccc|cc>{\columncolor{cyan!20}}c|>{\columncolor{gray!25}}c}
				1 & 0 & 0 & 0 & 0 & 0 & 0\\
				0 & 0 & 0 & 0 & 1 & 0 & 0\\
				0 & 0 & 0 & 0 & 0 & 1 & 0\\
			\end{array}
		}
		\right]
	\end{equation}
	
	
	
\end{example}

\subsection{Measurements in the Stabilizer Formalism}\label{sec:measurements}

A quantum measurement projects a qubit onto a chosen basis and yields a classical outcome determined probabilistically, irreversibly collapsing superpositions and leaving the system in an eigenstate of the measured observable. In this work, we restrict attention to \emph{projective Pauli measurements}, and in particular to computational-basis (Z-basis) measurements of individual qubits. Measurements of multiple qubits are performed by repeating the same single-qubit procedure for each measured qubit. More general measurements, such as POVMs \cite{NielsenChuang2010}, are out of scope.

A projective measurement is specified by a set of projectors $\{P_m\}$ corresponding to the possible outcomes. For a computational-basis measurement, these projectors are $P_m=\ket{m}\bra{m}$ (with $m\in\{0,1\}$). If the system is in state \ket{\phi}, the probability of observing outcome $m$ is given by the Born rule,
$$
p(m) = \bra{\phi}P_m\ket{\phi}
     = \bra{\phi}\bigl(\ket{m}\bra{m}\bigr)\ket{\phi}
     = \bigl|\langle m \mid \phi\rangle\bigr|^2.
$$
The computational-basis measurement is equivalently a measurement of the Pauli-$Z$ operator, whose eigenstates are $\ket{0}$ and $\ket{1}$, satisfying $Z\ket{0}=+1\ket{0}$, $Z\ket{1}=-1\ket{1}$, and $Z=\ket{0}\bra{0}-\ket{1}\bra{1}$.
Measuring qubit $q$ in the computational basis thus corresponds to measuring the Pauli observable
$$
Z_q \;\equiv\; I^{\otimes q}\otimes Z \otimes I^{\otimes (n-q)}.
$$
The two measurement outcomes correspond to the eigenvalues $+1$ and $-1$, which we encode as a classical bit
$$
m\in\{0,1\}\quad\text{with eigenvalue}\quad (-1)^m.
$$

Equivalently, writing the pre-measurement state as a density matrix $\rho = \ket{\phi}\bra{\phi}$, the outcome probabilities can be expressed as $p(m) = \operatorname{Tr}(\rho\, P_m)$.
Since all Pauli operators are traceless except the identity, this trace extracts precisely the coefficient of the projector $\ket{m}\bra{m}$ in $\rho$.
For a computational-basis measurement on qubit $q$, this gives
$$
p(m=0)=\bra{\phi}\, ( \ket{0}\!\bra{0}_q )\,\ket{\phi},\qquad
p(m=1)=\bra{\phi}\, ( \ket{1}\!\bra{1}_q )\,\ket{\phi}.
$$
Equivalently, in terms of the expectation value of $Z_q$,
$$
p(m)=\frac{1+(-1)^m \,\bra{\phi}Z_q\ket{\phi}}{2}.
$$
For a more elaborate introduction to quantum measurements, see \cite{NielsenChuang2010}. 
\myparagraph{Deciding whether the measurement is deterministic.}
Let $\ket{\tableauBlock}$ be the stabilizer state represented by a tableau $\tableauBlock$ with stabilizer generators
$\stab_1, \dots, \stab_n$, and let $\rho = \ket{\tableauBlock}\!\bra{\tableauBlock}$. The quantity $\langle Z_q\rangle \;=\; \mathrm{tr}(Z_q \rho)$ fully determines whether the outcome is deterministic or probabilistic. For stabilizer states
$\langle Z_q\rangle\in\{+1,-1,0\}$, it can be characterized purely by the stabilizer group
$\langle \stab_1,\dots,\stab_n\rangle$:

\begin{equation}\label{eq:meastab}
\mathrm{tr}(Z_q\rho)=
\begin{cases}
+1 &\text{if } +Z_q \in \langle \stab_1,\dots,\stab_n\rangle,\\
-1 &\text{if } -Z_q \in \langle \stab_1,\dots,\stab_n\rangle,\\
0  &\text{if } Z_q \text{ anti-commutes with at least one generator.}
\end{cases}
\end{equation}

Interpreting this in terms of the classical measurement outcome $m$:
\begin{itemize}
\item If $\langle Z_q\rangle=+1$, then $p(m=0)=1$ (deterministic outcome $m=0$).
\item If $\langle Z_q\rangle=-1$, then $p(m=1)=1$ (deterministic outcome $m=1$).
\item If $\langle Z_q\rangle=0$, then $p(m=0)=p(m=1)=\tfrac12$ (uniformly random).
\end{itemize}

At the tableau level, testing commutation with $Z_q$ is straightforward. A stabilizer generator has an $X$
(or $Y$) on qubit $q$ exactly when its $x$-bit at column $q$ is $1$, and this is precisely
the condition under which it anti-commutes with $Z_q$. Hence the measurement of $q$ is probabilistic iff \emph{at least one stabilizer
row has $x_{iq}=1$}.

\myparagraph{Updating the tableau after measurement.}
If the measurement is \emph{deterministic}, the post-measurement quantum state is unchanged
, since it is already an eigenstate of $Z_q$.

If the measurement is probabilistic, the stabilizer description must
be updated. Let $p$ (the pivot) be any index such that $\stab_p$ anti-commutes with $Z_q$ (i.e., $\stab_p$ has
$x_{pq}=1$). A standard stabilizer-update rule proceeds as follows:
\begin{enumerate}
\item For every $i\neq p$ such that $\stab_i$ anti-commutes with $Z_q$, replace
      $\stab_i \leftarrow \stab_i \stab_p$ so that $\stab_i$ now commutes with $Z_q$.
\item Sample a fresh random bit $m\in\{0,1\}$.
\item Replace the anti-commuting generator ($\stab_p$) by the measured observable with the sampled sign:
      $$\stab_p \leftarrow (-1)^m Z_q.$$
\end{enumerate}
After this update, all stabilizer generators commute with $Z_q$, and the resulting tableau represents the
post-measurement stabilizer state conditioned on the sampled outcome $m$.

In \cite{aaronson2008improved}, it is shown that GE can be performed more efficiently by introducing \concept{destabilizer} generators, which are Pauli operators that each anti-commute with exactly one stabilizer generator. The tool CHP implements the improved algorithm, which requires time $\bigO(\numQubits^2)$. Next, we explain the CHP algorithm in detail.

\paragraph{The CHP algorithm \cite{aaronson2008improved}.}\label{CHP alg}

The CHP algorithm operates on the forward-evolving \emph{state tableau}, explicitly maintaining both stabilizer and destabilizer generators.
We denote destabilizer and stabilizer matrices by
$\XD, \ZD$ and $\XS, \ZS$, respectively, where the subscripts
$\mathrm{D}$ and $\mathrm{S}$ are formatted in Roman font to avoid
ambiguity with mathematical indices. Further, destabilizer bits are marked with a prime (e.g., $x'_{ij}, z'_{ij}$) to distinguish them from stabilizer bits.
The extended tableau representation is

\begin{equation}\label{tableauFormat2}
	\tableauBlock =   
    \begin{pmatrix}
		\XD & \vert & \ZD & \vert & \SignsD \\
		\hline
		\XS & \vert & \ZS & \vert & \SignsS \\
	\end{pmatrix}
    =
	\left(
	\begin{array}{ccc|ccc|c}
		\x_{11}' & \cdots & \x_{1n}' & \z_{11}' & \cdots & \z_{1\numQubits}' & \sign_{1}' \\
		\vdots & \ddots & \vdots & \vdots & \ddots & \vdots & \vdots \\
		\x_{\numQubits1}' & \cdots & x_{\numQubits\numQubits}' & \z_{\numQubits1}' & \cdots & \z_{\numQubits \numQubits}' & \sign_{\numQubits}' \\\hline
        \x_{11} & \cdots & \x_{1n} & \z_{11} & \cdots & \z_{1\numQubits} & \sign_{1} \\
		\vdots & \ddots & \vdots & \vdots & \ddots & \vdots & \vdots \\
		\x_{\numQubits1} & \cdots & x_{\numQubits\numQubits} & \z_{\numQubits1} & \cdots & \z_{\numQubits \numQubits} & \sign_{\numQubits}
	\end{array}
	\right)
\end{equation}



\renewcommand{\braceY}{-1.}
\renewcommand{\raiseLabel}{10pt}
\renewcommand{\labelY}{-0.7cm}
If the initial state is a product state, for example $\ket{+01}$, 
one valid initialization of the tableau is shown in \cref{tableauExample2}. In this example, the $\X$-part of the stabilizer rows coincides with the $\Z$-part of the destabilizer rows, and vice versa, reflecting the standard CHP initialization.

\begin{equation}\label{tableauExample2}
 \begin{matrix}
    \text{Destabilizers} \left\{ 
	\begin{matrix}
    \hphantom 1\\
    \hphantom 1\\ 
    \hphantom 1
	\end{matrix}
    \right.
	\\
    \text{~~~Stabilizers} \left\{ 
	\begin{matrix}
    \hphantom 1\\
    \hphantom 1\\ 
    \hphantom 1
	\end{matrix}
    \right.
    \end{matrix}
    \hspace{-1em}
    \left(
		\begin{array}{ccc|ccc|c}
			0 & 0 & 0 & 1 & 0 & 0 & 0 \\
			0 & 1 & 0 & 0 & 0 & 0 & 0 \\
			0 & 0 & 1 & 0 & 0 & 0 & 1 \\\hline
			1 & 0 & 0 & 0 & 0 & 0 & 0 \\
			0 & 0 & 0 & 0 & 1 & 0 & 0 \\
			0 & 0 & 0 & 0 & 0 & 1 & 1 \\
		\end{array}
	\right)
	\begin{tikzpicture}[overlay, remember picture]
		\node at (0, \braceY) (X) {};
		\node at (0, \braceY) (Z) {};
		\node at (0, \braceY) (S) {};
		\draw[decorate,decoration={brace,amplitude=4pt,mirror,raise=\raiseLabel},yshift=0pt]
		(-4.2, \braceY) -- (-2.7, \braceY) node [black,midway,yshift=\labelY] {\scriptsize $\X$ part};
		\draw[decorate,decoration={brace,amplitude=4pt,mirror,raise=\raiseLabel},yshift=0pt]
		(-2.5, \braceY) -- (-1, \braceY) node [black,midway,yshift=\labelY] {\scriptsize $\Z$ part};
		\draw[decorate,decoration={brace,amplitude=4pt,mirror,raise=\raiseLabel},yshift=0pt]
		(-0.8, \braceY) -- (-0.45, \braceY) node [black,midway,yshift=\labelY] {\scriptsize Signs};
	\end{tikzpicture}
	\cong
    \begin{matrix}
	\left\langle
	\begin{matrix*}[c]
		&\phantom-\Z\I\,\I& \\
		&\phantom-\I\X\I& \\
		&-\I\I\X& 
	\end{matrix*}
	\right\rangle\\
	\left\langle
	\begin{matrix*}[c]
		&\phantom- \X\I\,\I& \\
		&\phantom- \I\Z\I& \\
		&-\I\I\Z& 
	\end{matrix*}
	\right\rangle
    \end{matrix}
\end{equation}
\vspace{1em}

\tikzset{meter/.append style={draw, inner sep=10, rectangle, font=\vphantom{A}, minimum width=30, line width=.8,  path picture={\draw[black] ([shift={(.1,.3)}]path picture bounding box.south west) to[bend left=50] ([shift={(-.1,.3)}]path picture bounding box.south east);\draw[black,-{Latex[length=8mm, width=1.8mm]}] ([shift={(0,.2)}]path picture bounding box.south) -- ([shift={(.4,-.1)}]path picture bounding box.north);}}}

\defmath\mmeter{\scalebox{.2}{\tikz{\node[meter]{};}}}
\defmath\Pm{P_{\mmeter}}

We now describe the CHP procedure for measuring qubit $q$ in the computational basis, i.e., the observable $Z_q$.
According to \cref{eq:meastab}, there are two possible cases: (1) the outcome is deterministic (the state does not change), or (2) the outcome is probabilistic (the state collapses).
To determine whether the measurement is deterministic, we check whether any stabilizer generator anti-commutes with $\Z_q$. As discussed above, this occurs iff the $q$-th column of $\XS$ contains a $1$. If no such row exists, the outcome is deterministic; otherwise it is probabilistic.
 

\myparagraph{Probabilistic case.} 
There is at least one $p\in [n]$ such that $\stab_p$ anti-commutes with $Z_q$. 
We call $\tilde{p}$, the smallest such index.
We update the state as follows:
\begin{enumerate}

	\item For the indices $i \in [n] \setminus \set{\tilde p}$ with $S_i$ anti-commuting with $Z_q$, update $\stab_i \leftarrow \stab_i \stab_{\tilde{p}}$ and $\destab_i\leftarrow \destab_i \destab_{\tilde{p}}$.
    
	\item  Set $\destab_{\tilde p}$ to $\stab_{\tilde p}$. 
    
	\item Flip an unbiased coin $m\in \set{0,1}$. 
            Set $\stab_{\tilde p}$ to $(-1)^m Z_q$. 
            Return outcome~$m$.
\end{enumerate} 

\noindent In CHP, enforcing commutation requires explicit  multiplication of Pauli generators, implemented as bitwise XOR over the full tableau rows. Since each such operation touches $\bigO(\numQubits)$ bits and may be applied  to up to $\bigO(\numQubits)$ rows, this leads to an overall quadratic cost.

\myparagraph{Deterministic case.}
For all  $p\in [n]$, we have that $\stab_p$ commutes with $Z_q$. The tableau requires no update in absence of the collapse of a superposition. 
We may avoid the GE by using the destabilizers to check membership of $\pm Z_q$ in the full stabilizer group $\gen{\stab_1, \dots, \stab_n}$ as required by \cref{eq:meastab} in this case.
As explained in~\cite{aaronson2008improved}, we can do this by observing that GE yields
$\pm Z_q = \prod_{i\in [n]} \stab_i^{c_i}$ for some  assignments to $c_i\in\{0,1\}$.
The destabilizers can identify the assignments directly since
$c_i = \prod_{j\in [n]}  (\destab_i \stab_j)^{c_j}  = \destab_i \prod_{j\in [n]}  \stab_j^{c_j} = \destab_i \pm Z_q $.

\begin{enumerate}
\item Let $A\subseteq [n]$ be the largest index set marking anti-commuting destabilizers, i.e.,\\
$ \destab_i$ and $Z_q$ anti-commute iff $i \in A$.

\item Let $(-1)^m Z_q = \prod_{i\in A} \stab_i$. Return outcome $m$.
\end{enumerate}

Gottesman and Aarsonson~\cite{aaronson2008improved} describe in detail how this CHP algorithm functions on the level of the bits of the tableau.

\paragraph{Innovations introduced by \stim simulator \cite{stim}.} 

The state-of-the-art tool for simulating measurements in stabilizer circuits is \stim \cite{stim} simulator.
\stim resolves measurements using the procedure detailed in \cref{alg:stim}. The algorithm builds and extends on the ideas from CHP \cite{aaronson2008improved}. We briefly outline the innovations of \stim that are most relevant to our GPU adaptation.


Instead of explicitly multiplying Pauli generators as in CHP, \stim replaces these row multiplications with conditional CX injections, which implement the same row-addition effect in the tableau during GE. These injections are applied only for probabilistic measurements and significantly reduce computational cost. \ourTool adopts the same measurement approach while operating on a data structure tailored for GPU architectures (see \cref{sec:tableauLayout}).

The key steps of \cref{alg:stim} are referenced by their line numbers (e.g., CX injection in \li{l:injectCX}). We use bracket notation to access individual generator rows of the tableau; for example, $\ZS[i]$ denotes the $i$-th stabilizer $Z$-row, and similarly $\XS[i]$, $\XD[i]$, and $\ZD[i]$.

\begin{algorithm}[htp]
	\setstretch{1.15}
	\DontPrintSemicolon
	\caption{Conceptual CPU-based probabilistic measurement in \stim.}
	\label{alg:stim}
	\midsmall
	\SetCommentSty{mycommfont}
	\SetArgSty{}
	\SetKwInput{Input}{Input~}
	\SetKwInput{Output}{Output~}
	\KwIn{$\tableauBlock := (\XD, \XS, \ZD, \ZS, \SignsD, \SignsS),\ q,\ \numQubits$ \tcp*[r]{$\tableauBlock \ (\text{Tableau}),\ q \ (\text{qubit index}),\ \numQubits \ (\text{Number of qubits})$.}}
	\KwOut{$m$ \tcp*[r]{Meaurement outcome}}

	\Function{\upshape{\routine{ProbabilisticMeasurement}}($\tableauBlock, \observable, \numQubits$)}{
        $p = \min\bigl(\{\,i\in [n] \mid \XS[i, q]=1\} \cup \{-1\}\bigr)$
        \label{l:findMinPivot} \tcp*[r]{Find a stabilizer $\pivot$ that anti-commutes with $\Z[q]$}
		\If(\tcp*[f]{Collapsing into \Z is needed.}){$\pivot \neq -1$}{
			${\GE}(\tableauBlock, \pivot, \observable, \numQubits)$\\             
            ${\swapAntiComm}(\tableauBlock, \pivot, \observable, \numQubits)$\;
			\tcp{If random measurement differs from observable's sign, invert signs by injecting an \X gate.}
			\If{$\SignsS[\observable] \neq \routine{random}(0,\, 1)$}{
                ${\injectX}(\,\SignsD,\, \ZD[\pivot]\,)$,\,\,
                ${\injectX}(\,\SignsS,\, \XD[\pivot]\,)$\;
			}
		}
        \textbf{return} $m \gets \SignsS[\observable]$
	}
    \tcc{Perform GE over stabilizer generators that anti-commute with $\observable$.}
	\Function{\upshape{\GE}($\tableauBlock, \pivot, \observable, \numQubits$)}{
		\For(\tcp*[f]{Targets must follow a strict order in $[\pivot+1, \numQubits-1]$}){$\target = \pivot+1 \ \KwTo\ \numQubits-1$}{ 
			\label{l:forTargets}
			\If{$\XS[\target, \observable] = 1$}{ \label{l:checkBeforeinjectingCX}
			${\injectCX}(\,\SignsD,\, \ZS[\pivot],\, \ZD[\pivot],\, \ZS[\target],\, \ZD[\target]\,)$, \,
            ${\injectCX}(\,\SignsS,\, \XS[\pivot],\, \XD[\pivot],\, \XS[\target],\, \XD[\target]\,)$\;
			}
		}
	}
	\tcc{Swap anti-commuting stabilizer generator for one that commutes with $\observable$.}
	\Function{\upshape{\swapAntiComm}($\tableauBlock, \pivot, \observable, \numQubits$)}{
		\If(\tcp*[f]{$\Y$ case.}){$\XD[\pivot, \observable] = 1$}{
			${\injectInvP}(\,\SignsD,\, \ZS[\pivot],\, \ZD[\pivot]\,)$, \, 
			${\injectInvP}(\,\SignsS,\, \XS[\pivot],\, \XD[\pivot]\,)$\; 
			
		}
		\Else(\tcp*[f]{$\X$ case.}){
			${\injectHad}(\,\SignsD,\, \ZS[\pivot],\, \ZD[\pivot]\,)$, \,
			${\injectHad}(\,\SignsS,\, \XS[\pivot],\, \XD[\pivot]\,)$\; 
		
		}
	}
\tcc{\x: stabilizer row, \z: destabilizer row, \sign: sign vector}
\lFunction{\upshape{\injectCX}{($\sign, \xc, \zc, \xt, \zt$)}}{ \label{l:injectCX}
        $\bigl(\sign,\ \xt,\ {\zc}\bigr)
        \gets \bigl(\sign\oplus\bigl(\xc\wedge\zt\wedge\neg({\zc}\oplus\xt)\bigr),\ \xt\oplus\xc,\ {\zc\oplus\zt}\bigr)$ \label{l:injectCXUpdate}
} 
	\lFunction{\upshape{\injectInvP}($\sign, \x, \z$)}{  \label{l:injectInvP}
        \quad\quad\quad
        $\bigl(\sign,\ \x,\ \z\bigr)
        \,\,\,\,\, \gets \bigl(\sign\oplus(\x\land\neg\z),\ \x \oplus \z,\ \z\bigr)$
	}
	\lFunction{\upshape{\injectHad}($\sign, \x, \z$)}{  \label{l:injectHad}
		\quad\quad\quad\quad\,
        $\bigl(\sign,\ \x,\ \z\bigr)
                \,\,\,\,\,\gets \bigl(\sign\oplus(\x\land\z),\ \z,\ \x\bigr)$
	}
    \lFunction{\upshape{\injectX}($\sign, \z$)}{  \label{l:injectx}
		\quad\quad\quad\quad\quad\,
        $\bigl(\sign,\ \z\bigr)   \,\,\,\,\,\,\,\,\,\,\,\,\,\gets \bigl(\sign\oplus\z,\ \z\bigr)$
	}
	
\end{algorithm}

\stim introduces the following implementation-level optimizations over CHP \cite{aaronson2008improved}.

\begin{itemize}
	
	\item It uses a Cache-friendly, bit-packed layout with 256-bit SIMD instructions where the tableau is stored in a bit-packed format that maximizes contiguous memory access. Tight loops apply 256-bit AVX2 instructions to entire rows or columns in one go, reducing the per-qubit Clifford-gate overhead.
	
	\item 
    It exploits \emph{Pauli frames}, a classical record of bit-flip and phase-flip corrections for each qubit.  
    Pauli frames can be used for both the simulation of errors and the randomness caused by measurements. Instead of re-updating the tableau as a whole for each error event or probabilistic occurrence, the tableau is calculated once (reference shot) and the appropriate bit- and phase-flips are propagated by Clifford conjugation. 
    
    

\end{itemize}
\subsection{GPU Programming}\label{sec:GPU}

CUDA~\cite{nvidia} is a programming interface that enables general-purpose programming for a GPU. It has been developed and continues to be maintained by NVIDIA since 2007.  In this work, we use CUDA with C\code{++}. Therefore, we use CUDA terminology when we refer to thread and memory hierarchies. 

\begin{figure}[htp]
	\centering
	\adjustbox{max width=0.8\linewidth} {
		\includegraphics[width=\textwidth]{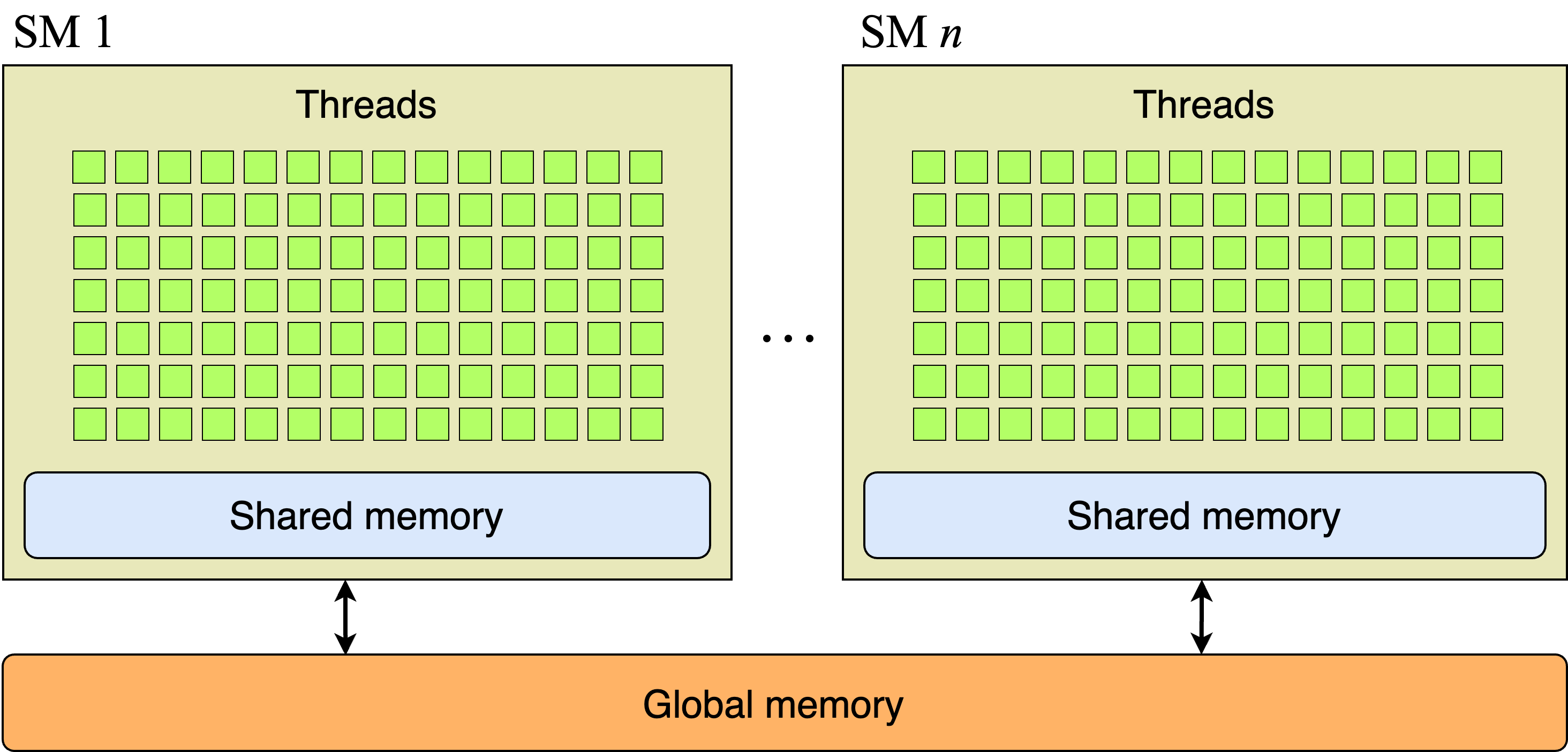}
	}
	\caption{GPU architecture \cite{hitchHike,multiGPU}.}
	\label{fig:gpu_architecture}
\end{figure}

\Cref{fig:gpu_architecture} gives an overview of a GPU architecture.
A GPU consists of a finite number of \emph{streaming multiprocessors} (SM), each containing hundreds of \emph{cores}. For instance, the RTX 4090, which we used for this work, has 128 SMs containing 16,384 cores in total. A GPU computation can be launched in a program by the \emph{host} (CPU side of a program), which calls a GPU function called a \emph{kernel}, executed by the \emph{device} (GPU side of a program).

When a kernel is called, the number of threads needed to execute it is specified. According to the CUDA paradigm \cite{nvidia}, these threads are partitioned into thread \emph{blocks}, i.e., 3-dimensional (3D) vectors grouping threads up to 1,024. Each thread block is assigned to an SM. All threads together form a 3D \emph{grid} where threads and blocks can be indexed by a 1D, 2D, or 3D unique identifier (ID) accessible within the kernel. This ID works similarly as the (\td{x}, \td{y}, \td{z}) coordinates in 3D space. 
With this ID, different threads in the same block can process multidimensional data (e.g., $\X$ and $\Z$ matrices). 

Within each block, threads are further grouped into fixed-size units of 32 threads called \emph{warps}, which are the fundamental units of scheduling and execution on the GPU. All threads in a warp execute the same instruction in lockstep. If threads within a warp follow different control-flow paths due to conditional branches (\emph{warp divergence}), execution is serialized across the different paths, causing stalls and degrading overall kernel performance. Avoiding warp divergence is therefore critical for achieving high throughput on GPUs.

Constant memory 
Shared memory is on-chip SRAM visible only to threads within the same block.  Each SM provides a limited amount (tens of kilobytes) of shared memory; it has much lower latency than global memory and is ideal for inter‐thread communication and tile‐based algorithms.  
Registers are the fastest storage, private to each thread and used for local variables.  Overuse of registers can limit occupancy, so we balance register usage against shared-memory needs.

\myparagraph{Memory Hierarchy.}
Concerning the memory hierarchy, a GPU has multiple types of memory:
\begin{itemize}
	\item[$\bullet$] \textit{Global memory} is the large, high-bandwidth DRAM that both CPU and GPU threads can access.  It has relatively high latency, so good performance requires coalesced, aligned accesses.  Memory accesses are coalesced when consecutive threads access consecutive memory addresses, allowing the hardware to combine them into fewer transactions.
	\item[$\bullet$] \textit{Constant memory} is a small, read-only cache for values that do not change during a kernel.  It offers lower latency when all threads read the same address simultaneously.  
	\item[$\bullet$] \textit{Shared memory} is on-chip SRAM visible only to threads within the same block.  Each SM provides a limited amount (tens of kilobytes) of shared memory; it has much lower latency than global memory and is ideal for inter‐thread communication and tile‐based algorithms.  
	\item[$\bullet$] \textit{Registers} are the fastest storage, private to each thread and used for local variables.  Overuse of registers can limit occupancy, therefore we balance register usage against shared-memory needs.
\end{itemize}

Regarding atomicity, a GPU is capable of executing \emph{atomic} operations on both global and shared memory. A GPU \emph{atomic} function typically performs a \emph{read-modify-write} memory operation on one 32-bit or 64-bit word. For example, $\atomicXOR(\var{addr},\; \var{val})$ reads the old value at \var{addr}, XORes with \var{val}, then writes back the new value at \var{addr}.

\subsection{Algorithmic Conventions}

Throughout this article, we follow a consistent notation.  Constants appear in all‐caps, for example \code{MASKS}.  Arrays and lists are written in lower‐case, such as \code{pivots} or \code{prefixes}.  Kernel and host‐side helper routines use the notation \routine{Function}, while local variables inside algorithm descriptions are typeset in an italic font via \var{variable}.  

Functions highlighted in \textcolor{gpu}{violet} are kernels executed on device. Memory qualifiers indicate the scope and latency of storage: data in on‐chip shared memory is marked with \sharedMem, and read‐only constants in the GPU’s constant cache with \constMem.  When we refer to thread dimensions, we use bold italic symbols \td{x}, \td{y}, and \td{z}.  In particular, the thread’s block‐local index in the \td{x} dimension is written as \tx.  

All parallel loops over a range of indices employ the grid‐stride pattern.  For example, we express grid-stride in \td{x}-dimension as `\textbf{for all} $\tidx \in [ 0, N -1 ]$ \textbf{do in parallel},' meaning that each thread begins with $\tidx=\tx$, processes that element, then increments $\tidx$ by the total number of threads and repeats until $\tidx \ge N$.  This ensures that any range larger than the number of available threads is still fully covered.  

%% file: challenges.tex
\section{Key Challenges for Parallel Measurement and Approach}

Our objective is to efficiently parallelize {\stim}’s measurement algorithm to realize full simulation of stabilizer circuits on GPUs. To achieve this, we first analyze the access patterns in {\stim}’s algorithm to identify parallelization opportunities. Before presenting our GPU-based measurement algorithm, we outline the key challenges and describe a data layout designed to maximize coalesced memory access and fully exploit modern GPU architectures.
\subsection{Identifying Probabilistic Measurements}

For each measurement operator in the input circuit, we must first identify a stabilizer generator that anti-commutes with the measurement (see \li{l:findMinPivot} in \cref{alg:stim}). Although probabilistic measurements lead to different post measurement states, we can precompute, independently of the measurement outcomes, which observable qubits will be probabilistic.

Prior to executing measurements, we launch a two-dimensional GPU kernel in which one dimension iterates over measurement observables and the other searches for at least one anti-commuting stabilizer generator, whose index we refer to as the pivot $\pivot$.
This precomputation allows us to skip deterministic measurements, as in \stim, while still correctly handling measurements that become probabilistic after earlier state updates.

Once a probabilistic measurement is performed and the quantum state is modified, the pivot must be recomputed with respect to the updated tableau. We address this efficiently using stream compaction \cite{streamCompact}, as described in the following subsections.

\subsection{Gaussian Elimination by Injecting CX Gates}

One might ask why we parallelize the collapse step of \stim instead of the CHP update in \cref{sec:measurements}. In the CHP rule $\stab_i \leftarrow \stab_i\stab_{\tilde p}$ and $\destab_i \leftarrow \destab_i\destab_{\tilde p}$, the pivot generator $\stab_{\tilde p}$ serves only as a fixed multiplier and is not overwritten until all anti-commuting rows have been updated. The updates for different indices $i$ are thus independent, so CHP can in principle be parallelized directly. Doing so is wasteful, however, because each update multiplies full Pauli generators and recomputes their signs bit by bit. \stim avoids this cost by replacing the row multiplications with word-level conditional CX injections (\cref{alg:stim}), which it applies only for probabilistic measurements. This same optimization is what makes parallelization harder. Instead of a fixed multiplier, the injections now share a single control generator, the pivot, whose $Z$-destabilizer word $\zc$ is updated in place as the targets are processed. We therefore parallelize the more efficient \stim-style algorithm and show how to recover the parallelism that the CHP formulation would otherwise provide.

The main obstacle to parallelizing the Gaussian elimination (GE) step in \stim (Function~\GE{} in \cref{alg:stim}) is a loop-carried dependency across target generators. In the target loop at \li{l:forTargets}, every CX injection shares the
same control generator indexed by $\pivot$, but each invocation of \injectCX updates the control’s $Z$-destabilizer word $\zc$ in place (\li{l:injectCXUpdate}). As a result, the sign update applied to a given target depends on the value of $\zc$ produced by all preceding targets.

This dependency enforces the strict execution order
$\target=\pivot+1,\dots,\numQubits-1$ in \stim and makes naive parallelization incorrect. Even with atomic operations, concurrent threads would observe
inconsistent values of $\zc$, breaking the required read-after-write dependence and producing incorrect sign histories. The challenge is therefore not merely avoiding write conflicts, but preserving the sequential evolution of $\zc$ that subsequent sign updates depend upon.

To expose parallelism, we unfold the sequential updates in \injectCX (\li{l:injectCX} in \cref{alg:stim}). Let $\sign^k$ denote the sign state after processing the
$k$-th target and $\zc^k$ the corresponding control $Z$-destabilizer word. For $k=1,2,\dots,n-\pivot-1$, the updates satisfy
\begin{align}\label{eq:CXSign}
	\sign^k & \gets \sign^{k-1} \oplus
	\bigl(\xc^{k-1} \wedge \zt^{k-1} \wedge
	\neg(\textcolor{blue}{\zc^{k-1}} \oplus \xt^{k-1})\bigr), \\
	\textcolor{blue}{\zc^{\,k}} & \gets
	\textcolor{blue}{\zc^{\,k-1}} \oplus \textcolor{blue}{\zt^{\,k-1}} .
\end{align}
Quantities highlighted in blue denote loop-carried state that evolves across successive targets and enforces the sequential dependency in \li{l:forTargets}. Unfolding the recurrence for $\zc^k$ yields
\begin{equation}\label{eq:zc}
	\zc^k = \zc^0 \oplus \bigotimes_{l=0}^{k-1} \zt^l ,
\end{equation}
which is a classic prefix-XOR (scan) over the sequence $\{\zt^l\}$. Intuitively, each target contributes its $\zt$ word to the evolving control state, and the value of $\zc$ observed by target $k$ is exactly the XOR of all prior contributions.

Once all $\zc^k$ values are computed in parallel via a prefix scan, each target’s sign contribution can be evaluated independently. The accumulated sign updates can then be combined using an XOR reduction, yielding the same final sign state as the sequential target loop. This reformulation eliminates the sequential bottleneck at \li{l:forTargets} and enables correct, fully parallel execution of CX injections on the GPU.

\subsection{Handling Sparse $X$-Stabilizers}

In \stim, a CX injection is required only for stabilizer generators that anti-commute with the measurement observable, as determined by the condition at \li{l:checkBeforeinjectingCX}. In practice, the corresponding $X$-stabilizer rows are often sparse. Naively scanning the full range $[\pivot+1,\numQubits-1]$ for each measurement therefore leads to unnecessary work, irregular memory accesses, and warp divergence on GPUs.

To address this, we first compact all target indices $t$ such that $\X_{S}[t,q]=1$ (i.e., generators that anti-commute with the observable) into a contiguous array before performing GE. This preprocessing step yields several benefits:
\begin{itemize}
	\item[$\checkmark$] \textbf{Reduced warp divergence}: Only targets that actually require
	a CX injection are processed, avoiding conditional branches over non-pivots.
	\item[$\checkmark$] \textbf{Coalesced memory access}: Consecutive threads operate on consecutive target indices, improving memory locality during tableau updates.
	\item[$\checkmark$] \textbf{Lower overhead}: The GE kernels bypass scanning the full $[\pivot+1,\numQubits-1]$ range, which is especially beneficial when stabilizers are sparse.
\end{itemize}

By combining prefix-XOR reformulation of CX injections (\cref{eq:CXSign}–\cref{eq:zc}) with compact indexing of sparse anti-commuting stabilizers, our algorithm preserves the semantics of \stim’s sequential GE while achieving high GPU utilization and substantially improved memory throughput.

%% file: simulation.tex
\section{Parallel Algorithm for Simulation}

In this section, we present the first algorithm for performing stabilizer‐based single-shot simulation in parallel on GPUs. Our parallel algorithm for Many-shot sampling is described separately in \cref{sec:sampling}. We begin with a high‐level overview---from the input circuit all the way to the recorded measurement outcomes---then explain in detail each key component in the subsections that follow.

\subsection{Overview of Our Simulator}

\Cref{alg:parallelSimulation} gives a high‐level view of our GPU‐based stabilizer simulator. It accepts as input a quantum circuit $\circuit$, the number of qubits $\numQubits$, word size $\wordSize$ (i.e., the machine word size), and an initial basis state $\initialState\in\{0,1\}^n$. 
It produces as output the final state tableau $\tableauBlock$ and, if any measurements occur, a bit‐string $\measured\in\{0,1\}^n$ recording the measured outcomes.

\begin{algorithm}[htp]
	\setstretch{1.15}
	\DontPrintSemicolon
	\caption{Top-level overview of parallel simulation on GPUs. Functions highlighted in \textcolor{gpu}{violet} denote GPU kernels.}
	\label{alg:parallelSimulation}
	\midsmall
	\SetCommentSty{mycommfont}
	\SetArgSty{}
	\SetKwInput{Input}{Input~}
	\SetKwInput{Output}{Output~}
	\KwIn{$\circuit,\; \numQubits,\; \wordSize,\; \initialState\in \{0,1\}^{\numQubits}$}
	\KwOut{$\tableauBlock,\; \measured\in \{0,1\}^\numQubits$}

	\tcc{Main procedure for simulation.}
	$\circuitScheduled \gets \scheduleParallelGates$($\circuit$) \label{l:schedule} \tcp*[r]{Schedule parallel gates on host into layers.}
	$\tableauBlock, \numWords \gets \allocate$($2 \numQubits(2\numQubits + 1),\,\wordSize$) \label{l:allocate}	\tcp*[r]{Allocate $2\numQubits(2\numQubits+1)$ bits in $\numWords=\lceil \numQubits/\wordSize\rceil$ words.}
	\begingroup \color{gpu}
	$\tableauBlock \gets {\initTableau}$($\initialState$) \label{l:initTableau} \tcp*[r]{Initialize tableau on device.}
	\endgroup
	\ForAll(\tcp*[f]{For all layers.}){$\circuitWindow \in \circuitScheduled$}{ \label{l:forSimulation} 
		$\circuitWindow_d \gets \copyWindowToDevice$($\circuitWindow$) \label{l:copyWindow} \tcp*[r]{Copy gate layer to device.}
        \If(\tcp*[f]{Does current window have measurement?}){$\hasMeasurements(\circuitWindow)$}{
			$\measureWindow(\tableauBlock, \circuitWindow_d, \measured, \numQubits,\wordSize, \numWords)$ \label{l:measure} \tcp*[r]{Dispatch measurement procedure.}
		}
		\Else {		
            \begingroup \color{gpu}
			${\simulateWindow}$($\tableauBlock, \circuitWindow_d, \numWords$) \label{l:applyGates} \tcp*[r]{Apply gate rules on device.}
			\endgroup
		}
	}
	\tcc{Main procedure for measurement.}
	\Function{\upshape{\measureWindow}($\tableauBlock, \circuitWindow_d, \measured,\numQubits,\wordSize, \numWords$)}{ \label{l:measureBegin}
		\begingroup \color{gpu}
		$\transposed \gets \transpose(\tableauBlock, \numQubits,\wordSize, \numWords)$ \label{l:transpose} \tcp*[r]{In-place row-major transpose.}
		$\allNondet_d \gets \findAllNondetQubits(\transposed, \circuitWindow_d, \numQubits,\wordSize, \numWords)$ \label{l:findAllNondet} \tcp*[r]{Find all probabilistic observables.}
		\endgroup
		$\allNondet \gets \copyListToHost(\allNondet_d)$ \label{l:copyListToHost} \tcp*[r]{Copy $\allNondet_d$ to host container \allNondet.}
		\ForAll(\tcp*[f]{For all (potential) probabilistic cases.}){$\observable \in \allNondet \wedge \observable \neq -1$}{ \label{l:forAllPivots} 
			\begingroup \color{gpu}
			$\pivots \gets \compactPivots(\transposed, \observable, \numQubits,\wordSize, \numWords)$ \label{l:compactPivots} \tcp*[r]{Filter anti-commuting pivots.}
			\endgroup
			$\pivot \gets \copyFirstPivot(\pivots)$ \tcp*[r]{Bring first pivot to host.}
			\If(\tcp*[f]{Collapsing into \Z is needed.}){$\pivot \neq -1$}{ \label{l:checkPivot}
				\begingroup \color{gpu}
				${\paraGE}(\transposed, \pivots, \numWords)$ \label{l:runGE} \tcp*[r]{Mutates $\transposed$ in place.}
				${\paraSwap}(\transposed, \pivot, \qubit, \numQubits, \numWords, \wordSize)$\;\label{l:runSwap}
				\endgroup
				$\sign_\observable \gets \routine{CopySignBitToHost}(\observable)$ \tcp*[r]{Copy \observable's sign bit to host.}
				\If{$\sign_\observable \neq \random(0, 1)$}{ \label{l:checkSign}
					\begingroup \color{gpu}
					${\paraX}(\transposed, \pivot, \numQubits, \numWords)$ \label{l:runX} \tcp*[r]{Mutates the sign words.}
					\endgroup
				}
			}
		}
		\begingroup \color{gpu}
		$\measured \gets \copySigns(\Signs)$ \label{l:copySigns} \tcp*[r]{Unpack stabilizer signs in $\Signs[\numWords \to \columnLen]$ to $\measured$.}
		$\tableauBlock \gets \transpose(\transposed, \numQubits,\wordSize, \numWords)$ \label{l:transposeBack} \tcp*[r]{In-place column-major transpose (restores column-major).}
		\endgroup
	}
\end{algorithm}

At \li{l:schedule}, We first decompose the input circuit into a sequence of windows \circuitScheduled. Each window contains either a single measurement or a group of parallel unitary gates acting on disjoint qubits, allowing  them to be applied simultaneously. The construction of these windows is described in \cref{sec:parallelGates}. At \li{l:allocate}, we allocate space for the $2\numQubits\times(2\numQubits+1)$ tableau, packed into $\wordSize$‐bit words and with $\numWords=\lceil \numQubits/\wordSize\rceil$ words per column. \Cref{sec:tableauLayout} describes this layout in detail. Then, at \li{l:initTableau}, the tableau is initialized to the basis state \initialState on the device.

At \li{l:copyWindow}, each scheduled window $\circuitWindow \in \circuitScheduled$ is copied to device memory as $\circuitWindow_d$ for use by subsequent GPU kernels. We then check whether the current window contains a measurement. If so, we invoke the host‐driven \measureWindow procedure (\li{l:measure}); otherwise, we launch the \simulateWindow kernel to apply all Clifford gates in parallel (\lis{l:copyWindow}{l:applyGates}). \Cref{sec:applyGates} describes the implementation of \simulateWindow in detail.

When a window contains a measurement, the \measureWindow procedure (\lis{l:measureBegin}{l:transposeBack}) is executed. First, the tableau is transposed in place from column-major to row-major order using the \transpose kernel (\li{l:transpose}), so that each generator’s qubit bits are contiguous in memory. Next, we identify qubits that will collapse non-deterministically by launching the \findAllNondetQubits kernel (\li{l:findAllNondet}), which collects all observables whose generators anti-commute. The resulting device array $\allNondet_d$ is copied back to the host into \allNondet (\li{l:copyListToHost}), allowing the CPU to iterate over each candidate qubit sequentially.

For each nondeterministic observable qubit $\observable \in \allNondet$, the GPU kernel \compactPivots filters and compacts the indices of anti-commuting generators into a dense array $\pivots \in [-1,\numQubits-1]^{\numQubits}$. The host then reads the first pivot $\pivot=\pivots[0]$. If $\pivot \neq -1$, the observable remains nondeterministic after prior tableau updates, and the GPU executes \paraGE at \li{l:runGE}. This kernel implements the three-pass prefix-XOR elimination described in \cref{sec:GEPrefix}, zeroing out all anti-commuting generators. Afterward, \paraSwap is invoked at \li{l:runSwap} to swap in a commuting generator at index $\pivot$.

Finally, the host reads back the sign bit corresponding to qubit $\observable$, compares it with a freshly generated random bit, and if they differ, triggers the \paraX kernel to inject an $\X$ gate and flip the measurement outcome.

Once all nondeterministic qubits have been collapsed, the GPU unpacks the final stabilizer sign bits into the output array $\measured \in \{0,1\}^{\numQubits}$ and applies \transpose again to restore the column-major layout of $\tableauBlock$. The procedure returns the updated tableau and the measured bit string, after which the simulation loop at \li{l:forSimulation} continues with the remaining windows.
\subsection{Parallel Gate and Measurement Scheduling on CPU}\label{sec:parallelGates}

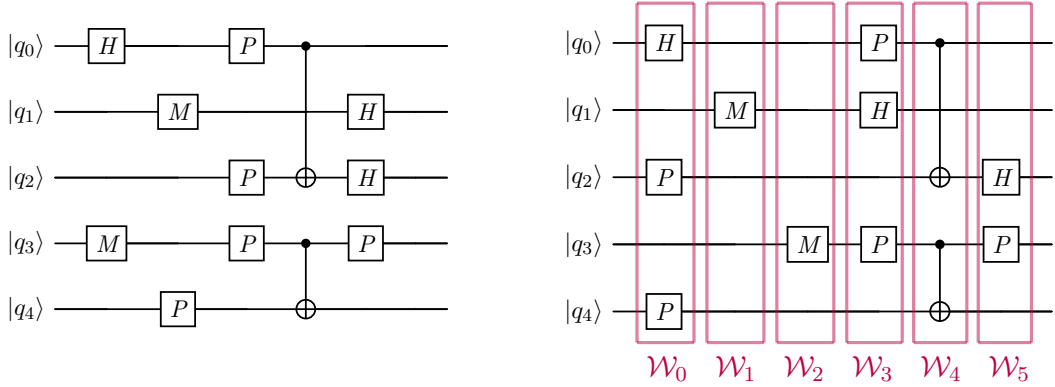
\begin{figure}[htp]
	\input{circuit.tex}
	\caption{Input circuit on the left. Scheduled circuit on the right.}
	\label{fig:circuit}
\end{figure}

We first partition the input circuit into \emph{maximal windows} on the host, where each window is the largest possible set of mutually independent gates (cf.\ definitions \ref{def:dependency}–\ref{def:MaximalWindow}). By doing so, we expose fine-grained parallelism before any GPU work begins. Further, we treat measurement operations as scheduling barriers: all unitary gates are
advanced as early as possible into maximal windows, while measurements are delayed and
executed only after the completion of their preceding window (see \cref{fig:circuit}).

\begin{definition}[Gate Dependency Relation]\label{def:dependency} 
	Given a circuit\, \circuit, 
	let $\newGate, \newGate'\in\circuit$ be two gates.  We say $\newGate\preceq \newGate'$ iff $\newGate'$ is causally dependent on $\newGate$.  This relation is a partial order over\, $\circuit$.
\end{definition}

\begin{definition}[Parallel Gates]\label{def:window} 
	A subset $W\subseteq\circuit$ is a set of \emph{parallel gates} (a \emph{window}) if 
	\[
	\forall\,\newGate,\newGate'\in W,\quad \neg\bigl(\newGate\preceq \newGate'\bigr)\ \land\ \neg\bigl(\newGate'\preceq \newGate\bigr).
	\]
\end{definition}

\begin{definition}[Maximal Window]\label{def:MaximalWindow} 
	A window $ W \subseteq \mathcal{C} $ is \emph{maximal} if there exists no strictly
	larger window $ W' \subseteq \mathcal{C} $ and
	$ W' $ contains up to one measurement operator.
\end{definition}

Henceforth, \emph{window} will always mean a maximal window, denoted \circuitWindow.

The above concepts underpin the \scheduleParallelGates routine in \cref{alg:parallelSimulation}. Starting from the input-circuit queue (see left of \cref{fig:circuit}), we iteratively:

\begin{enumerate}
	\item Advance gates from the queue while preserving dependencies, until either $\numQubits$ independent gates are collected or no further independent gates can be advanced. Measurement operations are treated as scheduling barriers and are therefore delayed and appended to the next available window.
	\item Group the collected gates into the next maximal window \circuitWindow.
	\item Repeat until the queue is empty.
\end{enumerate}

This produces the scheduled list \circuitScheduled (right of \cref{fig:circuit}), whose length scales with the original circuit depth. Each window can then be dispatched as a single batch to the GPU, maximizing parallel gate execution. Finally, measurements act as scheduling barriers: they cannot be included in a unitary
window and are therefore deferred and appended to the next available window after all
unitary gates in their causal past have been scheduled.

\paragraph{Remark on Sampling Mode.}
The above scheduling applies to single-shot simulation, where
measurements act as barriers because the tableau must collapse
after each non-deterministic outcome.

In contrast, for many-shot sampling (see \cref{sec:sampling}),
we do not collapse the tableau into the $Z$-basis.
Instead, we use Pauli frames to track measurement randomness.
As a result, all measurement operators within a layer can be
merged into a single measurement window and processed fully
in parallel across shots.
This removes the sequential barrier present in single-shot
simulation and enables constant-time sampling per qubit.
\subsection{Tableau Layout and Indexing}\label{sec:tableauLayout}

Upon designing the data structure for storing and updating the extended tableau on a GPU, several key considerations were made to ensure high efficiency of both the GPU memory and execution performance. GPUs operate on aligned memory, meaning data are stored and accessed in \emph{words} (e.g., 8, 32, or 64 bits). This allows multiple bits to be processed simultaneously with a single instruction (e.g. XOR), which is crucial for fast tableau evolution. 

Let $\wordSize$ be the GPU word size in bits (here $\wordSize=64$). We encode the extended tableau $\tableauBlock$ as a 2D array of $\lceil \numWords = \numQubits/w \rceil$ rows by $(2\numQubits+1)$ columns of words: 
$$
\tableauBlock\in\{0,1\}^{2\numQubits \times (2\numQubits+1)}
\;\cong\;
[0,2^\wordSize-1]^{2\numWords \times (2\numQubits+1)}
\quad\text{(packed $\wordSize$-bit words).}
$$

\noindent where each entry is a packed \wordSize-word. The first $2\numQubits$ columns encode the \X and \Z components of destabilizer and stabilizer generators, while the final column stores the sign vector \Signs, which encodes the global phase of each generator in column-major layout. Each word encodes \wordSize bits of a single tableau column across \wordSize Pauli generators.

Transposing \tableauBlock for measurement kernels switches to a row-major layout of $2\numWords$ rows by $2\numQubits + 1$ columns of words, so that each word packs $\wordSize$ qubit bits across $2\numQubits$ generators. In contrast to the column-major layout, where words pack generator bits for a single qubit, the row-major layout packs qubit bits for a single generator.
The size of the sign vector \Signs remains unchanged and consists of  $2\numWords$ words; however, in the transposed layout, each word encodes the sign bits of \wordSize qubits. After performing a measurement on an observable \observable, the corresponding sign bit encodes the measurement outcome for that qubit.

While \cref{alg:stim} presents the tableau in terms of explicit stabilizer ($\XS, \ZS$) and destabilizer ($\XD, \ZD$) blocks for clarity, the GPU implementation operates on a single packed 2D array of words. In this representation, both stabilizer and destabilizer rows are stored contiguously and accessed via linear word indices. Offsets are used to distinguish between the destabilizer and stabilizer halves of the tableau. This linear indexing enables coalesced memory access and simplifies parallel kernel design while remaining equivalent to the block-based formulation.

\begin{figure*}[htp]
	\centering
	\adjustbox{max width=0.9\linewidth} {
		\includegraphics[width=\linewidth]{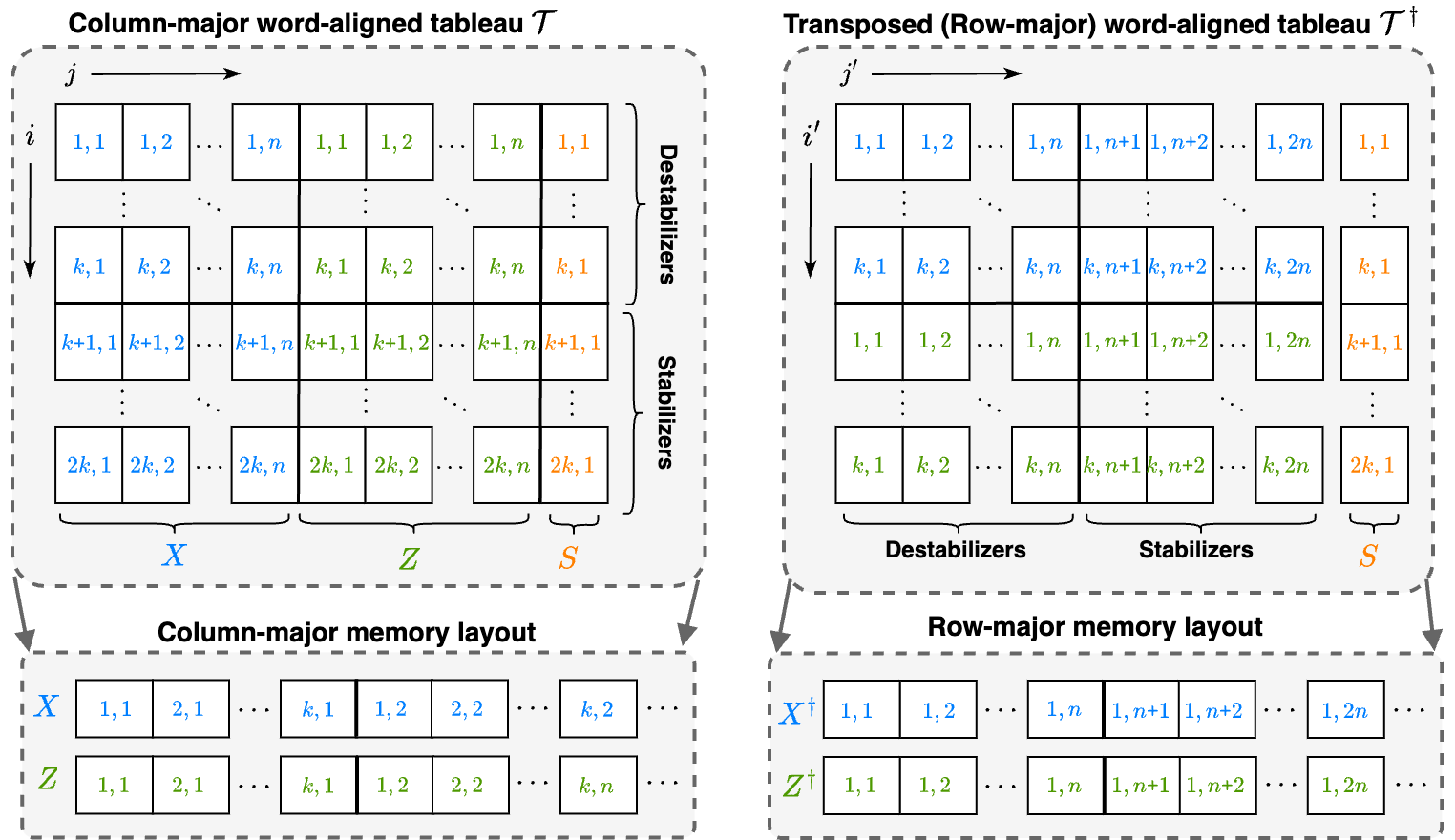}
	}
	\caption{Logical and physical formats of a word-aligned tableau.}
	\label{fig:tableau}
\end{figure*}

\paragraph{Memory Alignment.}  
We pack bits into 64-bit words—the maximum GPU’s native width—to maximize bitwise throughput. In our benchmarks, 64-bit words outperformed smaller sizes by updating more generators per instruction and reducing total memory traffic.

\paragraph{Memory Layout.}
Given the characteristics of the tableau evolution when applying Clifford gates, it is logical to store the $\X$ and $\Z$ in a column-major format (\cref{fig:tableau}, left). This ensures each column (flattening destabilizers and stabilizers) can be accessed contiguously in memory, improving locality during tableau evolution. 
Measurement routines perform pivot searches and Gaussian elimination on different generators, per measured qubit, therefore we use a row-major layout (\cref{fig:tableau}, right) to coalesce reads across generators. The sign vector \Signs is stored separately and remains word-aligned; it is updated consistently with both layouts and is not affected by the transpose.

\paragraph{Tableau Density.}
As circuit depth increases, the tableau densifies and probabilistic measurements become more frequent. Our dual-layout strategy ensures both gate application and measurement kernels maintain coalesced access even under high tableau density.
\subsection{Tableau Evolution under Clifford Gates}\label{sec:applyGates}

The kernel in \cref{alg:buildTableau} implements tableau evolution by applying parallel gates to the relevant tableau columns.
Within this process, threads in the \td{y}-dimension (\li{ll:forWords}) are responsible for fetching new words per column, whereas threads in the \td{x}-dimension (\li{ll:forGates}) handle the retrieval of parallel gates. The variables $\tidy$ and $\tidx$ hold the global thread ID in \td{y}- and \td{x}-dimension, respectively.
To branch over all supported gates by our algorithm, a switch statement is used at \li{ll:switch}. Here, we only show the \Had gate, but many more Clifford gates are directly supported without decomposition.
At \lis{ll:H}{ll:otherGates}, we implement the update rules executed for each gate modifying both the words encoding the Pauli strings and their signs. Recall that $\tableauBlock$ is stored in a column-major array when applying Clifford gates; thus, given a qubit $\qubit$, and the column length \columnLen, a column in the $\X$ or $\Z$ matrix can be accessed in position ($\qubit \cdot \columnLen$). To access words per column in parallel, the thread ID ($\tidy$) is used to offset the column index. Note that in this work, the tableau has been extended to include destabilizers; the algorithms below treat tableau evolution for both stabilizers and destabilizers, unlike \cite{paraEC} where only stabilizers were evolved.

The subscript $q$ in the gate's type refers to the connected wire or qubit. Qubit values are assumed to start from 0 to $\numQubits - 1$ to align with the array index.
At \li{ll:SUpdate}, the signs are calculated and temporarily stored in shared memory (denoted as \shared). The variable $\shared[\tx]$ holds the thread-local sign per thread block in the \td{x}-dimension. To ensure each thread block has written its signs to \shared before proceeding, we need to synchronize at \li{ll:syncBlock}. Once synchronized, these signs need to be collapsed into a final result, which will be used to update the global sign in the tableau. This collapsing process is handled by a specialized procedure, \collapseSigns , which we will describe in more detail next. 
The main tableau evolution to the tableau occurs at \li{ll:XZUpdate}, where the words in $\X$ and $\Z$ (representing the Pauli strings) are swapped in global memory.

\begin{algorithm}[t]
	\setstretch{1.15}
	\DontPrintSemicolon
	\caption{Tableau Evolution on GPU.}
	\label{alg:buildTableau}
	\midsmall
	\SetCommentSty{mycommfont}
	\SetArgSty{}
	\SetKwInput{Input}{Input~}
	\SetKwInput{Output}{Output~}
	\KwIn{$\tableauBlock := (X, Z, S),\; \circuitWindow_d,\; \numQubits,\; \wordSize,\; \numWords$}
	\KwOut{$\tableauBlock$}
	\begingroup \color{gpu}
	\Kernel{\upshape{\simulateWindow}($\tableauBlock := (\X, \Z, \Signs), \circuitWindow_d, \numWords$)}{  \label{ll:simWindowKernel} 
		\Forpary(\tcp*[f]{$\tidy$ fetches a new word per column.}){$0,\; \columnLen-1$}{ \label{ll:forWords}  
			\Forparx(\tcp*[f]{$\tidx$ fetches a new gate.}){$0,\; |\circuitWindow_d|-1$}{ \label{ll:forGates} 
				\Switch(\tcp*[f]{Switch over gate type.}){$\routine{Type}(\circuitWindow_d[\tidx])$}{ \label{ll:switch} 
					\Case(\tcp*[f]{$q\in \{ 0 , \cdots ,n-1 \}$.}){$\Had_\qubit$}{ \label{ll:H} 
						$\wordidx \gets \qubit \cdot \columnLen + \tidy$ \tcp*[r]{Linear word index in column-major $X,Z$.}
						$\sharedMem\ \shared[\tx] \gets \X[\wordidx] \wedge \Z[\wordidx]$ \tcp*[r]{Load local signs into shared memory.} \label{ll:SUpdate} 
						{{\Swap}}($\X[\wordidx], \Z[\wordidx]$)\; \label{ll:XZUpdate}
					}
					\lCase(\tcp*[f]{Other Clifford gates like \CX, \fontFormat{CY}, etc.}){\dots}{\dots} \label{ll:otherGates} 
				}
			}
			\syncThreads{(\:)} \label{ll:syncBlock} \tcp*[r]{Synchronize shared memory.}
			{{\collapseSigns}}($\Signs[\tidy], \shared$) \label{ll:collapseSigns} \tcp*[r]{Collapse signs in shared memory.}
		}
	}
	\tcc{A function to collapse all thread-local signs using tree structure.}
	\Devfunction{\upshape{\collapseSigns}($\sign, \shared$)}{  \label{ll:collapseFunction}
    \tcc{\textbf{Notation.} $\blockDim =$ threads per block along $x$; $\tx \in [0,\,B_x-1]$.}
		\For(\tcp*[f]{$\text{Loop over } \log(\blockDim) \text{ steps}$.}){$\blockidx := \blockDim / 2, \blockDim / 4, \ldots, 1$} { \label{ll:forShared} 
			\lIf{$\tx < \blockidx$} {
				$\shared[\tx] \gets \shared[\tx] \oplus \shared[\tx + \blockidx]$ \label{ll:collapse} 
			}
			\syncThreads{(\:)}\; \label{ll:syncCollapse}
		}
		\If(\tcp*[f]{Collapse root $\shared[0]$ with old signs-word $\sign$.}){$\tx = 0 \wedge \shared[0] = 1$} { 
			\atomicXOR{($\sign, \shared[0]$)} \label{ll:atomic}  
		}
	}
	\endgroup
\end{algorithm}

\paragraph{Signs Collapsing.}
In our initial attempt to parallelize the signs collapsing process, we employed two kernels. The first kernel handled the tableau evolution and the local sign calculations, storing the results in global memory. The second kernel read these local signs from global memory and performed the collapsing using the atomic operation \atomicXOR{}. Unfortunately, this trial proved to be considerably slow due to the excessive global memory accesses (approximately $\columnLen\cdot \numQubits$) and the high usage of atomic instructions ($\columnLen\cdot |\circuitWindow_d|$). 

To remedy the global memory access issue, we made a second attempt. In this trial, we eliminated the temporary storage of local signs in global memory. Instead, we collapsed all local signs atomically on-the-fly during the tableau evolution within the whole window. While this approach saved roughly ($\columnLen\cdot \numQubits$) memory accesses compared to the initial attempt, the number of atomic operations remained a bottleneck. This is where our tree-like approach comes into play. The \collapseSigns routine in \cref{alg:buildTableau} effectively reduces the number of atomics by collapsing signs per thread block using a bottom-up binary tree. The downside is the excessive usage of global memory during the tree build-up, which we mitigate by utilizing the fast on-chip shared memory. 

The loop at \li{ll:forShared} performs the collapsing using the XOR operator in shared memory by sweeping the binary tree in a bottom-up manner. Initially, $\blockidx$ is set to $\blockDim / 2$, where $\blockDim = 2^l$ and $l \in \{1, 2, \ldots, 10\}$. 
At \li{ll:collapse}, we XOR in parallel the first half in $\shared[\tx]$ with the second half $\shared[\tx + \blockidx]$. Afterward, threads are synchronized and $\blockidx$ is halved until only one element remains at $\shared[0]$ (resembles the root of the tree), which holds the collapsed block-local sign. Finally, at \li{ll:atomic}, we collapse the block-local signs stored at $\shared[0]$ atomically with the old global signs at $\Signs[\tidy]$.
This approach introduces an additional $log(\blockDim)$ steps to jump between tree levels (\li{ll:forShared}). Albeit, the number of atomic operations is reduced from $\columnLen\cdot |\circuitWindow_d|$ to $\columnLen\cdot (|\circuitWindow_d| / \blockDim)$.

\begin{figure}[!t]
	\centering
	\adjustbox{max width=.8\linewidth} {
		\includegraphics[width=\linewidth]{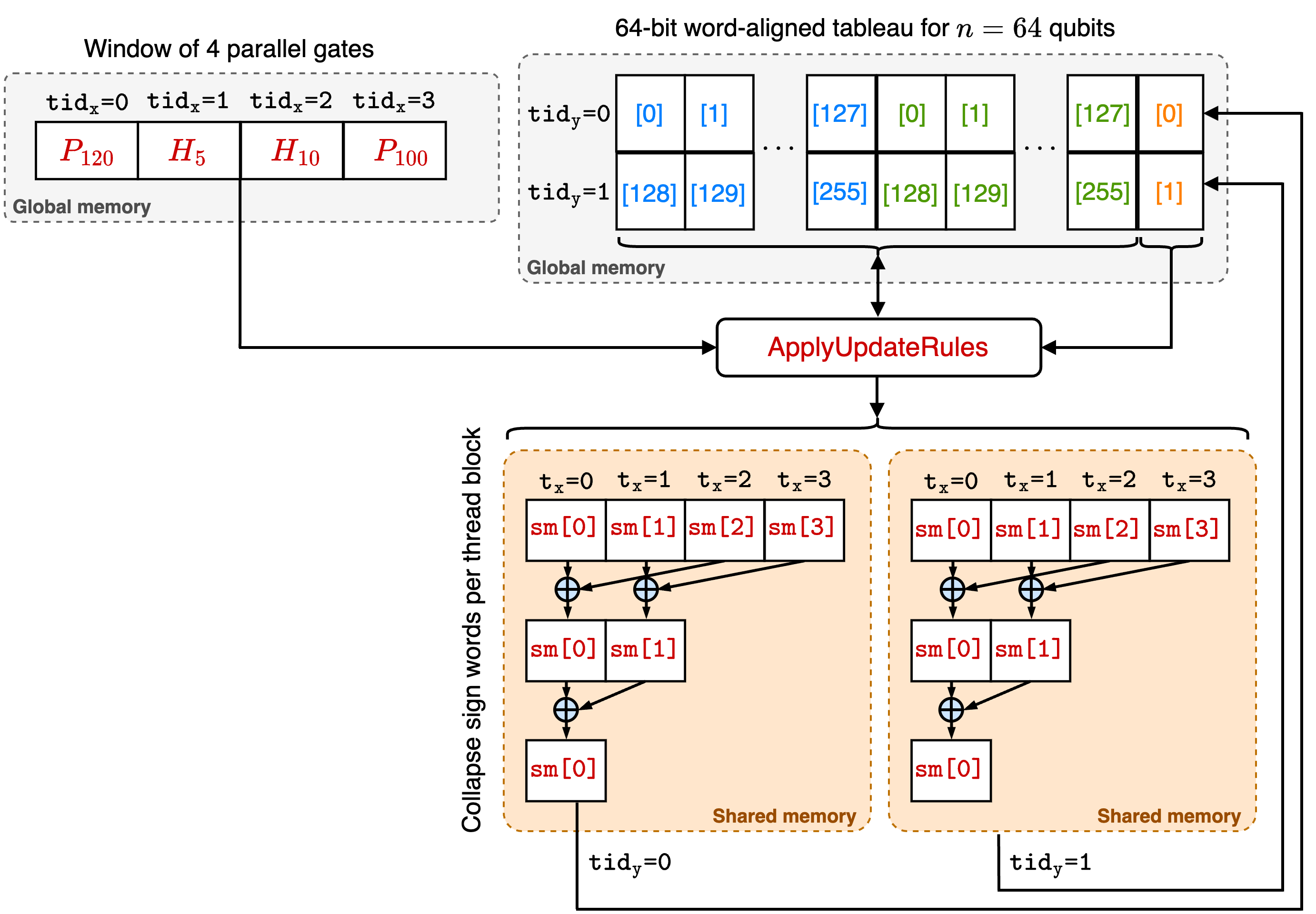}
	}
	\caption{Running example of \cref{alg:buildTableau} on 128-qubit system.}
	\label{fig:tableauUpdate}
\end{figure}

\Cref{fig:tableauUpdate} depicts a running example of \cref{alg:buildTableau} executed on the window $\circuitWindow = \{\Phase_{120}, \Had_{5}, \Had_{10}, \Phase_{100}\}$ with ($\numQubits = 64$). For that, given a word size of 64 bits, we require a tableau of dimensions ($\numWords = 64 / 64 = 1,\ \columnLen \cdot (2\numQubits + 1) = 2 \cdot (128 + 1) = 258$ words). As a result, the number of threads in the \td{y}-dimension is 2, and 4 in the \td{x}-dimension. In that case, one thread block for both dimensions is sufficient (e.g. \blockDim = 4, \yblockDim = 2). 
Once the collapsing process is complete and threads are synchronized per block, the root of the tree, denoted as $\shared[0]$, holds the final collapsed word. If multiple thread blocks are involved in the \td{x}-dimension, their results are combined using an \atomicXOR{} operation (as shown in \li{ll:atomic}). In this specific example, as we only have one thread block in the \td{x}-dimension, $\shared[0]$ can be directly written to $\Signs[\tidy]$ non-atomically.

\paragraph{Complexity.}\label{sec:complexity}
The total work performed by \cref{alg:buildTableau} is dominated by the nested loops over $\numWords$ words and $|\circuitWindow_d|$ gates, giving $\bigO(\numWords\times|\circuitWindow_d|)$ operations. In the worst case, both $\numWords$ and $|\circuitWindow_d|$ grow as $\bigO(\numQubits)$, so the total work is $\bigO(\numQubits^2)$.  These operations are spread across up to $\maxThreads$ hardware threads (the product of grid dimensions in $\td{x}$ and $\td{y}$), yielding an ideal parallel time of $\bigO(\frac{\numQubits^2}{\maxThreads})$.
In addition, each block executes the $\collapseSigns$ routine (\lis{ll:forShared}{ll:syncCollapse}), which is a binary‐tree XOR over $\blockDim$ threads.  That loop takes $O(\log\blockDim)$ sequential steps within each block.  Because $\blockDim$ is capped by the hardware’s maximum threads per block rather than by $\numQubits$ \cite{nvidia}, the overall worst‐case parallel span becomes
$\bigO(\frac{\numQubits^2}{\maxThreads} + \log\blockDim)$.

\subsection{In‐Place Transpose}\label{sec:transpose}

Before performing any measurement-related operations, we transpose the tableau layout from column-major to row-major order. Measurement procedures, such as collapsing a qubit into the $\Z$-basis, require applying CX gates to anti-commuting generators \emph{across all qubits}. To ensure that each generator’s words remain contiguous and coalesced in memory during these operations, we store the tableau in row-major order and therefore must transpose it from its natural column-major layout.

\begin{algorithm}[!t]
	\setstretch{1.15}
	\DontPrintSemicolon
	\caption{Tile‐wise In‐Place Transpose on GPU}
	\label{alg:transpose}
	\midsmall
	\SetCommentSty{mycommfont}
	\SetArgSty{}
	\KwIn{$\tableauBlock:=(X,Z,S),\; \numQubits,\; \wordSize,\; \numWords$}
	\KwOut{$\transposed$}
	\begingroup \color{gpu}
	\Function{\upshape{\transpose}($\tableauBlock, \numQubits,\wordSize, \numWords$) $\to \transposed$}{ \label{trans:transposeFn}
		$ {\shuffleTiles}(\X,\Z,\numQubits,\wordSize, \numWords)$
		${\swapTiles}(X, Z,\numWords)$\;
		return $\transposed$ \tcp*[f]{now row-major in memory (denoted $\transposed$); $S$ unchanged. }
	}
	\Kernel{\upshape{\shuffleTiles}($\X,\Z,\numQubits,\wordSize, \numWords$)}{ \label{trans:shuffleKernel}
		$\tabMatrix \gets (\tidz = 0)\ ?\ \X : \Z$ \tcp*[r]{$\tidz\in \{ 0  , 1 \}$ selects $\X$ vs $\Z$.} \label{trans:getMatrixShuffle}
        \tcc{Lookup table of masks in constant memory (listed at end of algorithm).}
$\constMem\ \lookuptab \gets \{\text{ bit-permutation masks }\}$\;
		\Forpary(\tcp*[f]{Tile-row index.}){$0,\;\numWords-1$}{  \label{trans:forAllTilesRow}
			\Forparx(\tcp*[f]{Tile-column index.}){$0,\;\columnLen-1$}{ \label{trans:forAllTilesCol} 
				$\tileidx \gets (\tidy \times \wordSize) \times 2\numWords + \tidx $ \label{trans:calcTileIndex} \tcp*[r]{Calculate linear tile index.} 
				$\sharedMem\ \shared[\tx] \gets \tabMatrix[\tileidx]$ \label{trans:loadShared} \tcp*[r]{ Load $\wordSize\times \wordSize$-bit words (tile) into \shared.}
				\syncThreads{(\:)} \label{trans:syncLoading} \tcp*[r]{Synchronize threads loading a tile.}
				\For(\tcp*[f]{For every pair of $2^l$-bit groups.}){$l\gets0$ \KwTo $\log(\wordSize)-1$}{ \label{trans:forPairofBitGroups}
					$\mask \gets \lookuptab[l],\ \offset \gets 2^l$\; \label{trans:loadMaskandOffset}
					\tcc{Shuffle (new) pairs of $2^l$-bit groups.}
					\If{$(\tx \bmod (2\cdot \offset)) = 0$}{ \label{trans:scanOnlyNewPair}
						$x\leftarrow \shared[\tx]$, 
						$y\leftarrow \shared[\tx+\offset]$\; \label{trans:startShuffle}
						$a\leftarrow x\land \mask$, 
						$b\leftarrow x\land\neg \mask$\;
						$c\leftarrow y\land \mask$, 
						$d\leftarrow y\land\neg \mask$\;
						$\shared[\tx]\gets a\lor\shiftLeft(c\,,\ \offset)$\;
						$\shared[\tx+\offset]\gets \shiftRight(b\,,\ \offset)\lor d$\;\label{trans:endShuffle}
					}
					\syncThreads{(\:)} \label{trans:syncShuffle} \tcp*[r]{Synchronize block threads shuffling a pair.}
				}
				$\tabMatrix[\tileidx]\gets \shared[\tx]$ \label{trans:writeBackShuffled} \tcp*[r]{Write new shuffled word.}
			}
		}
	}
	\Kernel{\upshape{\swapTiles}($\X,\Z,\numWords$)}{  \label{trans:swapKernel}
		$\tabMatrix \gets (\tidz = 0)\ ?\ \X : \Z$\;
		\Forpary(\tcp*[f]{Word-row index.}){$0,\;\numWords-1$}{  \label{trans:scanTilesPerRowSwap}
			\Forparx(\tcp*[f]{Word-column index.}){$0,\numWords-1$}{  \label{trans:scanTilesPerColSwap}
				\If(\tcp*[f]{If \tidx is above the diagonal.}){$\tidx>\tidy$}{ \label{trans:ifAboveDiagonal}
					$i_1\gets \tidy \times \columnLen + \tidx$ \tcp*[r]{Word index above the diagonal.}
					$i_2\gets \tidx \times \columnLen + \tidy$ \tcp*[r]{Word index below the diagonal.}
					${\Swap}(\tabMatrix[i_1],\ \tabMatrix[i_2])$ \label{trans:swapDestabs} \tcp*[r]{Swap destabilizers off the diagonal.}
					${\Swap}(\tabMatrix[i_1 + \numWords],\ \tabMatrix[i_2 + \numWords])$ \label{trans:swapStabs} \tcp*[r]{Swap stabilizers off the diagonal.}
				}
			}
		}
	}\
    \tcc{\textbf{Bit-permutation masks:} \{\texttt{0x5555555555555555}, \texttt{0x3333333333333333}, \texttt{0x0F0F0F0F0F0F0F0F}, \texttt{0x00FF00FF00FF00FF}, \texttt{0x0000FFFF0000FFFF}, \texttt{0x00000000FFFFFFFF}\}}

	\endgroup
\end{algorithm}

We implement a two-stage, in-place GPU algorithm to transpose the tableau while concurrently converting from word-packed generators to bit-packed qubits. Because multiple threads may swap the same memory locations, we coordinate shared-memory shuffling and per-word swaps carefully to avoid data races.

\paragraph{In-Register Bit Shuffle.}  
As shown in the \shuffleTiles{} kernel (\lis{trans:shuffleKernel}{trans:writeBackShuffled}),
we launch a 3D grid with two threads in \td{z} so that \tidz{} selects between the $\X$ and $\Z$ matrices.  Recall matrices are stored in column-major format as 1D arrays; therefore, an index of the form $i \times \columnLen + j$ expresses the location of the $i$-th qubit’s $j$-th generator word.  We then partition each matrix into $\wordSize\times \wordSize$-word tiles, flattening the triple $(\text{tile-row},\text{tile-col},\text{word-in-tile})$ into a single offset in the word array (see \li{trans:calcTileIndex}).

Within each tile, we first load all \wordSize words into shared memory (\li{trans:loadShared}) and synchronize (\li{trans:syncLoading}).  Then we run the bit-shuffling \cite{hackersDelight, knuthArt} loop at \li{trans:forPairofBitGroups} which, for each $l=0,\dots,5$, swaps bit-groups of width $2^l$ using the precomputed \wordSize-bit masks in \lookuptab{}, inserting a barrier after each iteration to avoid races.  Finally, we write back the now-transposed words to global memory (\li{trans:writeBackShuffled}).

\paragraph{Off‐Diagonal Word Swaps.}  
After bit‐transposing each tile of $\wordSize\times\wordSize$-bit words in place, the entire tableau still needs its off‐diagonal tiles exchanged. We sweep the $\numWords\times \numWords$ word grid and, for each $\tidx > \tidy$, we swap locations $i_1$ and $i_2$ once in the destabilizer half (\li{trans:swapDestabs}) and once (offset by \numWords) in the stabilizer half (\li{trans:swapStabs}). Because each pair $(i_1, i_2)$ is visited exactly once, no further synchronization is needed.

\paragraph{Complexity.}
This two‐stage approach achieves a fully in‐place transpose in
$\bigO(\frac{\numQubits^2}{\maxThreads})$ work, with perfect coalescing and no extra output buffers.

\subsection{Finding Maximal Nondeterministic Measurements }

Once the tableau has been transposed and prepared for measurement, we must determine which qubits require a probabilistic measurement, i.e., collapsing into the $\Z$-basis. Scanning each qubit sequentially on the CPU would be prohibitively slow; instead, we launch a single 2D GPU kernel over the transposed tableau that tests all qubits in parallel and records only those whose stabilizer generators anti-commute with the measured observable.

\begin{algorithm}[htp]
	\setstretch{1.15}
	\DontPrintSemicolon
	\caption{Finding Maximal Nondeterministic Measurements on GPU}
	\label{alg:findingAllNondeter}
	\midsmall
	\SetCommentSty{mycommfont}
	\SetArgSty{}
	\KwIn{$\transposed,\; \circuitWindow_d,\; \numQubits,\; \wordSize,\; \numWords $ }
	\KwOut{$\allNondet_d \equiv (q_i)_{i=0}^{|\circuitWindow_d|-1},\;\; q_i \in \{0,\dots,\numQubits-1\}\cup\{-1\}$}

	\begingroup \color{gpu}
	\Kernel{\upshape{\findAllNondetQubits}($\transposed, \circuitWindow_d, \numQubits,\wordSize, \numWords$) $\to \allNondet_d$}{ \label{findAll:start}
		\Forpary(\tcp*[f]{Gate index.}){$0,\;|\circuitWindow_d|-1$}{ 	
			$\observable \gets \routine{Qubit}(\circuitWindow_d[\tidy])$ \label{findAll:getObservable} \tcp*[r]{Get observable $\qubit\in \{ 0 , \cdots ,n-1 \}$.}
			$\allNondet_d[\tidy] \gets -1$ \tcp*[r]{Initialize device list.}
			\Forparx(\tcp*[f]{Generator index.}){$0,\;\numQubits-1$}{ 
				$\wordidx \gets \qWord  \times \rowLen + \tidx + \numQubits $ \tcp*[r]{Stabilizer index for observable \qubit.}
				\If(){$\getBit(\transposedX[\wordidx], \qubit) = 1 \wedge \allNondet_d[\tidy] = -1$}{ \label{findAll:test}
					$\allNondet_d[\tidy] \gets \observable$ \label{findAll:storeObservable}
				}
			}
		}
	}
	\endgroup
\end{algorithm}

\Cref{alg:findingAllNondeter} launches a 2D grid in which the $\td{y}$–dimension checks every measurement in the current window, and the $\td{x}$–dimension scans through all $\numQubits$ stabilizer generators. For each $(\tidx,\tidy)$ pair, we compute the
word index $\wordidx$ that holds the corresponding stabilizer bit in $\transposedX$, where $\qWord$ is the observable’s word index and $\numQubits$ is an offset used to access the stabilizer half of the row-major layout.
At \li{findAll:test}, we test whether the stabilizer anti-commutes with the observable via the condition $\getBit(\transposedX[\wordidx], \observable)=1$. As soon as one anti-commuting generator is found, we write $\observable$ into $\allNondet_d[\tidy]$ and prevent further overwrites by checking that $\allNondet_d[\tidy] = -1$ beforehand. 

In practice, multiple threads may attempt to write the observable index $\observable$ into $\allNondet_d[\tidy]$ if more than one anti-commuting generator exists. This benign race condition is harmless, since the algorithm only requires the existence of at least one anti-commuting generator, not its specific index. If none is found, the observable $\observable$ is deterministic and $\allNondet_d[\tidy]$ remains $-1$.

\paragraph{Complexity.}
Each of the $|\circuitWindow_d| \times \numQubits$ bit tests is a constant-time operation; therefore, the total work is $\bigO(\numQubits^2)$. Since these tests are distributed across
up to $\maxThreads$ hardware threads (the product of the grid’s $\td{x}$– and $\td{y}$–dimensions), the parallel time is
$\bigO\!\left(\frac{\numQubits^2}{\maxThreads}\right)$.

\subsection{Anti-Commuting Pivots Compaction}\label{sec:compact}

In order to eliminate warp divergence and maximize coalesced memory access, we compact all anti-commuting pivots into a one-dimensional array called \pivots. \Cref{alg:compactPivots} performs this compaction in two steps. First, we launch a one-dimensional grid (\li{compact:callScatter}) that scans all generators and scatters the indices of anti-commuting ones into \pivots, writing $-1$ for non pivots and the generator index (i.e., thread ID \tidx) for pivots.

\begin{algorithm}[htp]
	\setstretch{1.15}
	\DontPrintSemicolon
	\caption{Pivot Stream Compaction on GPU}
	\label{alg:compactPivots}
	\midsmall
	\SetCommentSty{mycommfont}
	\SetArgSty{}
	\KwIn{$\transposed,\; \qubit, \; \numQubits, \; \wordSize,\; \numWords$}
	\KwOut{$\pivots\in [-1, \cdots , n-1 ]^\numQubits$ \tcp*[r]{Device array, compacted in place.}}
	\begingroup \color{gpu}
	\Function{\upshape{\compactPivots}($\transposed, \qubit, \numQubits,\wordSize,\numWords$) $\to \pivots$}{ \label{compact:start}
		$ \pivots \gets \scatter(\transposedX, \qubit, \numQubits,\wordSize, \numWords)$\; \label{compact:callScatter}
		$ {\compactCub}(\pivots, \numQubits)$\label{compact:callCompact}
	}
	\Kernel{\upshape{\scatter}($\transposedX, \qubit, \numQubits,\wordSize, \numWords$)}{ \label{compact:scatterKernel}
		\Forparx(\tcp*[f]{Generator index.}){$0,\;\numQubits-1$}{ 
			$\wordidx \gets \qWord  \times \rowLen + \tidx + \numQubits $ \tcp*[r]{Stabilizer index for observable \qubit.}
			\lIf{$\getBit(\transposedX[\wordidx], \qubit) = 1$}{
				$\pivots[\tidx] \gets \tidx$ \label{compact:storeIndex}
			}
			\lElse {
				$\pivots[\tidx] \gets -1$ \label{compact:storeEmpty}
			}
		}
	}
	\endgroup
\end{algorithm}
\begin{figure*}[htp]
	\centering
	\adjustbox{max width=.7\linewidth}{
		\includegraphics[width=\linewidth]{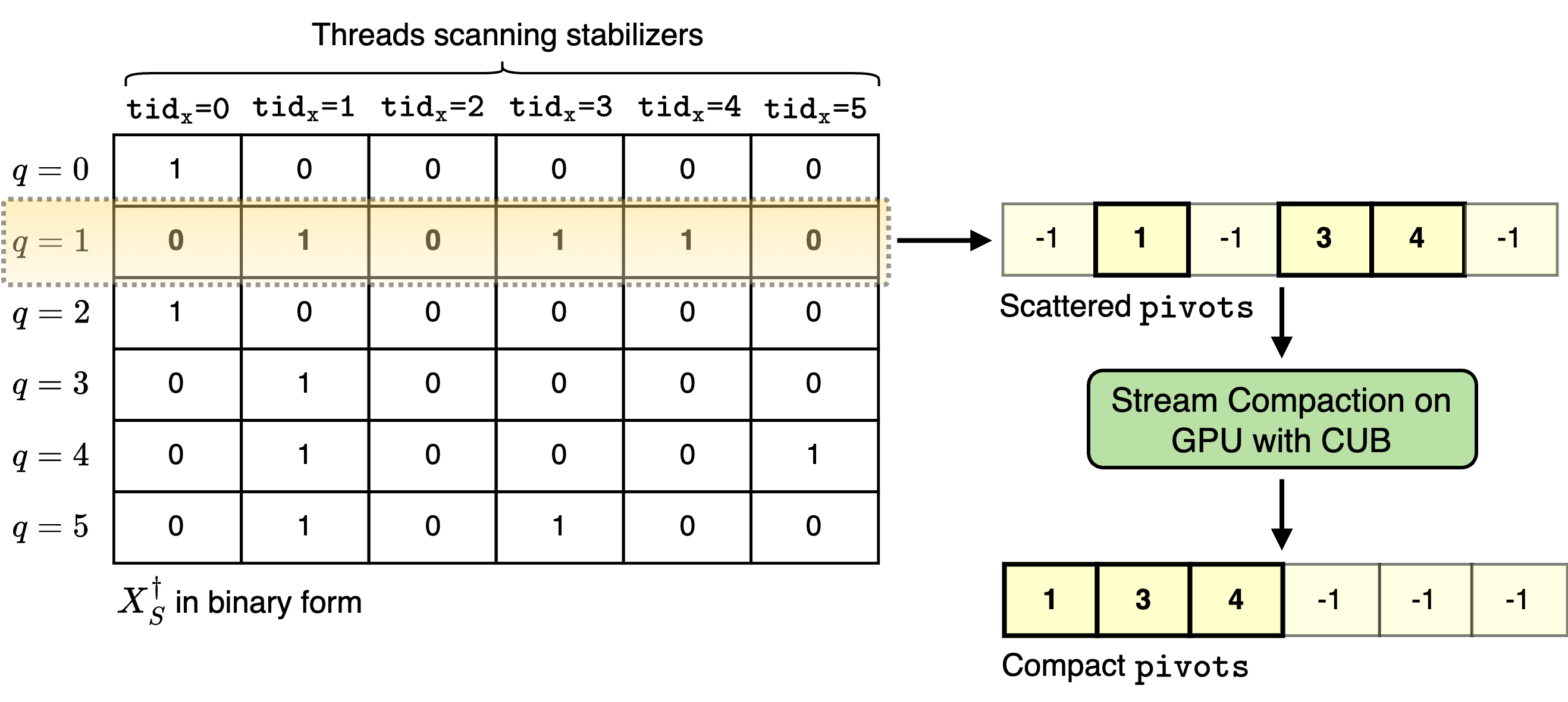}
	}
	\caption{Compacting pivots on a 6-qubit tableau.}
	\label{fig:compactPivots}
\end{figure*}

As in \cref{alg:findingAllNondeter}, a generator \tidx is considered a pivot for the observable $\qubit$ if and only if $\getBit(\transposedX[\wordidx], \qubit)=1$. Each thread therefore checks the $\X$-stabilizer entry corresponding to the observable $\qubit$: if it anti-commutes, it writes $\pivots[\tidx]=\tidx$; otherwise, it writes a negative value $\pivots[\tidx]=-1$, which lies outside the valid generator index range $[0,\numQubits)$ (\lis{compact:storeIndex}{compact:storeEmpty}). 

Next, at \li{compact:callCompact}, we invoke CUB’s \routine{DeviceSelect::If} to perform \emph{stream compaction} on \pivots in place, moving all entries $\ge 0$ to the front of the array while preserving their relative order. We refer to the CUB documentation\footnote{\url{https://nvidia.github.io/cccl/unstable/cub/device_wide.html\#device-module}} for details on \routine{DeviceSelect::If} and stream compaction \cite{streamCompact}. This produces a dense list of pivot indices $\{p_0, p_1, \dots\}$ followed by $-1$ fillers,
allowing subsequent kernels to iterate only over valid pivots without gaps.

While stream compaction itself is a standard GPU primitive, its application to anti-commuting pivot selection in stabilizer-tableau simulation is, to our knowledge, new. In contrast to the \stim approach, our compaction removes warp divergence and ensures fully coalesced memory access during Gaussian elimination.

\Cref{fig:compactPivots} illustrates the process for a 6-qubit system:
first scatter into \{-1,\,1,\,-1,\,3,\,4,\,-1\}, then compact to
\{1,\,3,\,4,\,-1,\,-1,\,-1\}.

\paragraph{Complexity.}
The \scatter kernel in \cref{alg:compactPivots} launches a grid that scans $\numQubits$ generators using up to \maxThreads threads, resulting in a parallel span of $\bigO\!\left(\frac{\numQubits}{\maxThreads}\right)$. The stream compaction step implemented by CUB is based on multi-level prefix sums \cite{streamCompact} and has a worst-case parallel time of $\bigO\!\left(\frac{\numQubits}{\maxThreads} + \log \numQubits\right)$. Therefore, the overall worst-case parallel time of \cref{alg:compactPivots} is $\bigO\!\left(\frac{\numQubits}{\maxThreads} + \log\numQubits\right)$.
\subsection{Three-Pass Prefix-XOR Elimination}\label{sec:GEPrefix}

Once the pivot indices have been compacted, we can perform Gaussian elimination without any further search for anti‐commuting generators.  \Cref{alg:paraGE} summarizes our three‐pass, prefix‐XOR approach to injecting CX gates in parallel for each pivot in the compacted array \pivots.  The inputs are the transposed tableau $\transposed$ (obtained by \cref{alg:transpose}), the compacted pivot list \pivots, and the number of qubit words per row $\numWords$.  

\begin{figure*}[htp]
	\centering
	\adjustbox{max width=.9\linewidth}{
		\includegraphics[width=\linewidth]{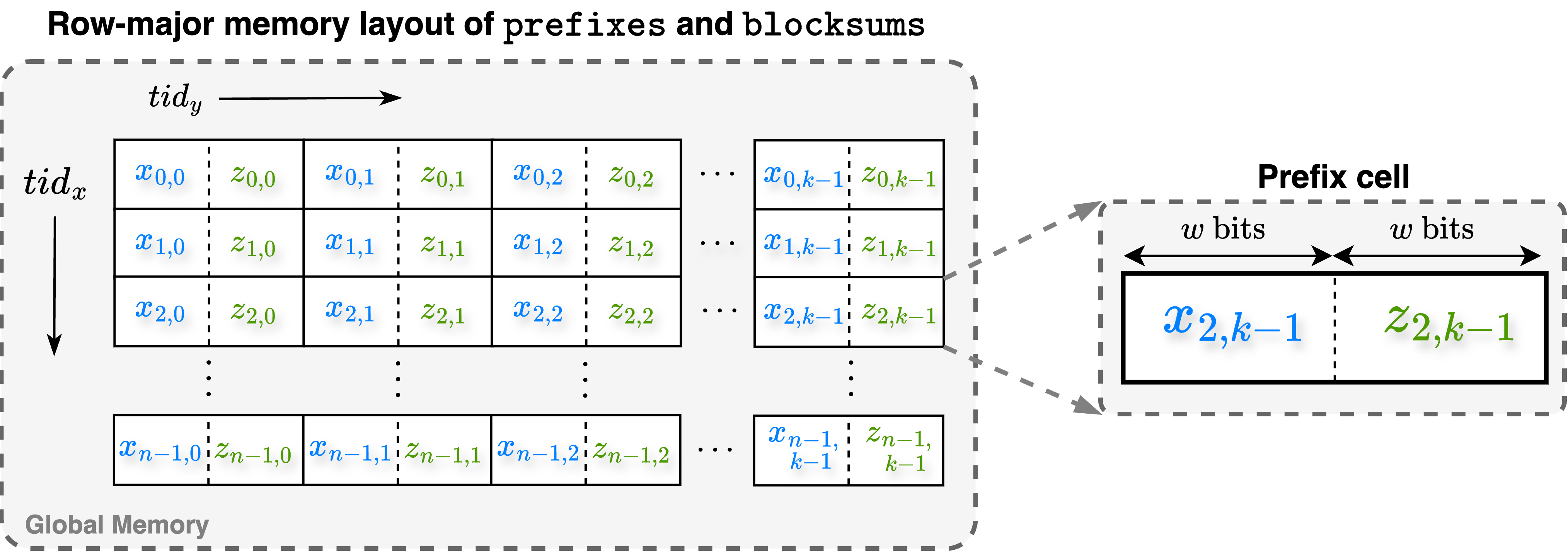}
	}
	\caption{Memory layout of \prefixes and \blocksums.}
	\label{fig:prefixData}
\end{figure*}

Because the number of pivots can significantly exceed a single thread‐block’s capacity, we split the computation into multiple CUDA kernel launches.  To carry intermediate results between passes, we allocate two arrays on the device, called \prefixes and \blocksums. Each measured qubit can have up to $\numQubits$ anti‐commuting pivots, therefore we provision these arrays with dimensions $\numQubits\times\numWords$.  Although only the first $|\pivots|$ rows are populated in any given measurement, allocating $\numQubits\times\numWords$ guarantees sufficient storage.  Each entry is a $2\wordSize$-bit cell that interleaves the $\Z$-prefix and $\X$-prefix words; by packing both $\wordSize$-bit values into one cell, every thread’s global‐memory access fetches exactly the paired prefixes it needs, yielding fully coalesced loads and stores when scanning \transposedZ\ and \transposedX.  \Cref{fig:prefixData} illustrates this layout.


The outer driver \paraGE (\lis{ge:paraGE}{ge:callPass3}) invokes three
kernels in succession:
\begin{enumerate}
	\item At line \li{ge:callPass1}, the kernel \prefixPassOne is launched. Inside \prefixPassOne, each thread block handles one contiguous block of pivots for a fixed qubit‐word.	For each pivot in the block, the kernel loads the corresponding target‐destabilizer bit $\zt$ into shared memory.	It then invokes CUB’s \prefixBlock routine, which implements Blelloch’s up/down sweep, to compute two values per block: the exclusive prefix (stored as $\localprefix_z$) and the total block‐sum (stored as $\blocksum_z$) of those $\zt$ words.
	The same process runs in parallel on the $\xt$ words to produce $\localprefix_x$ and $\blocksum_x,$.
	\item At \li{ge:callPass2}, \prefixPassTwo performs a device‐wide exclusive prefix‐XOR using the CUB's routine \routine{DeviceScan::ExclusiveSum}  over the accumulated \blocksums using \prefixPassTwo, producing global block‐prefix corrections that will be applied to every pivot’s per‐block prefixes.
	\item At \li{ge:callPass3}, \prefixPassThree merges the two levels of prefix data--combining per‐block prefixes with their block‐prefix corrections--then, for each pivot and qubit‐word, computes the final CX‐injection update	on the tableau via the device function \injectCXByPrefix (implementing 	equations~\cref{eq:CXSign} and~\cref{eq:zc}).
\end{enumerate}
\begin{algorithm}[!t]
	\setstretch{1.15}
	\DontPrintSemicolon
	\caption{Three‐Pass Prefix‐XOR Gaussian Elimination}
	\label{alg:paraGE}
	\midsmall
	\SetCommentSty{mycommfont}
	\SetArgSty{}
	\KwIn{$\transposed,\; \pivots,\; \numQubits,\; \wordSize,\; \numWords$}
	\KwOut{$\transposed$}
	\begingroup \color{gpu}
	\Function{\upshape{\paraGE}($\transposed,\pivots,\numWords$)}{ \label{ge:paraGE}
		$\prefixes,\, \blocksums \gets {\prefixPassOne}(\transposed,\pivots,\numWords)$\; \label{ge:callPass1}
		$\blocksums \gets {\prefixPassTwo}(\blocksums)$ \label{ge:callPass2} \tcp*[r]{Calls a CUB's device-wide routine.}
		$\transposed \gets {\prefixPassThree}(\pivots,\prefixes,\blocksums,\transposed,\numWords)$\; \label{ge:callPass3}
	}
	\Kernel{\upshape{\prefixPassOne}($\transposed, \pivots, \numWords$) $\to \prefixes, \blocksums$}{  \label{ge:pass1Kernel}
		\Forpary(\tcp*[f]{Qubits‐word index.}){$0,\;\numWords-1$}{
			\Forparx(\tcp*[f]{\pivots index.}){$0,\;|\pivots|-2$}{
				$\Control\gets \pivots[0],\ \Target\gets \pivots[\tidx+1]$\tcp*[r]{Fetch control and target pivots.}
				$\wordidx_\Target\gets \tidy \times \rowLen + \Target$\tcp*[r]{Target‐word index.}
				$\sharedMem\ \sharedz[\tx]\gets \transposedZ[\wordidx_\Target],\ 
				\sharedx[\tx]\gets \transposedX[\wordidx_\Target]$ \label{ge:loadTargetWords} \tcp*[r]{Load destabilizer words.}
				\syncThreads{(\:)} \tcp*[r]{Synchronize threads loading into \shared.}
				\tcc{Compute prefixes $\bigotimes_{{\tx}=0}^{\blockDim-1} \transposedZ[\wordidx_\Target]^{\tx}$ and $\bigotimes_{{\tx}=0}^{\blockDim-1} \transposedX[\wordidx_\Target]^{\tx}$ per thread block.}
				$(\localprefix_z,\; \blocksum_z)\gets\prefixBlock(\sharedz[\tx])$\; \label{ge:computePrefixZ}
				$(\localprefix_x,\; \blocksum_x)\gets\prefixBlock(\sharedx[\tx])$\; \label{ge:computePrefixX}
				$\wordidx_\Control\gets \tidy \times \rowLen + \Control$\tcp*[r]{Control‐word index.}
				$\prefixes[\tidx \times \numWords + \tidy]\gets
				(\localprefix_x \oplus \transposedX[\wordidx_\Control],\ 
				\localprefix_z \oplus \transposedZ[\wordidx_\Control])$\; \label{ge:storePrefixes}
				\tcc{Last thread per block stores \blocksum and updates \zc w.r.t. \cref{eq:zc}.}
				\If{$\tx = \blockDim - 1$}{
					$\blocksums[\lfloor \tidx/\blockDim\rfloor \times \numWords + \tidy]\gets
					(\blocksum_x,\; \blocksum_z)$\;
					$\atomicXOR(\transposedX[\wordidx_\Control],\; \blocksum_x),\
					\atomicXOR(\transposedZ[\wordidx_\Control],\; \blocksum_z)$\; \label{ge:atomicZC}
				}
			}
		}
	}
	\tcc{Compute the final prefixes and finalize CX injections.}
	\Kernel{\upshape{\prefixPassThree}($\pivots, \prefixes,\blocksums,\transposed,\numWords$)}{  \label{ge:pass3Kernel}
		\Forpary(\tcp*[f]{Qubits-word index.}){$0,\; \numWords-1$}{
			\Forparx(\tcp*[f]{\pivots index.}){$0,\; |\pivots|-2$}{
				\tcc{Finalize prefixes by applying the exclusive scan results of block-sums.}
				$(\localprefix_x,\; \localprefix_z) \gets \prefixes[ \tidx \times \numWords + \tidy] \oplus \blocksums[\lfloor \tidx/\blockDim\rfloor \times \numWords + \tidy] $\; \label{ge:finalizePrefixes}
				\tcc{Update (de)stabilizers signs using computed prefixes $(\localprefix_x,\; \localprefix_z)$. }
				$\Control \gets \pivots[0],\ \Target \gets \pivots[\tidx+1]$\;
				$\wordidx_\Control \gets \tidy \times \rowLen + \Control, \ \wordidx_\Target\gets \tidy \times \rowLen + \Target$\;
				$\injectCXByPrefix(\Signs[\tidy],\; \transposedZ[\wordidx_\Control + \numQubits],\; \localprefix_z,\; \transposedZ[\wordidx_\Target + \numQubits],\; \transposedZ[\wordidx_\Target],\;sm)$\;\label{ge:pass3End1}
				$\injectCXByPrefix(\Signs[\tidy + \numWords],\; \transposedX[\wordidx_\Control + \numQubits],\; \localprefix_x,\; \transposedX[\wordidx_\Target + \numQubits],\; \transposedX[\wordidx_\Target],\;sm)$\;
 \label{ge:pass3End2}
			}
		}
	}
	\tcc{Update (de)stabilizers and signs. 
		\xc: $\Control$-stabilizer, 
		\localprefix: $\Control$-destabilizer prefix, \\
		\xt: $\Target$-stabilizer, and 
		\zt: $\Target$-destabilizer}
	\Devfunction{\upshape{\injectCXByPrefix}($\sign, \xc, \localprefix, \xt, \zt, \shared$)}{ \label{l:injectCXByPrefix}
		$\sharedMem\ \shared[\tx] \gets \xc \wedge \zt \wedge \neg(\localprefix \oplus \xt)$\; \label{ge:historyLoad}
		$\collapseSigns(\sign, \shared)$ \label{ge:collapse} \tcp*[r]{Collapse signs in \shared with a binary tree.}
		$\xt \gets \xt \oplus \xc$\; \label{ge:updateTargets}
	}
	\endgroup
\end{algorithm}
\paragraph{Pass One: Block‐Local Prefix‐XOR.}  
In \prefixPassOne, we launch a 2D grid of thread‐blocks sized $\blockDim\times\yblockDim$ in
$(\td{x},\td{y})$-dimensions, where $\td{y}$ covers qubit‐word indices $\tidy\in[0,\numWords-1]$, and $\td{x}$ covers pivot indices $\tidx\in[0,|\pivots|-2]$.  Each thread loads the target’s destabilizer words $\transposedZ[\tidy\times\rowLen + \Target]$ and $\transposedX[\tidy\times\rowLen + \Target]$ into the shared memory array \shared at \li{ge:loadTargetWords}, then synchronizes.

Next, we invoke CUB’s \prefixBlock routine (\lis{ge:computePrefixZ}{ge:computePrefixX}), which performs an up‐sweep to build a full binary tree over the $\blockDim$ threads and captures the block‐sum at the root, then a down‐sweep to compute every thread’s exclusive prefix.  Thread \tx (per block) thus obtains $\localprefix_z$ and $\blocksum_z$, and analogously $\localprefix_x$ and
$\blocksum_x$.  We record the combined words $(\localprefix_x \oplus \transposedX[\wordidx_\Control], \localprefix_z \oplus \transposedZ[\wordidx_\Control])$ in the global \prefixes array at \li{ge:storePrefixes}, where $\transposedX[\wordidx_\Control],\transposedZ[\wordidx_\Control]$ are the control‐pivot’s
destabilizer words stored in \transposedX and \transposedZ, respectively.  

Finally, the last thread in each block (when $\tx=\blockDim-1$) writes the block-sums pair $(\blocksum_x,\blocksum_z)$ to the global array \blocksums at position $\bigl \lfloor\tidx / \blockDim \bigr \rfloor \times \numWords + \tidy$, where $\lfloor\tidx/\blockDim\rfloor$ is the block index (i.e.\ which block of pivots this thread belongs to). Immediately afterward, at \li{ge:atomicZC}, the same thread applies two atomic XORs to globally update the $\Control$-destabilizers $\transposedX[\wordidx_\Control]$ and $\transposedZ[\wordidx_\Control]$.  Thereby, accumulating the block‐sums into the global tableau entries according to \cref{eq:zc}.

\paragraph{Pass Two: Block‐Sum Prefix‐XOR.}  
In the second pass, we compute an exclusive prefix over the block‐sum pairs stored in \blocksums using a device-wide scan implemented in \prefixPassTwo. This operation combines the partial block-sums produced in \prefixPassOne into global block‐prefix corrections, without imposing any restrictions over the block size.

The resulting scanned corrections are stored back in \blocksums in place and are subsequently consumed by \prefixPassThree to finalize the prefix values for each pivot. 

\paragraph{Pass Three: Final CX Injection.}  
Finally, in the third kernel, \prefixPassThree (\li{ge:pass3Kernel}), we launch a 2D grid with size similar to the one in \prefixPassOne. Each thread first reads its per‐block prefix from the global array \prefixes at position $(\tidx,\tidy)$ and then reads the corresponding block‐prefix correction from \blocksums at position $\bigl(\lfloor \tidx/\blockDim\rfloor, \tidy\bigr)$. It computes the final exclusive prefix as
\[
\localprefix\;=\;\prefixes[\tidx, \tidy]\;\oplus\;\blocksums[\lfloor \tidx/\blockDim\rfloor, \tidy].
\]
With this prefix in hand, the thread calls the device function \injectCXByPrefix. Inside that function it loads four 64-bit words: the $\Control$-stabilizer $\xc$, the $\Target$‐destabilizer $\zt$, the $\Target$‐stabilizer $\xt$, and the computed prefix word. It then calculates the sign history $\xc \wedge \zt \wedge \neg(\localprefix\ \oplus\ \xt)$
and stores it into shared memory at index $\tx$. Next, all threads in the block cooperate in a binary‐tree reduction to collapse those history values into a single 64-bit result at the root $\shared[0]$. That result is atomically XORed into the global \Signs array at the entry corresponding to the current \X-signs word index $\tidy$ (or $\tidy+\numWords$ for the \Z-signs), finalizing the sign update for this CX injection. Afterward, \injectCXByPrefix finishes the CX injection by merging the \Control‐stabilizer word into the \Target‐stabilizer entry. This completes the in‐place injection of the CX gate for each pivot, with correct sign handling and coalesced memory access throughout.

\paragraph{Complexity.}  
Overall, this three‐pass scheme performs  $|\pivots|\times\numWords$
injections of the CX gate in parallel.  Noting that $|\pivots|$ and $\numWords$ in the worst case can grow to $\bigO(\numQubits)$, therefore, the total work is $\bigO(\numQubits^2)$. These $\bigO(\numQubits^2)$ operations are distributed across up to $\maxThreads$ hardware threads (the product of the grid’s $\td{x}$– and $\td{y}$–dimensions), giving an ideal parallel time of  $\bigO(\frac{\numQubits^2}{\maxThreads})$.  
Passes One and Three include an intra‐block exclusive scan over
\blockDim threads (e.g. lines~\ref{ge:computePrefixZ} and
\ref{ll:forShared}), which costs $\bigO(\log \blockDim)$ steps.  Since \blockDim is limited by the hardware’s maximum threads per block rather than by \numQubits \cite{nvidia}, the overall worst‐case parallel time is  $\bigO(\frac{\numQubits^2}{\maxThreads} + \log \blockDim)$.

The algorithm thus achieves a fully in‐place, lock‐free, coalesced elimination that scales to large pivot sets on GPUs.
\subsection{Final Measurement Assembly}\label{sec:assembly}

\paragraph{Swapping the Anti‐commuting Generator.}  
After collapsing non‐deterministic measurements into the $Z$-basis (via \cref{alg:paraGE}), we must replace the first anti‐commuting generator at index $\pivot$ with one that commutes with the measured observable~$\observable$.
The kernel \swapAntiComm (\cref{alg:assembly},
\lis{final:swapKernel}{final:updateXHad}) launches a one-dimensional grid of $\numWords$ threads, each responsible for a single \wordSize-bit word of the pivot row.

\begin{algorithm}[htp]
	\setstretch{1.15}
	\DontPrintSemicolon
	\caption{Final Measurement Assembly on GPU}
	\label{alg:assembly}
	\midsmall
	\SetCommentSty{mycommfont}
	\SetArgSty{}
	\KwIn{$\transposed,\; \pivot,\; \qubit,\; \numQubits,\; \wordSize,\; \numWords$}
	\KwOut{\transposed}
	\begingroup \color{gpu}
	\Kernel{\upshape{\paraSwap}($\transposed,\pivot,\qubit,\numQubits, \numWords,\wordSize$)}{  \label{final:swapKernel} 
		\Forparx(\tcp*[f]{For all qubit-word indices.}){$0,\;\numWords-1$}{
			\If{$\tidx=0$}{  \label{final:firstThread}
				$\isYPauli \gets \getBit\bigl(\transposedX[\qWord\times\rowLen + \pivot],\qubit\bigr)$
				\tcp*[r]{Observable’s bit.}
			}
			\syncGrid{(\:)}  \label{final:syncGrid} \tcp*[r]{Ensure \isYPauli is visible to all.}
			$\wordidx \gets \tidx\times\rowLen + \pivot$\;
			\If(\tcp*[f]{If Y-Pauli, inject $P^\dagger$.}){$\isYPauli=1$}{
				$\injectInvP(\transposedZ[\wordidx],\transposedZ[\wordidx+\numQubits],\Signs[\tidx])$\;  \label{final:updateZ}
				$\injectInvP(\transposedX[\wordidx],\transposedX[\wordidx+\numQubits],\Signs[\tidx+\numQubits])$\;  \label{final:updateX}
			}
			\Else(\tcp*[f]{If X-Pauli, inject $H$.}){
				$\injectHad(\transposedZ[\wordidx],\transposedZ[\wordidx+\numQubits],\Signs[\tidx])$\;  \label{final:updateZHad}
				$\injectHad(\transposedX[\wordidx],\transposedX[\wordidx+\numQubits],\Signs[\tidx+\numQubits])$\;  \label{final:updateXHad}
			}
		}
	}
	
	\Kernel{\upshape{\paraX}($\transposed,\pivot,\numWords, \numQubits$)}{  \label{final:injectXKernel}
		\Forparx(\tcp*[f]{For all qubit-word indices.}){$0,\;\numWords-1$}{
			$\wordidx \gets \tidx\times\rowLen + \pivot$\;
			$\Signs[\tidx]\gets \Signs[\tidx]\oplus\transposedZ[\wordidx]$\;
			$\Signs[\tidx+\numWords]\gets \Signs[\tidx+\numWords]\oplus\transposedX[\wordidx]$\;  \label{final:injectXKernelEnd}
		}
	}
	
	\endgroup
\end{algorithm}

Thread~0 first extracts the observable’s Pauli type at the pivot position $\qWord\times\rowLen + \pivot$ into the flag $\isYPauli$ (\li{final:firstThread}). A grid-wide barrier \syncGrid{} (\li{final:syncGrid}) ensures that all threads observe the same value.

\paragraph{Randomizing the Measurement.}  
Finally, if the coin flip disagrees with the observable’s sign (\li{l:checkSign} in \cref{alg:parallelSimulation}), we inject an $X$ gate via the kernel \injectX (\lis{final:injectXKernel}{final:injectXKernelEnd}). This one-dimensional grid XORs each word of the pivot’s $\Z$- and $\X$ stabilizer rows into the corresponding entries of the sign array, completing the measurement update.

\paragraph{Complexity.}
Both \paraSwap and \paraX launch $\numWords$ parallel threads, each performing a constant number of word-level bitwise operations. Since $\numWords = \bigO(\numQubits)$, the total work is $\bigO(\numQubits)$, yielding an ideal parallel time of $\bigO(\numQubits/\maxThreads)$. The single grid synchronization in \paraSwap incurs only constant overhead.

%% file: circuit.tex
\begin{minipage}{.48\textwidth}
\resizebox{.85\textwidth}{!}{ 
	\begin{quantikz}
		\lstick{$\ket{q_0}$} & \gate{\Had} & \qw & \gate{\Phase{}}  & \ctrl{2} & & \qw & \qw   \\
		\lstick{$\ket{q_1}$} & \qw &\gate{\M{}} & \qw  & \qw &   \gate{\Had{}} & \qw  & \qw  \\
		\lstick{$\ket{q_2}$} & \qw   & \qw   & \gate{\Phase{}}  & \targ{0}  & \gate{\Had{}}   & \qw & \qw \\
		\lstick{$\ket{q_3}$} & \gate{\M{}}  & \qw & \gate{\Phase{}} & \ctrl{1} & \gate{\Phase{}} & \qw   & \qw\\
		\lstick{$\ket{q_4}$} & \qw  & \gate{\Phase{}} & \qw  &\targ{3} & \qw  & \qw  & \qw \\
	\end{quantikz}
}
\end{minipage}
\begin{minipage}{0.5\textwidth}
\resizebox{0.9\textwidth}{!}{ 
		\begin{quantikz}
			\lstick{$\ket{q_0}$} & \gate{\Had{}} & \qw & \qw & \gate{\Phase{}} & \ctrl{2} & \qw & \qw   \\
			\lstick{$\ket{q_1}$} & \qw &  \gate{\M{}} & \qw & \gate{\Had{}} &  \qw &\qw & \qw  \\
			\lstick{$\ket{q_2}$} & \gate{\Phase{}} & \qw & \qw & \qw &\targ{0} & \gate{\Had{}} & \qw \\
			\lstick{$\ket{q_3}$} & \qw & \qw & \gate{\M{}} & \gate{\Phase{}} & \ctrl{1} & \gate{\Phase{}} & \qw  \\
			\lstick{$\ket{q_4}$} & \gate{\Phase{}} & \qw & \qw & \qw & \targ{3} & \qw & \qw \\
		\end{quantikz}
}
\end{minipage}%

\newcommand{\wWidth}{0.75}
\newcommand{\wGap}{0.25}
\newcommand{\wHeight}{4.5}
\newcommand{\lGap}{.8}
\newcommand{\xCordLeft}{8.5}
\newcommand{\yCordLow}{0.39}
\newcommand{\yLabelLow}{0.34}
\pgfmathsetmacro{\xLabelLeft}{\xCordLeft+0.4}
\pgfmathsetmacro{\xCordRight}{\xCordLeft + \wWidth}
\pgfmathsetmacro{\yCordHigh}{\yCordLow + \wHeight}
\newcommand{\lineWidth}{1.2pt}
\newcommand{\lineOpac}{0.5}
\begin{tikzpicture}[overlay]
	\draw[ purple,line width=\lineWidth, opacity=\lineOpac] (\xCordLeft ,   \yCordHigh) -- (\xCordLeft,    \yCordLow); 
	\draw[ purple,line width=\lineWidth, opacity=\lineOpac] (\xCordRight,  \yCordHigh) -- (\xCordRight,  \yCordLow); 
	\draw[ purple,line width=\lineWidth, opacity=\lineOpac] (\xCordLeft ,   \yCordHigh) -- (\xCordRight,  \yCordHigh); 
	\draw[ purple,line width=\lineWidth, opacity=\lineOpac] (\xCordLeft ,   \yCordLow) -- (\xCordRight,   \yCordLow); 
	\node[anchor=north, text=purple, opacity=1] at (\xLabelLeft,\yLabelLow) {$\circuitWindow_0$};
	
	\pgfmathsetmacro{\xCordLeft}{\xCordRight + \wGap - .08}
	\pgfmathsetmacro{\xCordRight}{\xCordLeft + \wWidth}
    \pgfmathsetmacro{\xLabelLeft}{\xLabelLeft + \lGap + .1}
	
	\draw[ purple,line width=\lineWidth, opacity=\lineOpac] (\xCordLeft ,   \yCordHigh) -- (\xCordLeft,    \yCordLow); 
	\draw[ purple,line width=\lineWidth, opacity=\lineOpac] (\xCordRight,  \yCordHigh) -- (\xCordRight,  \yCordLow); 
	\draw[ purple,line width=\lineWidth, opacity=\lineOpac] (\xCordLeft ,   \yCordHigh) -- (\xCordRight,  \yCordHigh); 
	\draw[ purple,line width=\lineWidth, opacity=\lineOpac] (\xCordLeft ,   \yCordLow) -- (\xCordRight,   \yCordLow); 
	\node[anchor=north, text=purple, opacity=1] at (\xLabelLeft,\yLabelLow) {$\circuitWindow_1$};

    \pgfmathsetmacro{\xCordLeft}{\xCordRight + \wGap - .08}
	\pgfmathsetmacro{\xCordRight}{\xCordLeft + \wWidth}
    \pgfmathsetmacro{\xLabelLeft}{\xLabelLeft + \lGap + .1}
	
	\draw[ purple,line width=\lineWidth, opacity=\lineOpac] (\xCordLeft ,   \yCordHigh) -- (\xCordLeft,    \yCordLow); 
	\draw[ purple,line width=\lineWidth, opacity=\lineOpac] (\xCordRight,  \yCordHigh) -- (\xCordRight,  \yCordLow); 
	\draw[ purple,line width=\lineWidth, opacity=\lineOpac] (\xCordLeft ,   \yCordHigh) -- (\xCordRight,  \yCordHigh); 
	\draw[ purple,line width=\lineWidth, opacity=\lineOpac] (\xCordLeft ,   \yCordLow) -- (\xCordRight,   \yCordLow); 
	\node[anchor=north, text=purple, opacity=1] at (\xLabelLeft,\yLabelLow) {$\circuitWindow_2$};

	\pgfmathsetmacro{\wGap}{\wGap}
	\pgfmathsetmacro{\lGap}{\lGap}
	\pgfmathsetmacro{\xCordLeft}{\xCordRight + \wGap - .08}
	\pgfmathsetmacro{\xCordRight}{\xCordLeft + \wWidth - .02}
    \pgfmathsetmacro{\xLabelLeft}{\xLabelLeft + \lGap + .1}
	
	\draw[ purple,line width=\lineWidth, opacity=\lineOpac] (\xCordLeft ,   \yCordHigh) -- (\xCordLeft,    \yCordLow); 
	\draw[ purple,line width=\lineWidth, opacity=\lineOpac] (\xCordRight,  \yCordHigh) -- (\xCordRight,  \yCordLow); 
	\draw[ purple,line width=\lineWidth, opacity=\lineOpac] (\xCordLeft ,   \yCordHigh) -- (\xCordRight,  \yCordHigh); 
	\draw[ purple,line width=\lineWidth, opacity=\lineOpac] (\xCordLeft ,   \yCordLow) -- (\xCordRight,   \yCordLow); 
	\node[anchor=north, text=purple, opacity=1] at (\xLabelLeft,\yLabelLow) {$\circuitWindow_3$};

	\pgfmathsetmacro{\wGap}{\wGap}
	\pgfmathsetmacro{\lGap}{\lGap}
	\pgfmathsetmacro{\xCordLeft}{\xCordRight + \wGap - .1}
	\pgfmathsetmacro{\xCordRight}{\xCordLeft + \wWidth - .05}
    \pgfmathsetmacro{\xLabelLeft}{\xLabelLeft + \lGap + .1}
	
	\draw[ purple,line width=\lineWidth, opacity=\lineOpac] (\xCordLeft ,   \yCordHigh) -- (\xCordLeft,    \yCordLow); 
	\draw[ purple,line width=\lineWidth, opacity=\lineOpac] (\xCordRight,  \yCordHigh) -- (\xCordRight,  \yCordLow); 
	\draw[ purple,line width=\lineWidth, opacity=\lineOpac] (\xCordLeft ,   \yCordHigh) -- (\xCordRight,  \yCordHigh); 
	\draw[ purple,line width=\lineWidth, opacity=\lineOpac] (\xCordLeft ,   \yCordLow) -- (\xCordRight,   \yCordLow); 
	\node[anchor=north, text=purple, opacity=1] at (\xLabelLeft,\yLabelLow) {$\circuitWindow_4$};

    \pgfmathsetmacro{\wGap}{\wGap}
	\pgfmathsetmacro{\lGap}{\lGap}
	\pgfmathsetmacro{\xCordLeft}{\xCordRight + \wGap - .1}
	\pgfmathsetmacro{\xCordRight}{\xCordLeft + \wWidth - .05}
    \pgfmathsetmacro{\xLabelLeft}{\xLabelLeft + \lGap + .1}
	
	\draw[ purple,line width=\lineWidth, opacity=\lineOpac] (\xCordLeft ,   \yCordHigh) -- (\xCordLeft,    \yCordLow); 
	\draw[ purple,line width=\lineWidth, opacity=\lineOpac] (\xCordRight,  \yCordHigh) -- (\xCordRight,  \yCordLow); 
	\draw[ purple,line width=\lineWidth, opacity=\lineOpac] (\xCordLeft ,   \yCordHigh) -- (\xCordRight,  \yCordHigh); 
	\draw[ purple,line width=\lineWidth, opacity=\lineOpac] (\xCordLeft ,   \yCordLow) -- (\xCordRight,   \yCordLow); 
	\node[anchor=north, text=purple, opacity=1] at (\xLabelLeft,\yLabelLow) {$\circuitWindow_5$};
 
\end{tikzpicture}

%% file: sampling.tex
\section{Sampling}\label{sec:sampling}

One often needs to run many \emph{shots} to evaluate the probability distribution of the state generated by the circuit, i.e., independent simulations of the same circuit, to estimate output probabilities \cite{AaronsonShadowTomogr20}. Naively, this would require reapplying all gates, including  probabilistic measurements, once per shot, which can be prohibitively expensive. \stim instead uses the \emph{Pauli‐frame} trick to amortize this cost. Each qubit maintains a bit-string that tracks phase flips, and measurements simply update that frame instead of re‐collapsing the full tableau state.

Regarding scheduling in sampling mode,
the window structure $\circuitScheduled$ constructed in
\cref{sec:parallelGates} is reused for sampling. However, measurement windows are interpreted differently. In single-shot simulation, measurements act as barriers because the tableau must collapse after each probabilistic outcome. In contrast, during sampling we do not collapse the tableau. Instead, Pauli frames absorb the randomness, allowing all measurement operators within a window to be processed fully in parallel across all shots.

Compared to state tableau in \cref{tableauFormat}, frames replace Pauli generators and their signs. 
Concretely, for $\numShots$ shots per frame (column), the stabilizer tableau $\tableauTilde=(\Xtilde, \Ztilde\,)$ is represented as two bit‐matrices of size $\numShots \text{ (rows)} \times \numQubits \text{ (columns)}$, but packs each row of \numQubits into words of size $\wordSize$. Recall $\numWords = \lceil \numQubits / \wordSize\rceil$ be the number of machine words needed per row. Then we store
$
\Xtilde, \Ztilde \in \{0,1\}^{\numShots\times\numQubits}
\equiv
\{0,\dots,2^{\wordSize}-1\}^{\numShots\times \numWords}
$
so that each word holds $\wordSize$ independent shots for a single qubit.

During sampling, every Clifford gate on shot‐bundle words simply updates these packed rows with bitwise operations, and each measurement step toggles bits in the Pauli frame randomly rather than collapsing the full tableau.  

\Cref{alg:sampling} exploits our frame simulation algorithm pulling out the randomness into those $\wordSize$‐bit flips, allowing us to process $\wordSize$ shots in parallel for all measured qubits simultaneously, achieving a dramatic speedup over conventional methods. 
We first allocate two packed matrices, $\Xtilde$ and $\Ztilde$, of size $\numShots\times\numWords$ (\cref{sam:allocateTableau}), plus an output buffer \record of the same shape (\cref{sam:allocateRecord}).
On the device, at  \li{sam:initTableau}, we reset $\Xtilde$ and fill $\Ztilde$ with independent Philox‐generated random words via cuRAND \cite{curand}, ensuring each word encodes $\wordSize$ fresh random bits.
The Philox family is well suited for parallel platforms as it can generate independent streams of random numbers without the need of synchronization \cite{parallelNumbers}.

\begin{algorithm}[htp]
	\setstretch{1.15}
	\DontPrintSemicolon
	\caption{Sampling on GPU}
	\label{alg:sampling}
	\midsmall
	\SetCommentSty{mycommfont}
	\SetArgSty{}
	\KwIn{$\circuitScheduled,\; \numQubits,\;\wordSize,\; \numShots$\tcp*[r]{\numShots (number of shots).}}
	\KwOut{\record \tcp*[r]{$\record \in \{0,\dots,2^{\wordSize}-1\}^{\numShots\times \numWords}$.}}
	\tcc{Main procedure for sampling.}
	$\tableauTilde, \numWords \gets \allocate$($\numShots \times 2\numQubits,\ \wordSize$) \label{sam:allocateTableau}	\tcp*[r]{Allocate $\numShots \times 2\numQubits$ bits in $\numWords=\lceil \numQubits/\wordSize\rceil$ words.}
	$\record \gets \allocate$($\numShots\times\numQubits,\ \wordSize$) \label{sam:allocateRecord}	\tcp*[r]{Allocate $\numShots\times\numQubits$ bits in \numWords words.}
	\begingroup \color{gpu}
	$\Xtilde \gets \{0\},\; \Ztilde \gets \randomize$() \label{sam:initTableau} \tcp*[r]{Initialize tableau on device.}
	\endgroup
	\ForAll(\tcp*[f]{For all windows.}){$\circuitWindow \in \circuitScheduled$}{ \label{sam:forSimulation} 
		$\circuitWindow_d \gets \copyWindowToDevice$($\circuitWindow$) \label{sam:copyWindow} \tcp*[r]{Copy gate layer to device.}
		\If(\tcp*[f]{Is \circuitWindow a measuring window?}){$\hasMeasurements(\circuitWindow)$}{
			\begingroup \color{gpu}
			$ \record \gets {\measureSample}(\tableauTilde, \circuitWindow_d, \numWords, \numQubits)$\;\label{sam:measure}
			\endgroup
		}
		\Else {
			\begingroup \color{gpu}
			$\simulateWindow$($\tableauTilde, \circuitWindow_d, \numWords$) \label{sam:applyGates} \tcp*[r]{Apply gate rules on device (\cref{alg:buildTableau}).}
			\endgroup
		}
	}
	\begingroup \color{gpu}
	\Kernel{\upshape{\measureSample}($\tableauTilde, \circuitWindow, \numWords, \numQubits$) $\to \record$}{  \label{sam:measureKernel}
		\Forpary(\tcp*[f]{Measuring-operator index.}){$0,\;|\circuitWindow_d|-1$}{ 	
			$\qubit \gets \routine{Qubit}(\circuitWindow_d[\tidy])$ \label{measure:getQubit} \tcp*[r]{Get measurable qubit.}
			\Forparx(\tcp*[f]{Word-shot index.}){$0,\;\numWords-1$}{ 
				$\wordidx \gets \qubit\times\numWords + \tidx$\;
				$\record[\wordidx] \gets \Xtilde[\wordidx]$ \tcp*[r]{Record \qubit's frame from \(\Xtilde\).}
				$\Ztilde[\wordidx] \gets \randomizeWord(\wordidx)$ \tcp*[r]{Randomize \qubit's frame in \(\Ztilde\).}
			}
		}
	}
	\endgroup
\end{algorithm}

Next, we simulate each scheduled window in turn. Non‐measurement windows invoke our standard $\simulateWindow$ kernel to update $\Xtilde,\Ztilde$ under Clifford gates (\li{sam:applyGates}). Measurement windows run the $\measureSample$ (\li{sam:measure}) kernel. This kernel launches a 2D grid: for every measured qubit and each frame-word index, we copy the current $\Xtilde$ word (outcome of \wordSize shots) into \record, then re‐randomize the corresponding $\Ztilde$ word for subsequent measurements. 

\paragraph{Complexity.}
For each window in $\circuitScheduled$, sampling performs either
(1) a call to \simulateWindow\ (Algorithm~\ref{alg:buildTableau}) for
Clifford gates, or
(2) a call to \measureSample\ (Algorithm~\ref{alg:sampling}) for
measurement windows.

Recall from \cref{sec:applyGates} that the cost of \simulateWindow is
$\bigO(\numQubits^2)$ total work per window, yielding an ideal
parallel time of
$\bigO(\frac{\numQubits^2}{\maxThreads} + \log\blockDim)$
due to the binary-tree sign aggregation.

For measurement windows, the \measureSample\ kernel performs
$\bigO(\numWords \times |\circuitWindow_d|)$ work.
In the worst case, both scale as $\bigO(\numQubits)$,
giving $\bigO(\numQubits^2)$ total work and an ideal parallel time of
$\bigO(\frac{\numQubits^2}{\maxThreads})$.
Unlike gate application, no logarithmic synchronization is required.

Because windows are maximal, the number of scheduled windows is
minimized. Independent Clifford gates are applied together via \simulateWindow, and independent measurement operators are processed
together via \measureSample, each in a single kernel launch.
Therefore, the overall sampling complexity scales with the circuit
depth rather than with the total number of gates or measurements.

\myparagraph{Comparison with \stim.}
In \stim, measurements are processed sequentially over qubits.
Each frame update touches $\mathcal{O}(\numQubits)$ bits,
and over $\mathcal{O}(\numQubits)$ measured qubits this results in
$\mathcal{O}(\numQubits^2)$ work per shot batch.
Moreover, the qubit-wise iteration is not parallelized across shots,
so increasing $\numShots$ increases total work proportionally.

In contrast, our $\measureSample$ kernel parallelizes both the
qubit dimension and the shot-word dimension.
Given sufficient hardware threads to cover
$\numShots \times \numWords$ words,
each measurement window has constant parallel span,
independent of the number of shots.

%% file: experiments.tex
\section{Experiments and Discussion}\label{sec:experiments}

\paragraph{Setup.}
We implemented \cref{alg:parallelSimulation} in a new CUDA C\code{++} tool called \ourTool\footnote{\url{https://github.com/System-Verification-Lab/QuaSARQ}}, compiled with
CUDA~12.6 targeting compute capability~8.9. GPU experiments were run with a timeout of 4~hours on a machine running \textsc{Ubuntu~22.04} and equipped with an RTX~4090 GPU
(16{,}384 cores at 2.23~GHz and 24~GB of global memory).

We compare \ourTool against six leading simulators:  \stim\footnote{\url{https://github.com/quantumlib/Stim}}, \qiskitAer (both CPU and GPU modes)\footnote{\url{https://github.com/Qiskit/qiskit-aer}}, 
\qibo\footnote{\url{https://github.com/qiboteam/qibo}}, \cirq\footnote{\url{https://github.com/quantumlib/Cirq}}, and \pennylane\footnote{\url{https://github.com/PennyLaneAI/pennylane}}, with their recommended Clifford-simulation settings.
Note that {\pennylane}’s Lightning‐GPU plugin and {\cirq}’s GPU alternative currently do not scale beyond tenths of qubits without exhausting device memory.
While \stim is CPU-only, it exploits SIMD instruction sets
(SSE/AVX) to process 128–512 bits in parallel, significantly accelerating tableau operations.

All CPU experiments were conducted separately on the DAS-6 cluster~\cite{DAS6}, whose
nodes are equipped with AMD EPYC~7282 CPUs (2.8~GHz) and 256~GB of memory. Each circuit
was executed in isolation on a dedicated node with a 4~hour timeout. Notably, the DAS-6
CPU clock is approximately 20\% faster than the GPU clock used by \ourTool. 

\paragraph{Benchmarks.} 
Since the computational cost of stabilizer-circuit simulation is determined entirely by the circuit structure (qubit count, gate count, and measurement locations), performance can be characterized using randomly generated circuits without evaluating a broad range of quantum algorithms~\cite{aaronson2008improved,stim,qiskit2024}. Accordingly, we constructed two benchmark suites, each comprising three sets of random OpenQASM circuits with depths $d\in\{100,500,1000\}$, generated using IBM’s Qiskit.\footnote{\url{https://github.com/Qiskit/qiskit}}

The \emph{light} suite contains 100 circuits per depth, with qubit counts ranging from
100 to 10{,}000 in steps of 100. This suite targets simulators with modest resource
requirements (e.g., \pennylane, \cirq, and \qiskitAer) while also evaluating \ourTool on
smaller instances. The \emph{heavy} suite spans 1{,}000 to 180{,}000 qubits in steps of
1{,}000, yielding 180 circuits per depth and stressing large-scale simulators such as
\stim. Larger circuits would exceed the 24~GB GPU memory limit, and \stim also times out
beyond this scale. Across both suites, circuit sizes range from 7K to 130M gates. By default, Qiskit generates uniformly distributed Clifford gates from the set
$\{\X,\Y,\Z,\Had,\Phase,\Phase^\dagger,\CX,\fontFormat{CY},\fontFormat{CZ},
\fontFormat{SWAP},\fontFormat{ISWAP}\}$, all supported by the evaluated simulators and decomposable into the basic Clifford gates $\{\X,\Y,\Z,\Had,\Phase,\CX\}$.

To assess measurements performance (runtime and energy consumption) across all simulators, we augmented each circuit with a randomized set of measurement operators.  We then executed a single simulation run for each circuit, which suffices to capture the impact of our GPU‐based measurement kernels on overall performance without the need for multiple repetitions. For sampling performance, we compared \stim’s built‐in high‐performance sampler (configured for 1,024 shots) against our Pauli‐frame sampler outlined in \cref{alg:sampling}.
\paragraph{Experiments.} 

\begin{figure*}[htp]
	\centering
	\captionsetup[subfloat]{labelfont=normal,textfont=normal}
	
	\adjustbox{max width=.9\linewidth}{
		\subfloat[Runtime cactus ($\depth=100$)]{
			\label{fig:time_others_d100}
			\includegraphics[width=\linewidth]{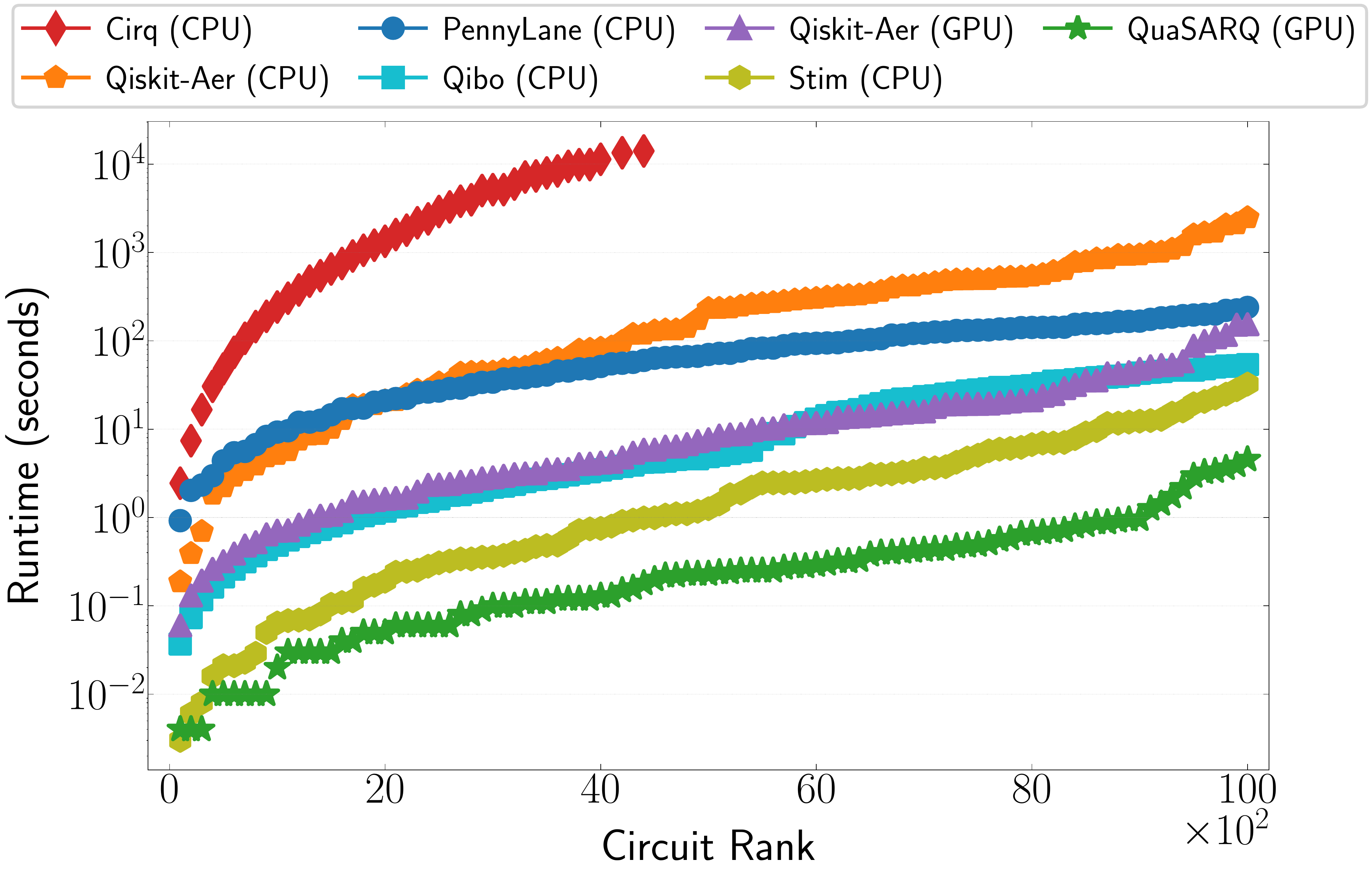}
		}\quad
		\subfloat[Energy cactus ($\depth=100$)]{
			\label{fig:energy_others_d100}
			\includegraphics[width=\linewidth]{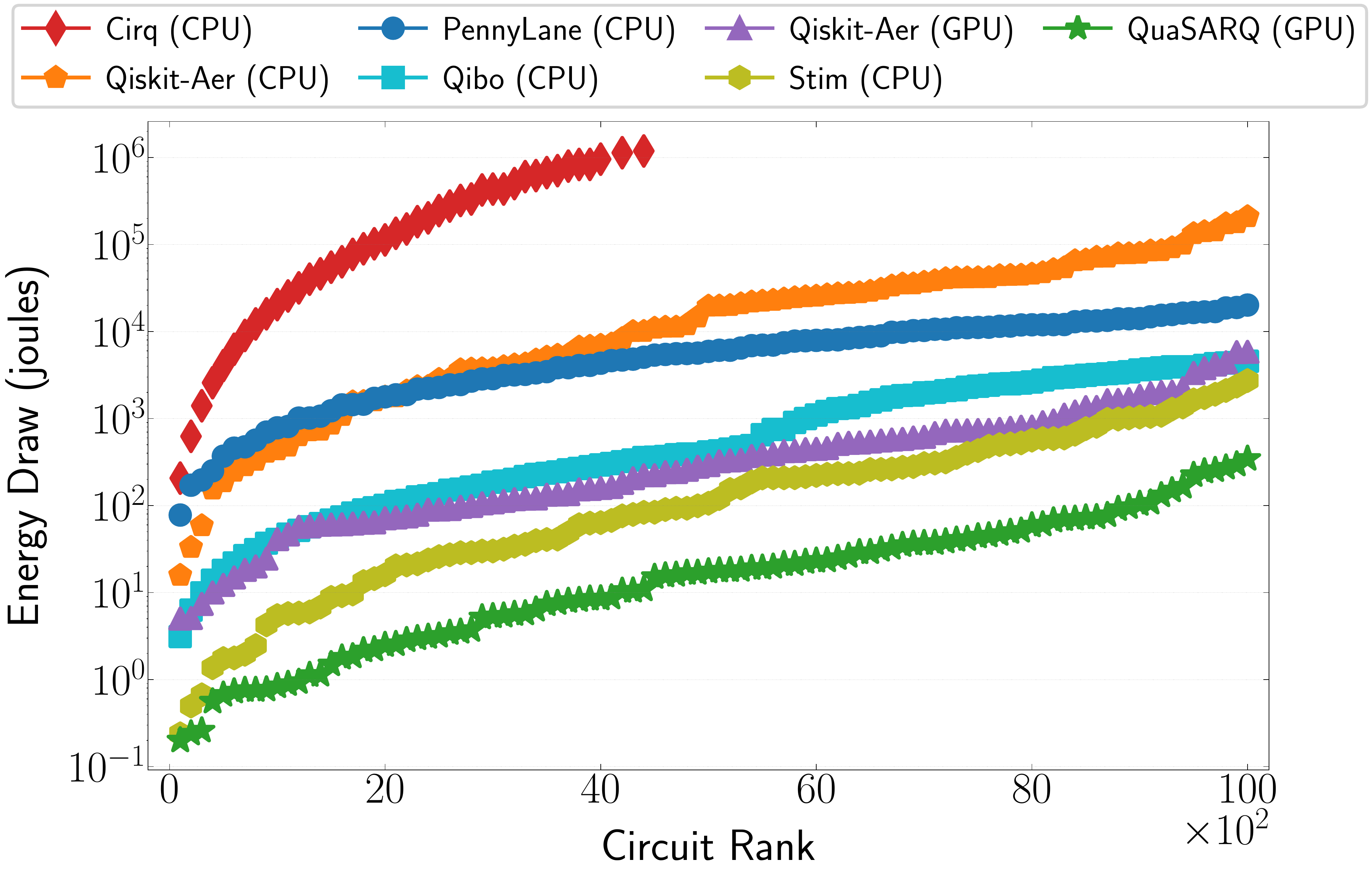}
		}
	}
	\\[.5em]
	
	\adjustbox{max width=.9\linewidth}{
		\subfloat[Runtime cactus ($\depth=500$)]{
			\label{fig:time_others_d500}
			\includegraphics[width=\linewidth]{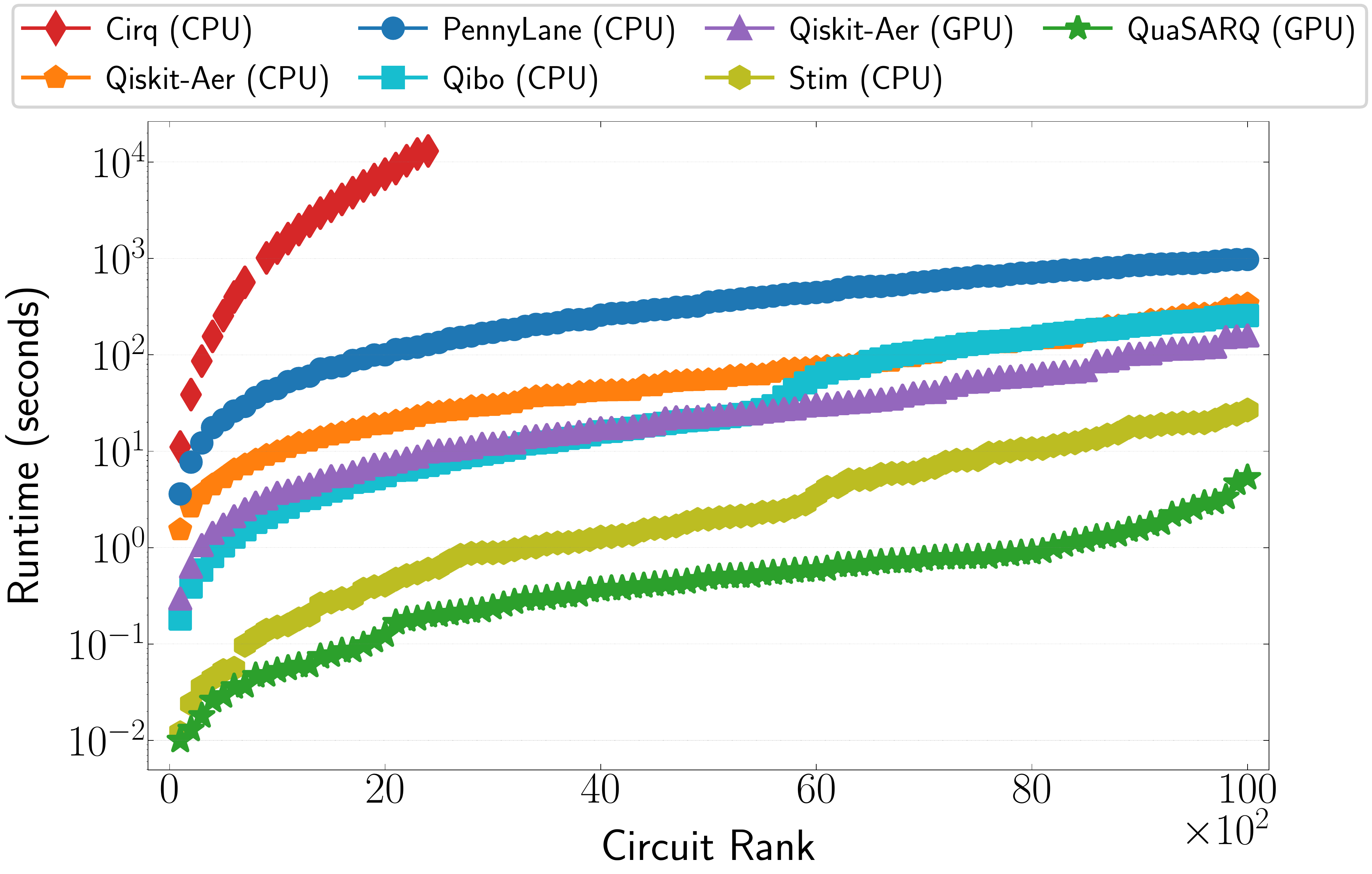}
		}\quad
		\subfloat[Energy cactus ($\depth=500$)]{
			\label{fig:energy_others_d500}
			\includegraphics[width=\linewidth]{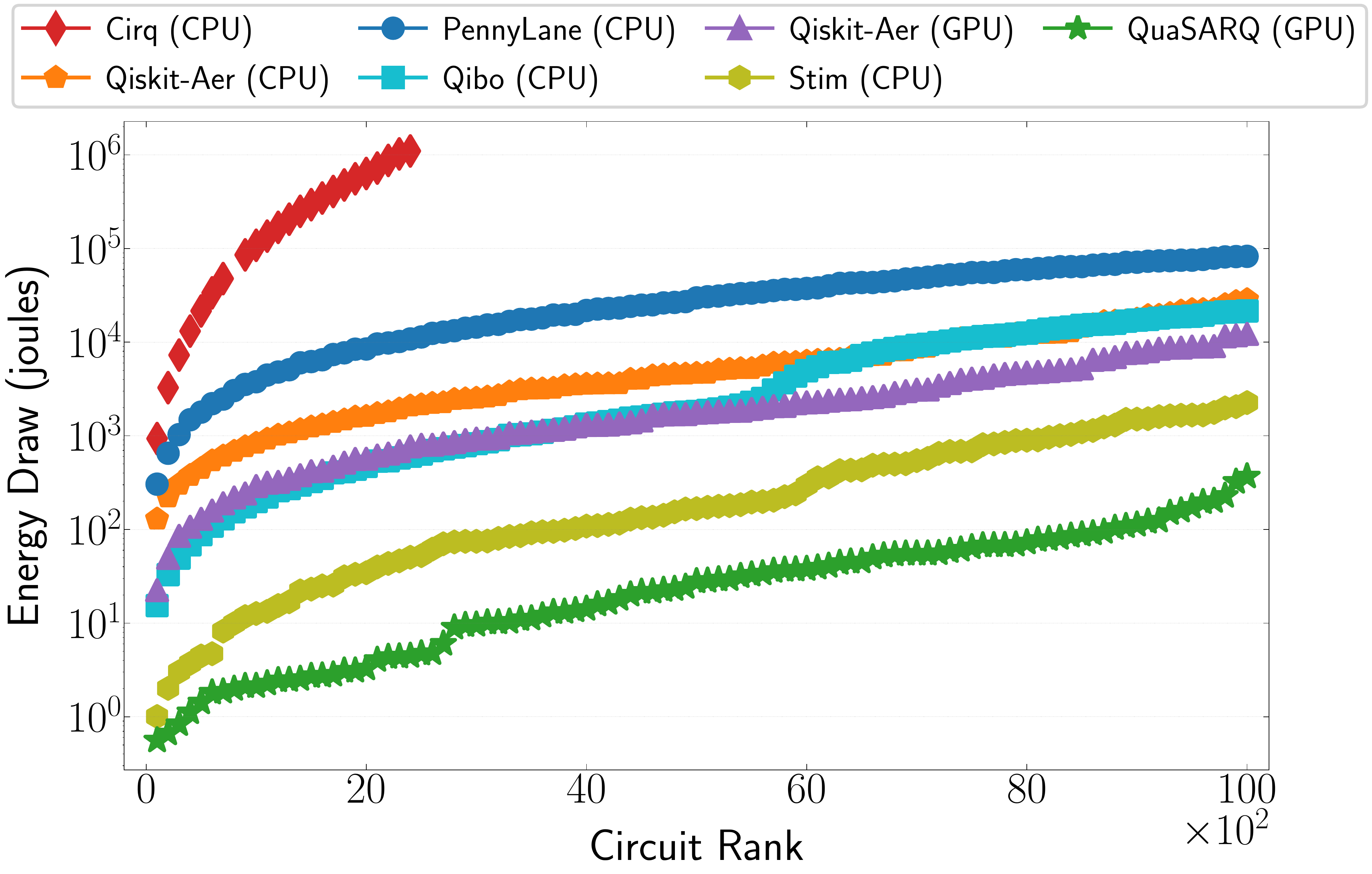}
		}
	}
	\\[.5em]
	
	\adjustbox{max width=.9\linewidth}{
		\subfloat[Runtime cactus ($\depth=1000$)]{
			\label{fig:time_others_d1000}
			\includegraphics[width=\linewidth]{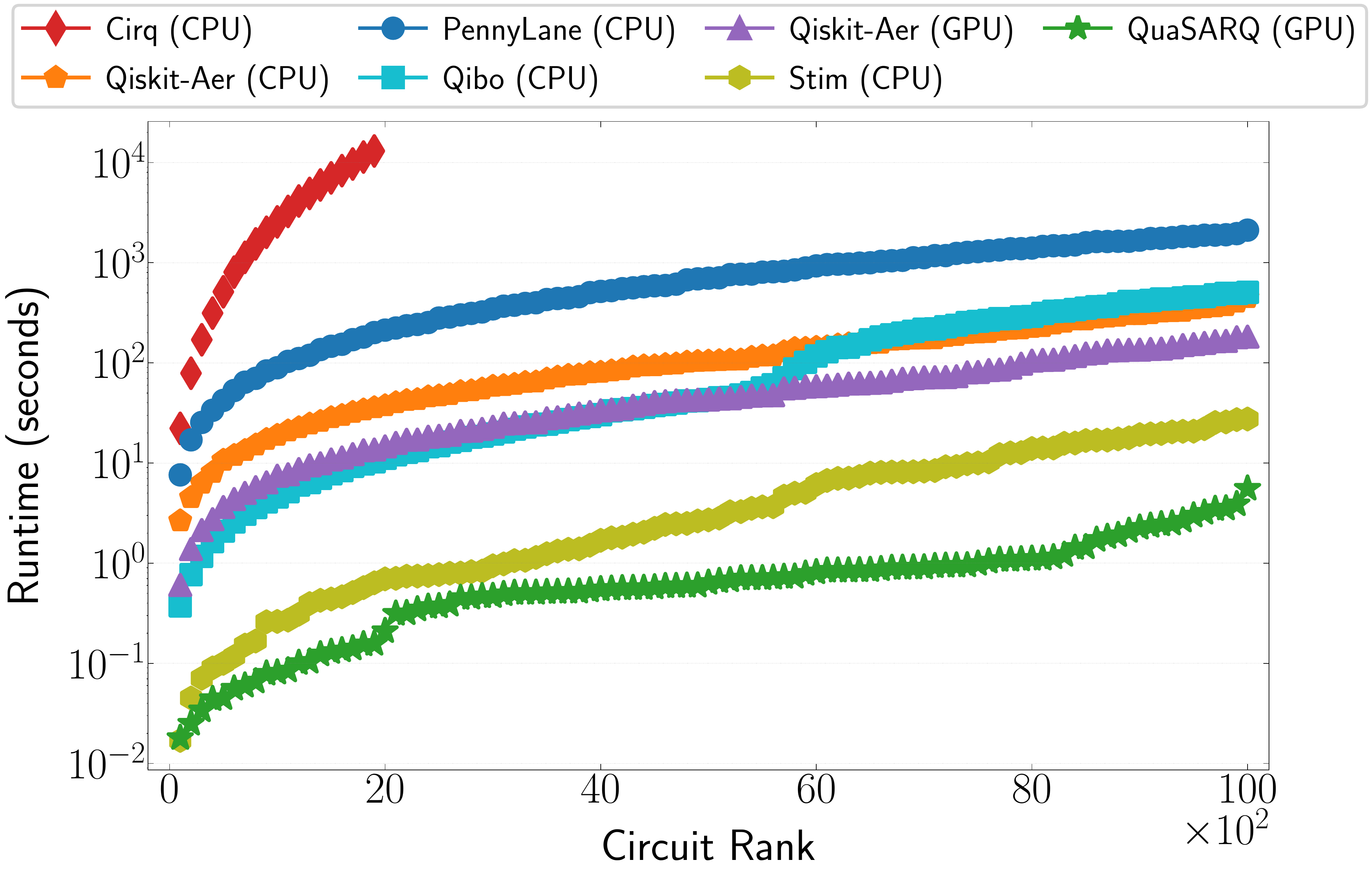}
		}\quad
		\subfloat[Energy cactus ($\depth=1000$)]{
			\label{fig:energy_others_d1000}
			\includegraphics[width=\linewidth]{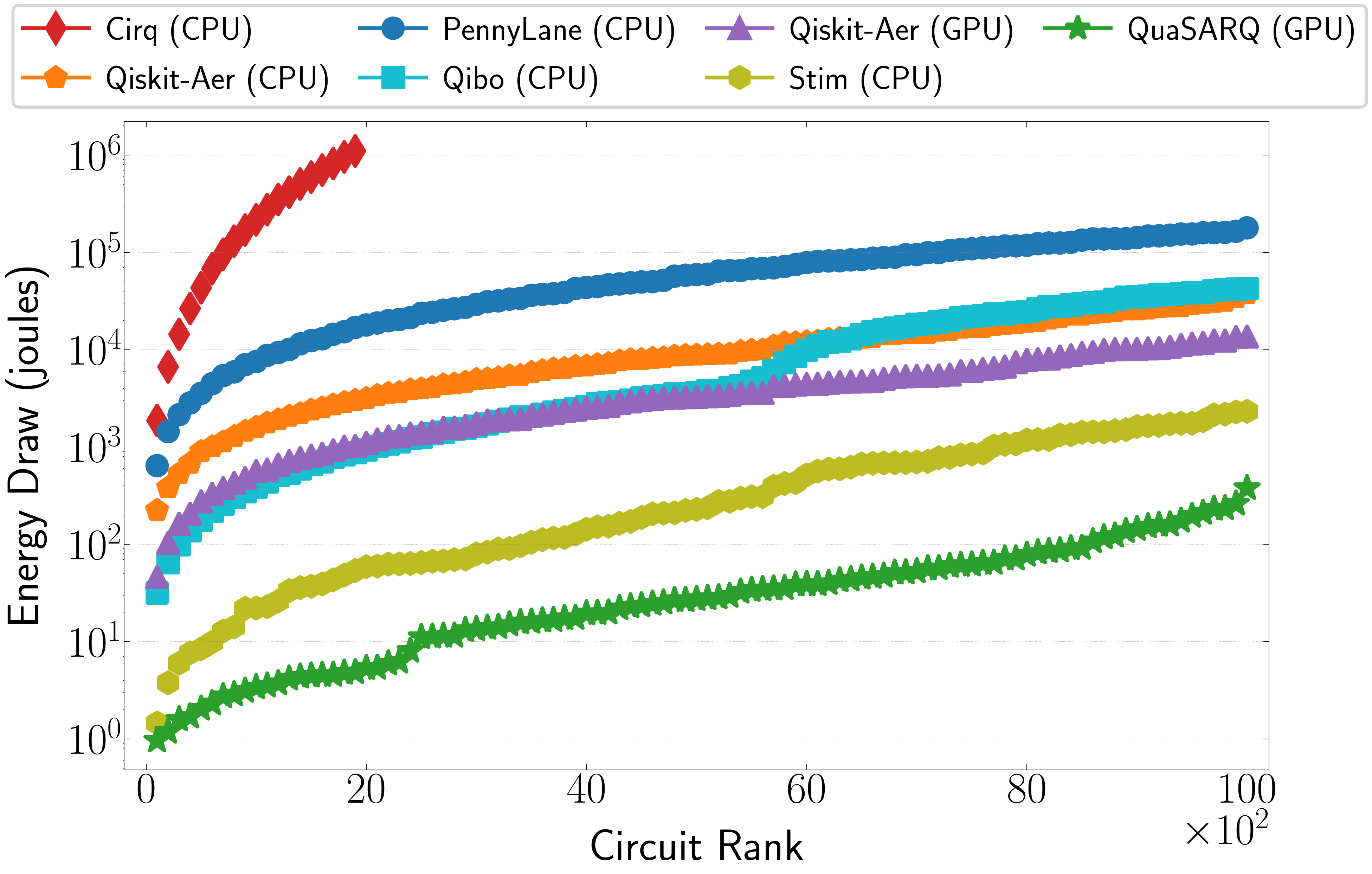}
		}
	}
	
	\caption{Simulation performance of \ourTool vs. \pennylane, \qibo, \qiskitAer (both CPU and GPU versions), \cirq, and \stim on 100 circuits (qubits \(\in [100, 10K]\), depth (\depth) \(\in \{100, 500, 1000\}\)).}
	\label{fig:performanceOthers}
\end{figure*}

\Cref{fig:performanceOthers} presents cactus plots comparing runtime and energy consumption of our GPU‐based \ourTool against six state-of-the-art simulators: \pennylane, \qibo, \stim, \qiskitAer (GPU), \qiskitAer (CPU), and \cirq on the light benchmark suite. In these plots, the x-axis shows the circuit rank, defined as the position of a circuit after sorting all benchmarks by increasing runtime or energy. 

The runtime in \ourTool includes the initial time (memory allocations, parsing, etc.), the scheduling time to obtain \circuitScheduled, the transfer time to send \circuitScheduled to the device, and the simulation time of \cref{alg:parallelSimulation}. To measure the energy draw, we multiplied the runtime to the average power consumed by the processing unit over the life cycle of running tools.

At depth 100, \pennylane outperforms both \qiskitAer (CPU) and \cirq on nearly all circuits. As qubit counts exceed roughly 4,000, \qiskitAer (GPU) begins to rival \qibo and \stim, occasionally surpassing them on the largest instances. Even so, \ourTool consistently outperforms all competitors by a substantial margin across the entire qubit range.

At depths 500 and 1,000, \qiskitAer (CPU) overtakes \pennylane on all circuits, while \qiskitAer (GPU) surpasses \qibo beyond approximately 6,000 qubits. In this regime, \stim regains an advantage over \qiskitAer (GPU). Nevertheless, \ourTool remains the fastest simulator at every tested instance, maintaining a clear lead that becomes increasingly pronounced beyond 5,000 qubits.

Regarding energy consumption, each point in the right-hand scatter plots reports total energy usage in joules (runtime multiplied by average power). \ourTool reduces energy consumption by approximately the same factors as its runtime speedups. Across all depths and qubit counts, it consistently consumes far less energy than both CPU- and GPU-based alternatives, reflecting not only its shorter runtimes but also the efficiency of executing measurement-intensive operations on the GPU.

\begin{figure*}[htp]
	\centering
	\captionsetup[subfloat]{labelfont=normal,textfont=normal}
	
	\adjustbox{max width=.9\linewidth}{
		\subfloat[Runtime Cactus ($\depth=100$)]{
			\label{fig:time_stim_d100}
			\includegraphics[width=\linewidth]{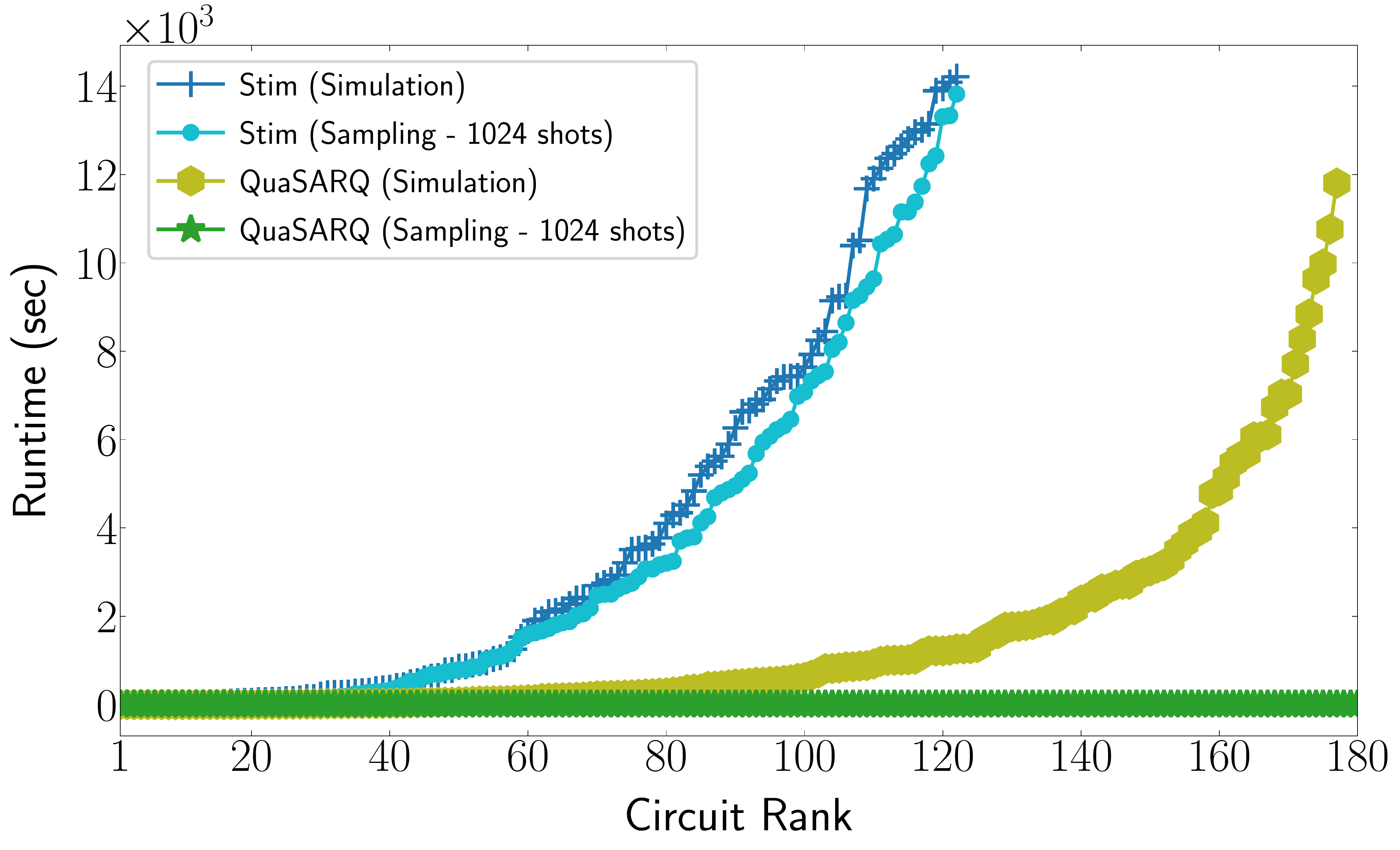}
		}\quad
		\subfloat[Energy vs qubits count ($\depth=100$)]{
			\label{fig:energy_stim_d100}
			\includegraphics[width=\linewidth]{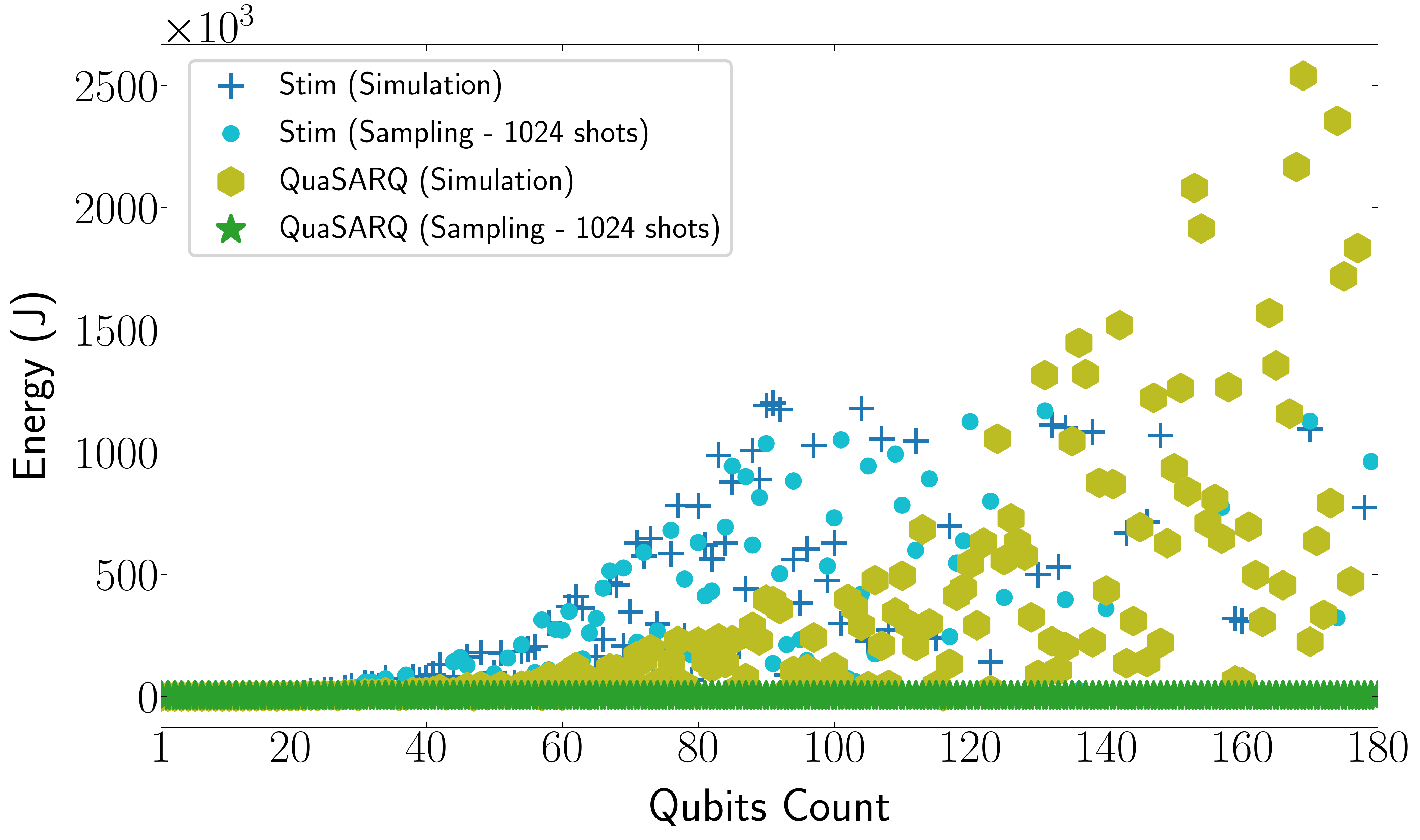}
		}
	}
	\\[.5em]
	
	\adjustbox{max width=.9\linewidth}{
		\subfloat[Runtime Cactus ($\depth=500$)]{
			\label{fig:time_stim_d500}
			\includegraphics[width=\linewidth]{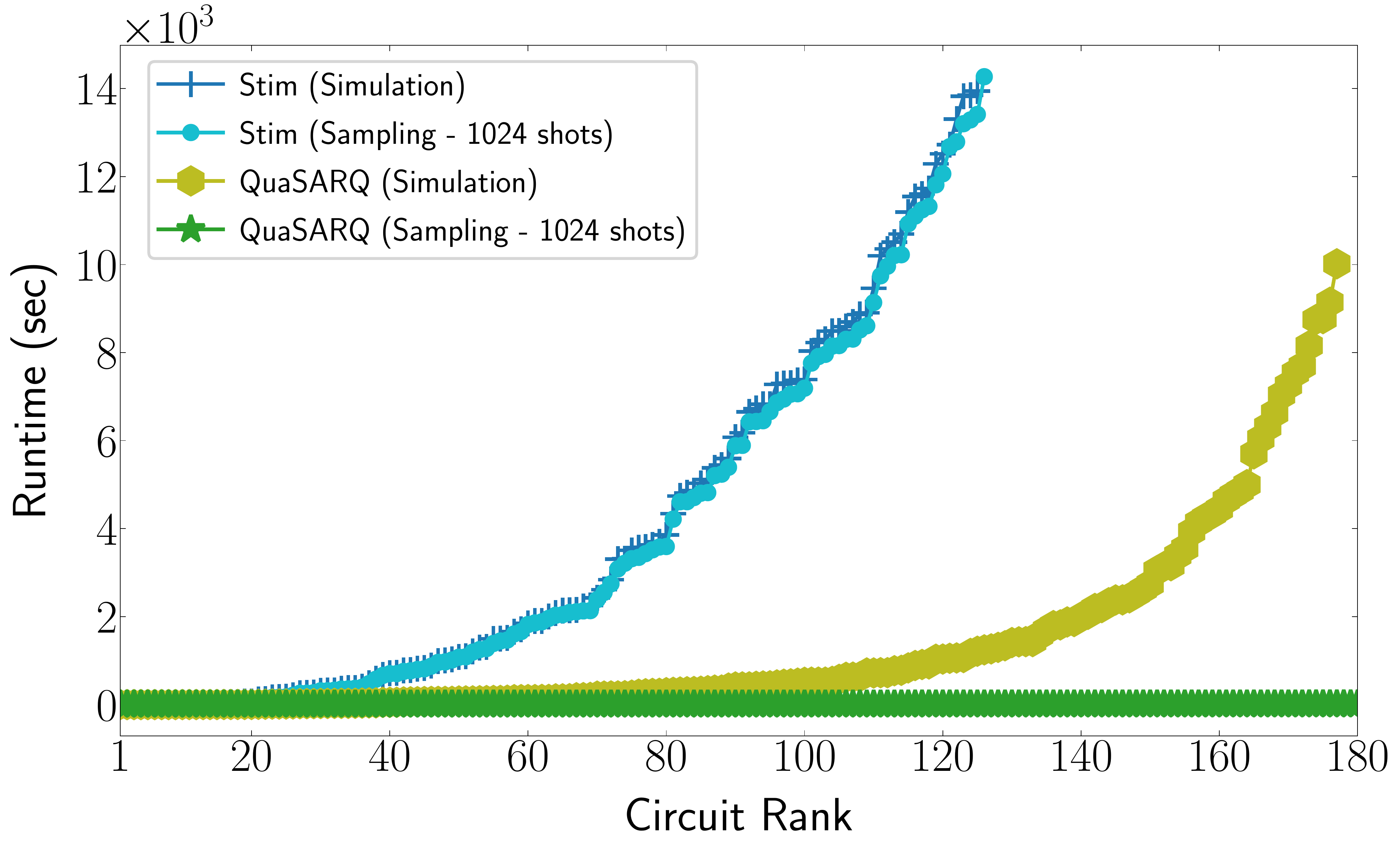}
		}\quad
		\subfloat[Energy vs qubits count ($\depth=500$)]{
			\label{fig:energy_stim_d500}
			\includegraphics[width=\linewidth]{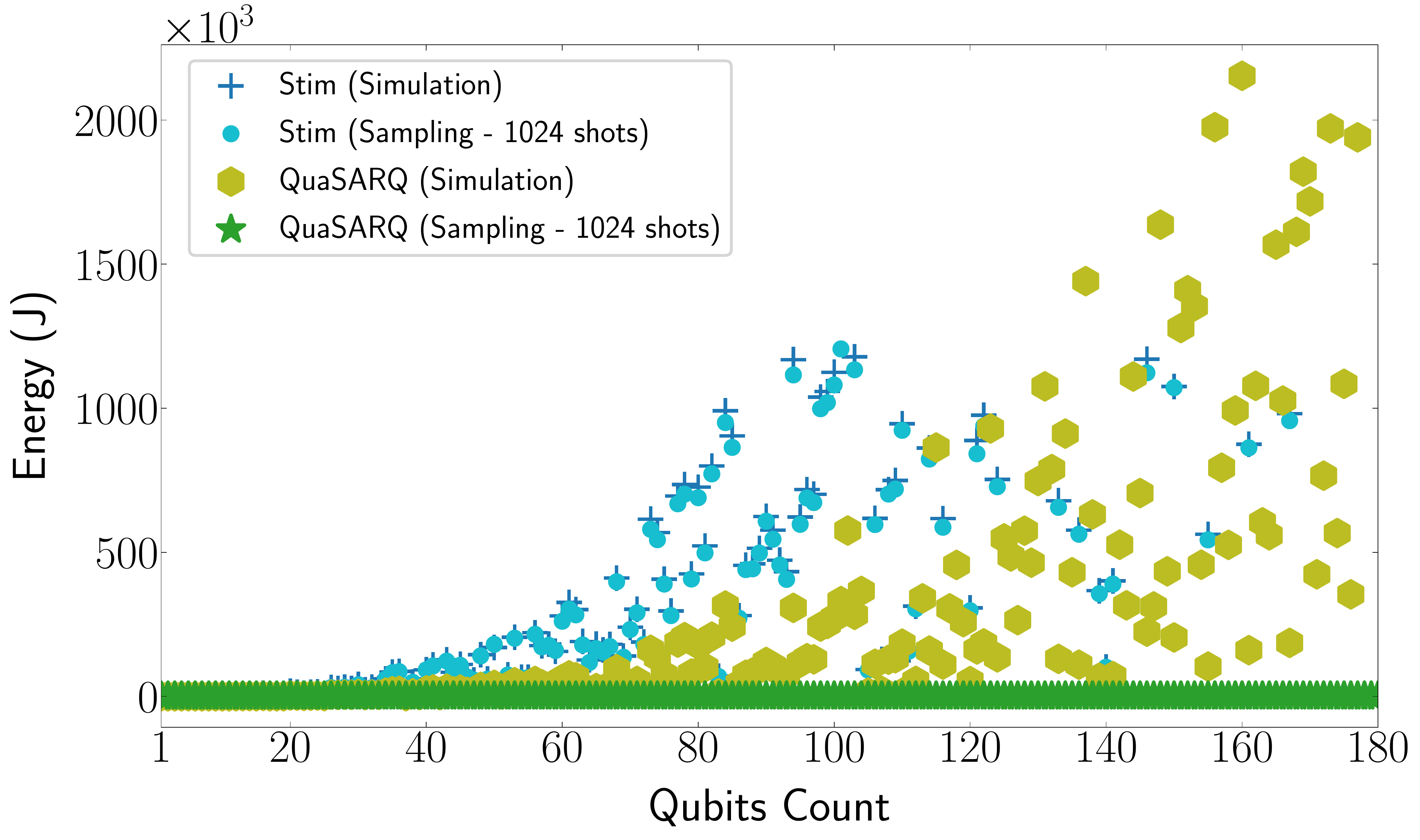}
		}
	}
	\\[.5em]
	
	\adjustbox{max width=.9\linewidth}{
		\subfloat[Runtime Cactus ($\depth=1000$)]{
			\label{fig:time_stim_d1000}
			\includegraphics[width=\linewidth]{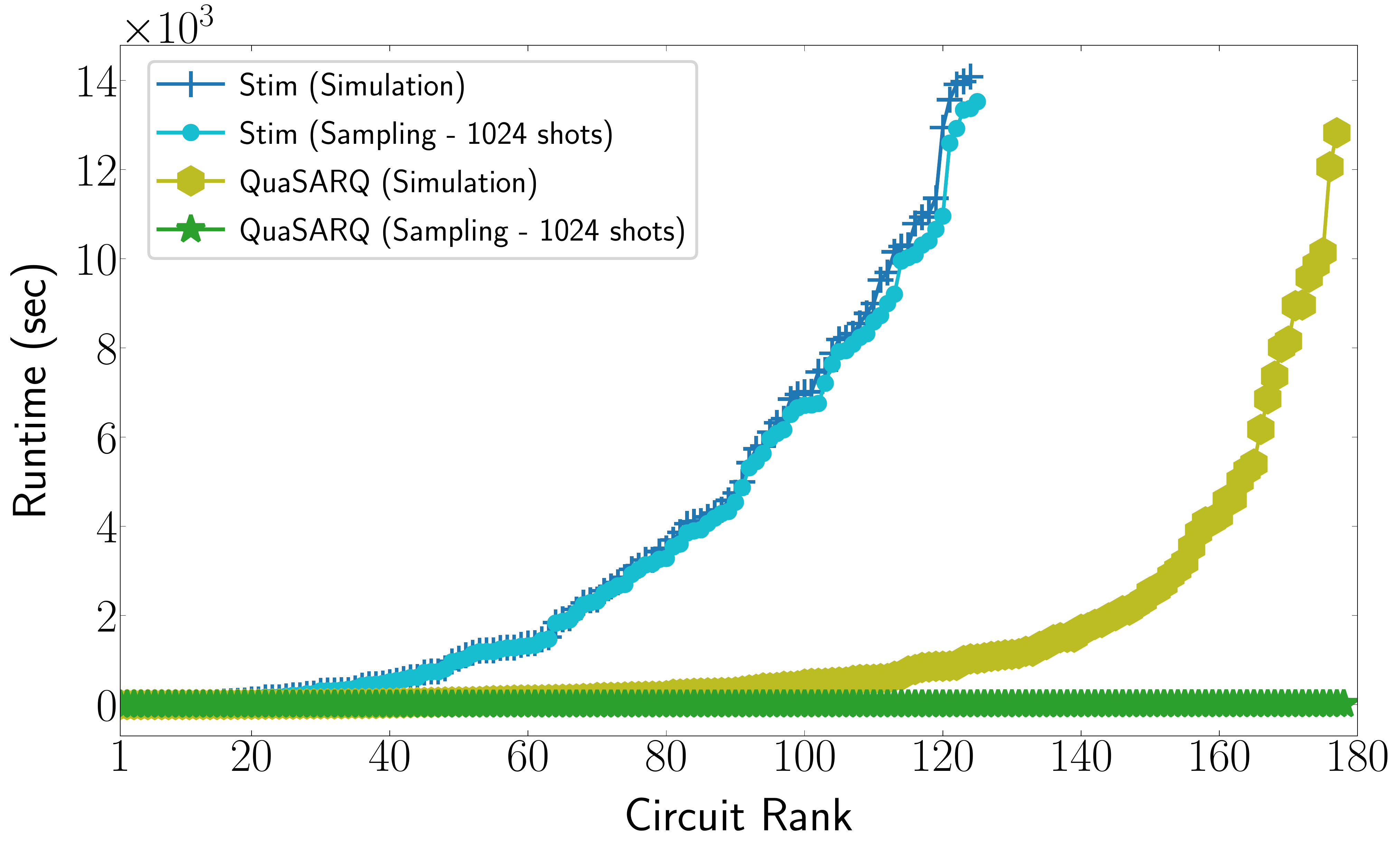}
		}\quad
		\subfloat[Energy vs qubits count ($\depth=1000$)]{
			\label{fig:energy_stim_d1000}
			\includegraphics[width=\linewidth]{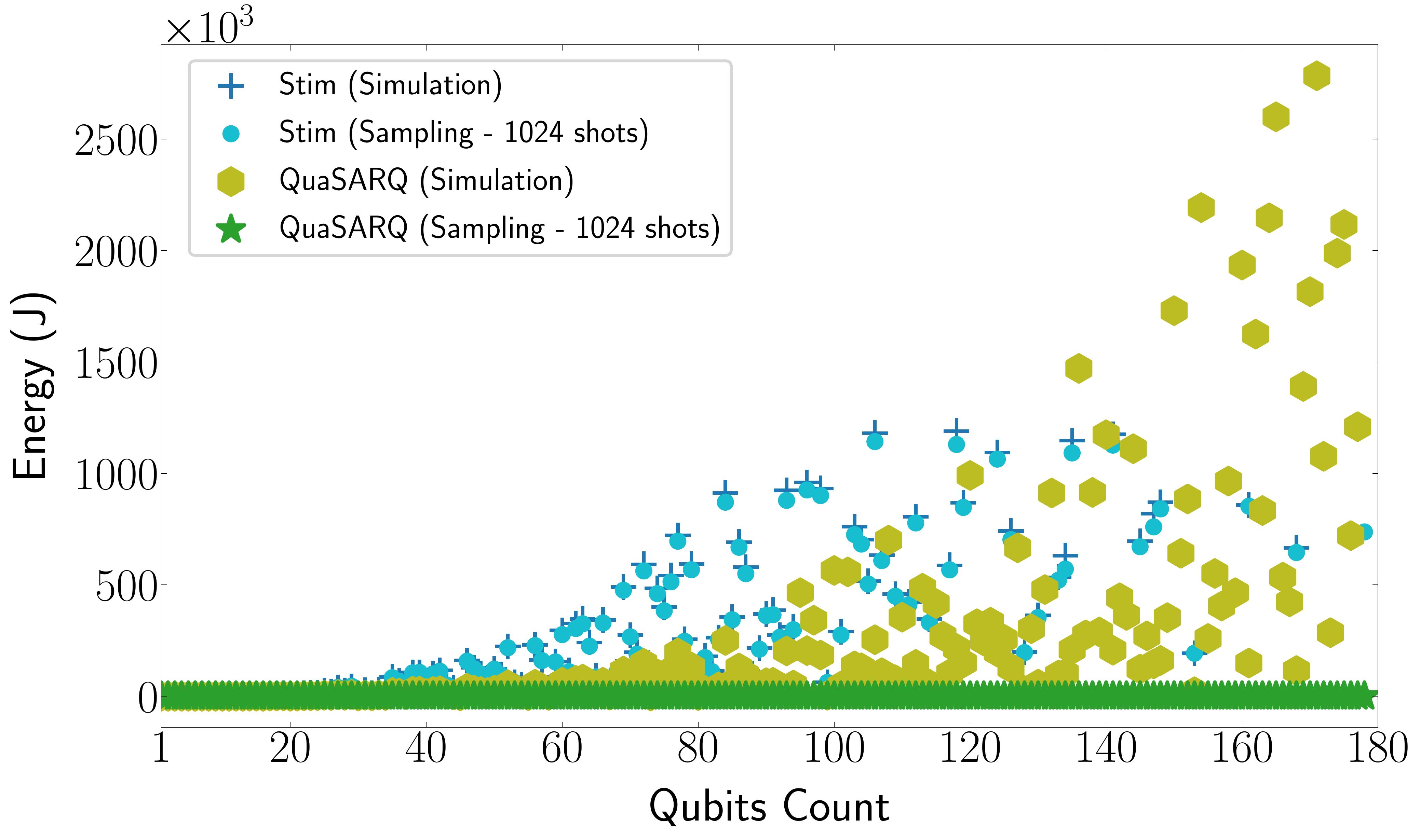}
		}
	}
	
	\caption{Simulation and sampling performance of \ourTool\ vs.\ \stim\ on 180 circuits (qubits \(\in [1K, 180K]\), \depth \(\in \{100, 500, 1000\}\)).}
	\label{fig:performance}
\end{figure*}

Based on \cref{fig:performanceOthers}, we selected \stim as the strongest baseline for a detailed comparison with \ourTool on the heavy benchmark suite. \Cref{fig:performance} compares the two simulators on this suite, reporting both single-shot simulation and many-shot sampling (1,024 shots) for circuits with 1K–180K qubits and depths up to 1,000. In the left-hand cactus plots, circuits are ordered by total runtime. \ourTool completes 177 circuits within 72 hours, whereas \stim completes only 125 circuits in 132 hours, demonstrating a substantial performance advantage at large scales.

This gap prevails for sampling: \ourTool’s Pauli-frame sampler exhibits flat runtime across circuit sizes, while \stim’s sampling cost increases steeply, resulting in speedups of up to several orders of magnitude. The right-hand scatter plots further show that these runtime improvements translate directly into energy savings, with \ourTool reducing energy consumption by over 80\% on average and by more than 90\% for sampling circuits. 

Overall, these graphs demonstrate that our GPU‐based algorithm delivers both faster runtime and dramatically lower energy consumption by over 90\% savings
On the heavy suite, \ourTool completes 42 more circuits within the same time window, extending the clear lead already observed on the light suite (100–10K qubits) over \pennylane, \cirq, and both CPU and GPU variants of \qiskitAer. Together, these findings highlight the superior scalability and efficiency of \ourTool across problem sizes ranging from a few hundred to nearly two hundred thousand qubits.

\begin{figure*}[htp]
	\centering
	\captionsetup[subfloat]{labelfont=normal,textfont=normal}
	
	\adjustbox{max width=.9\linewidth}{
		\subfloat[\stim ($\depth=100$)]{
			\label{fig:measure_stim_d100}
			\includegraphics[width=\linewidth]{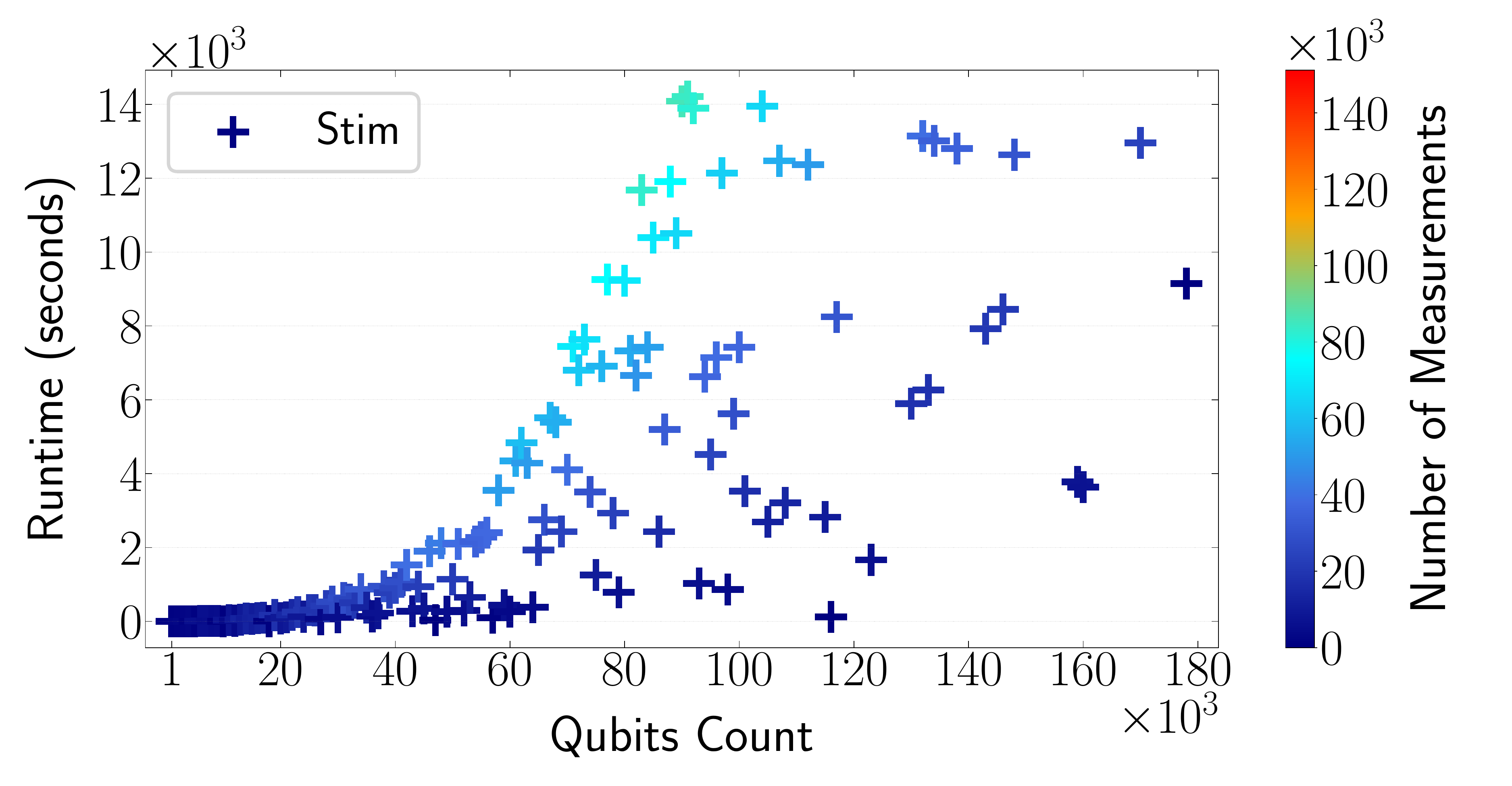}
		}\quad
		\subfloat[\ourTool ($\depth=100$)]{
			\label{fig:measure_quasarq_d100}
			\includegraphics[width=\linewidth]{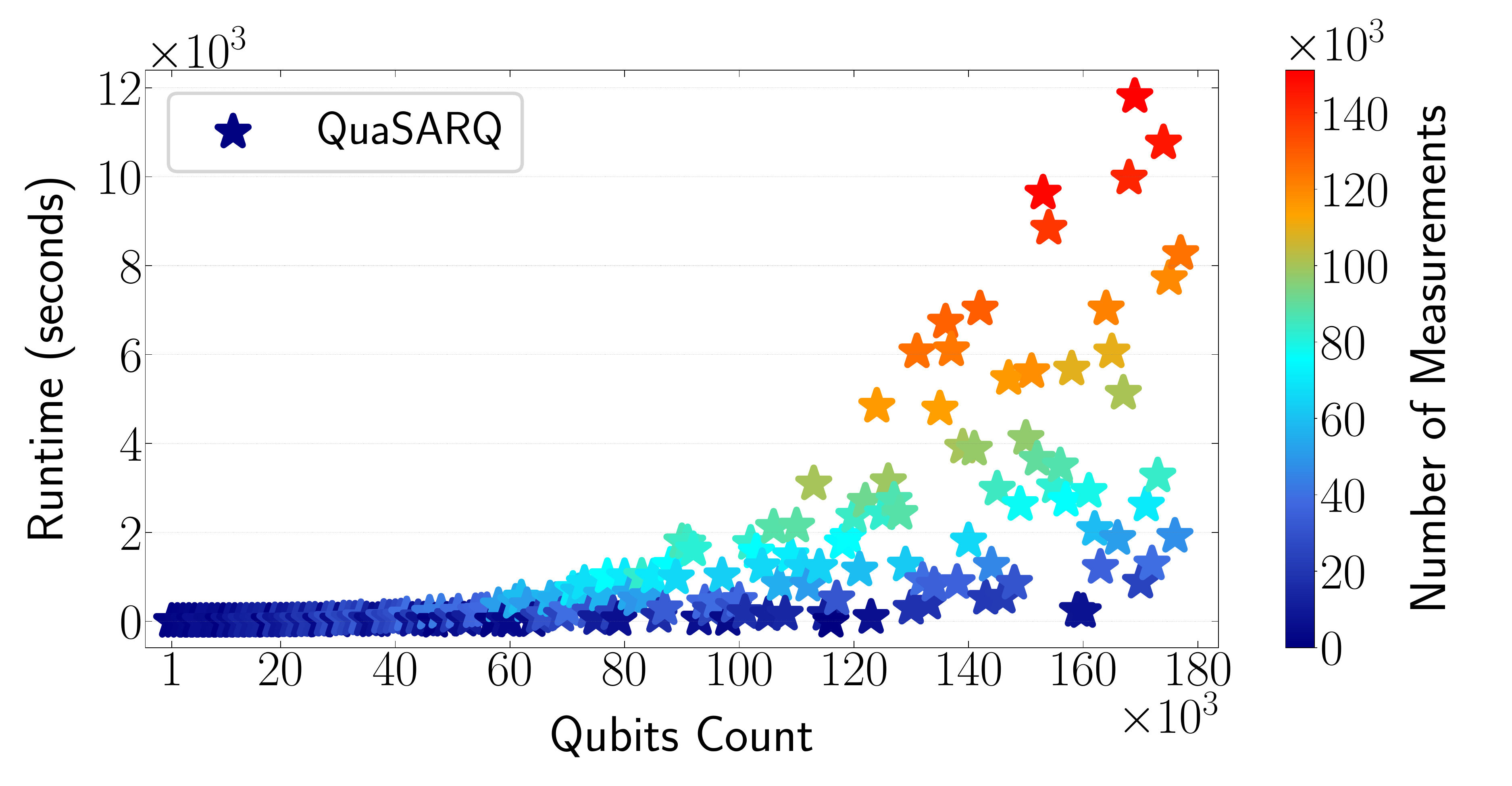}
		}
	}
	\\[.5em]
	
	\adjustbox{max width=.9\linewidth}{
		\subfloat[\stim ($\depth=500$)]{
			\label{fig:measure_stim_d500}
			\includegraphics[width=\linewidth]{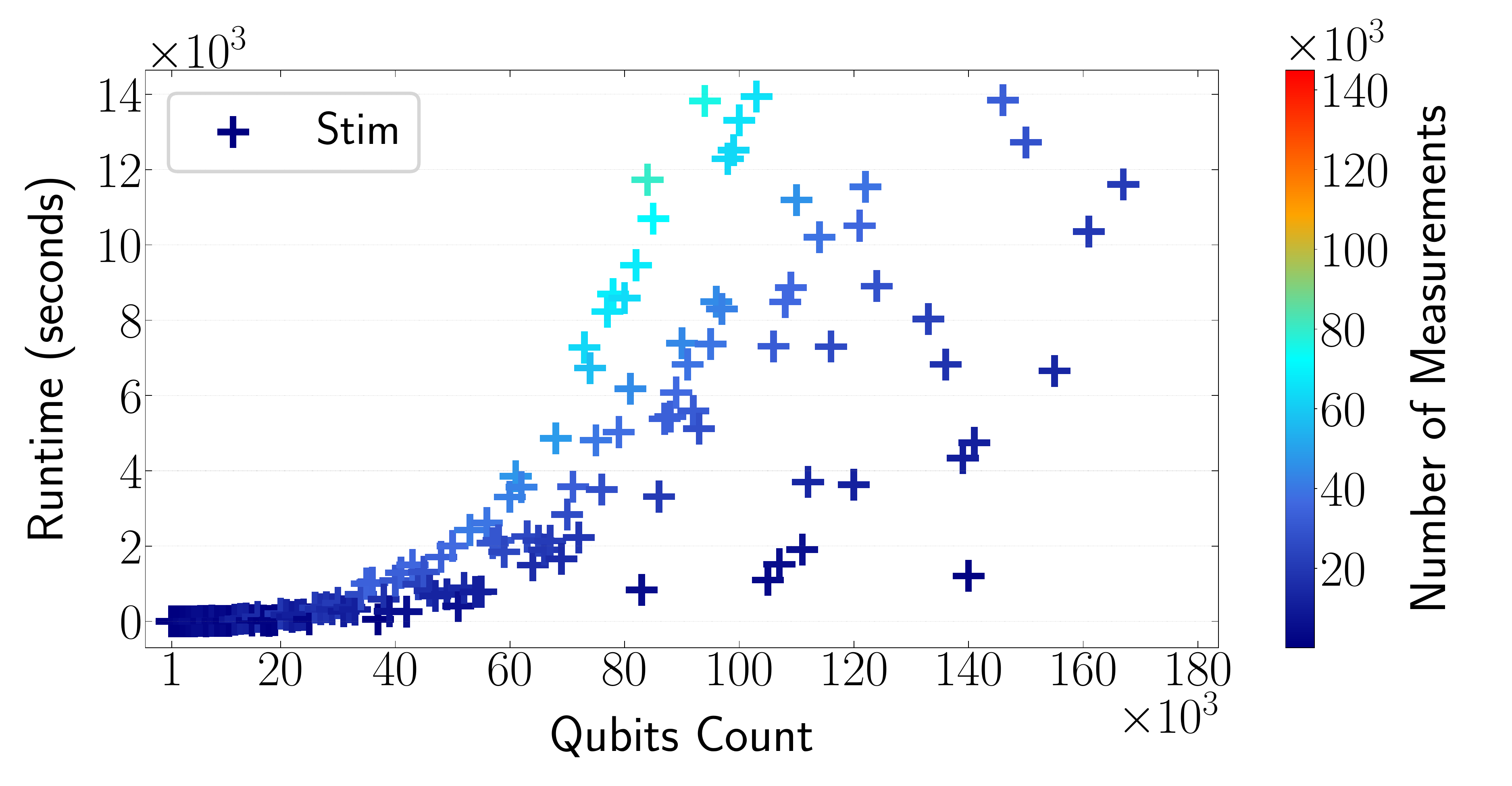}
		}\quad
		\subfloat[\ourTool ($\depth=500$)]{
			\label{fig:measure_quasarq_d500}
			\includegraphics[width=\linewidth]{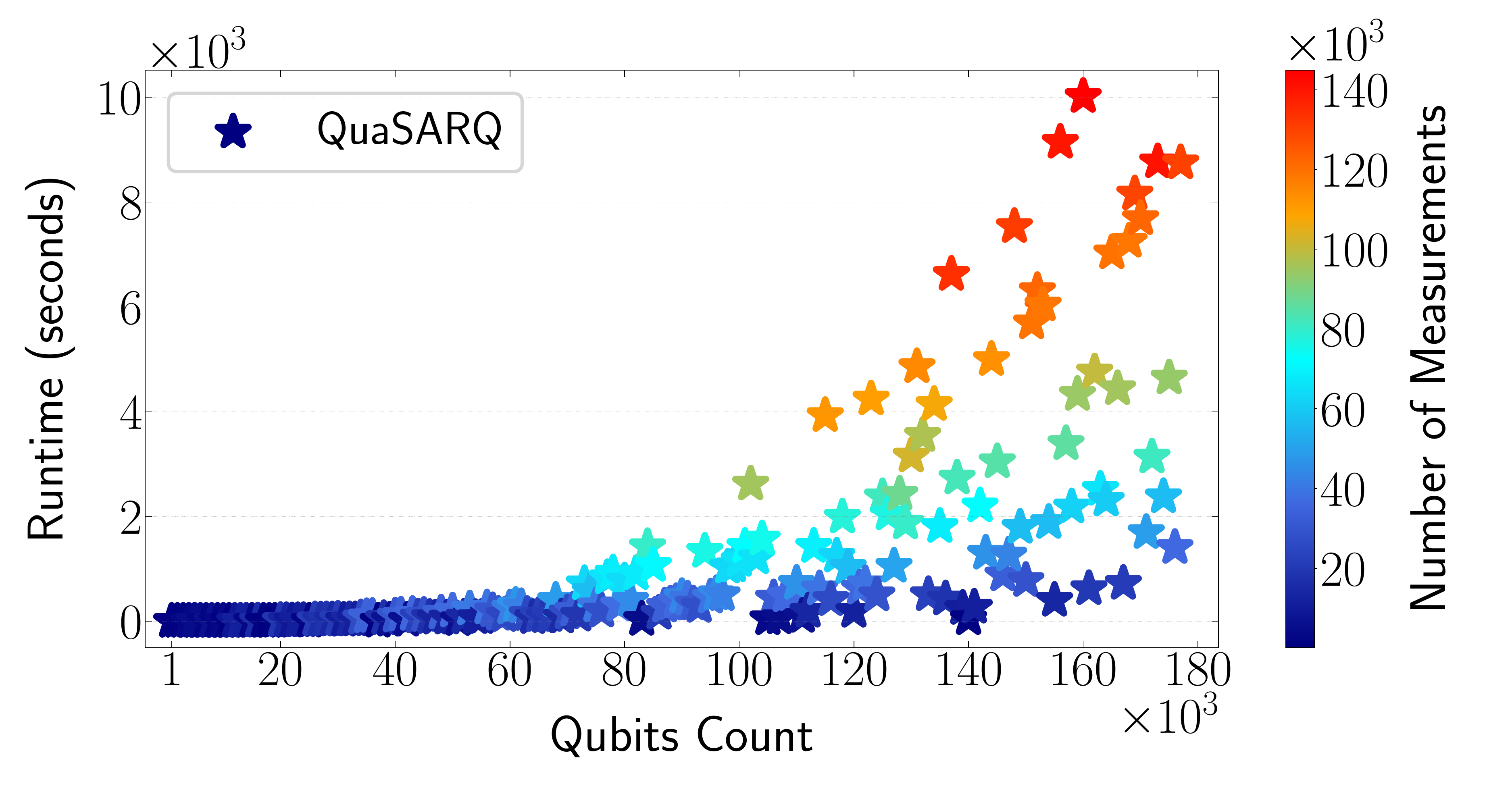}
		}
	}
	\\[.5em]  
	
	\adjustbox{max width=.9\linewidth}{
		\subfloat[\stim ($\depth=1000$)]{
			\label{fig:measure_stim_d1000}
			\includegraphics[width=\linewidth]{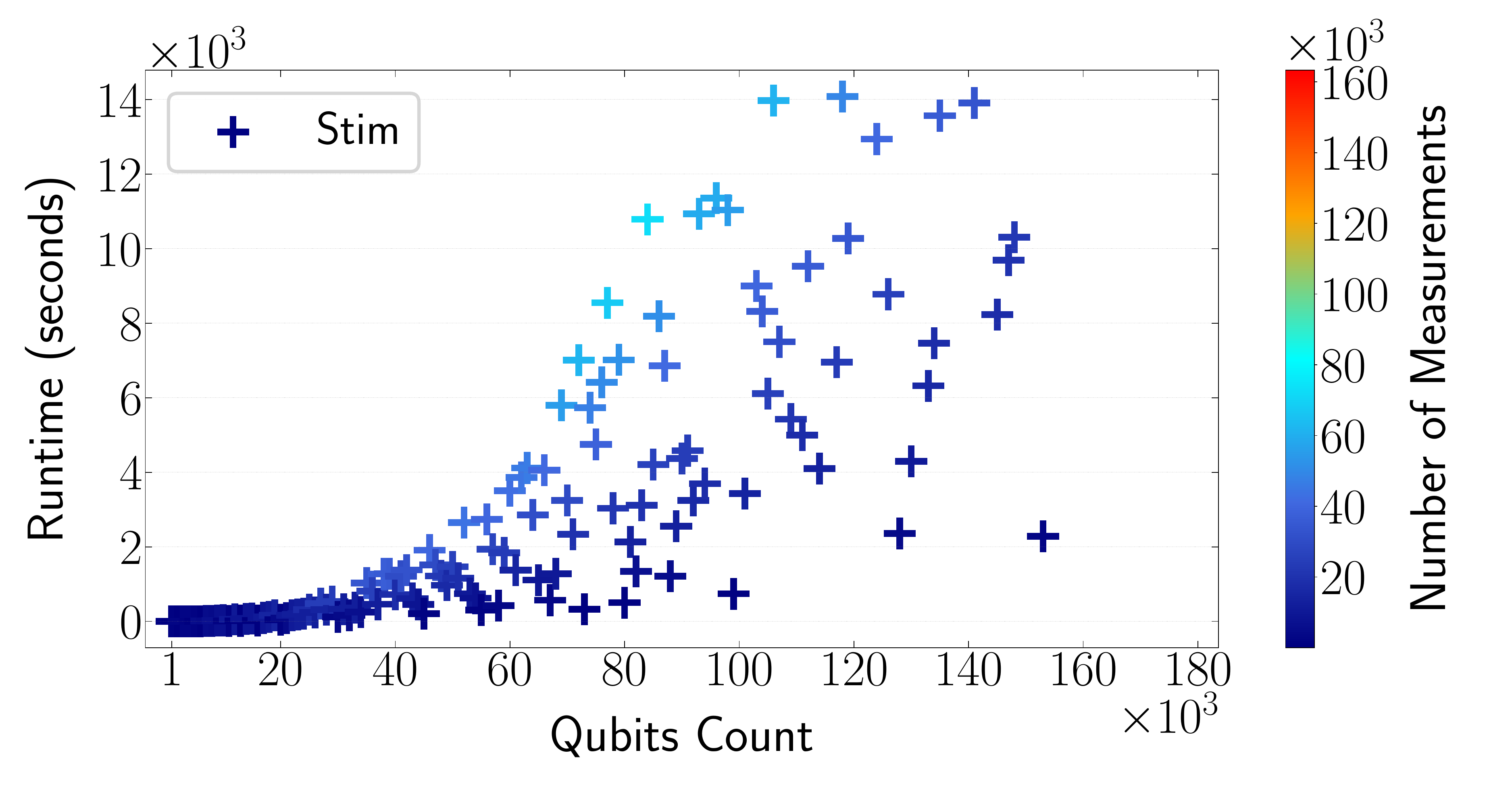}
		}\quad
		\subfloat[\ourTool ($\depth=1000$)]{
			\label{fig:measure_quasarq_d1000}
			\includegraphics[width=\linewidth]{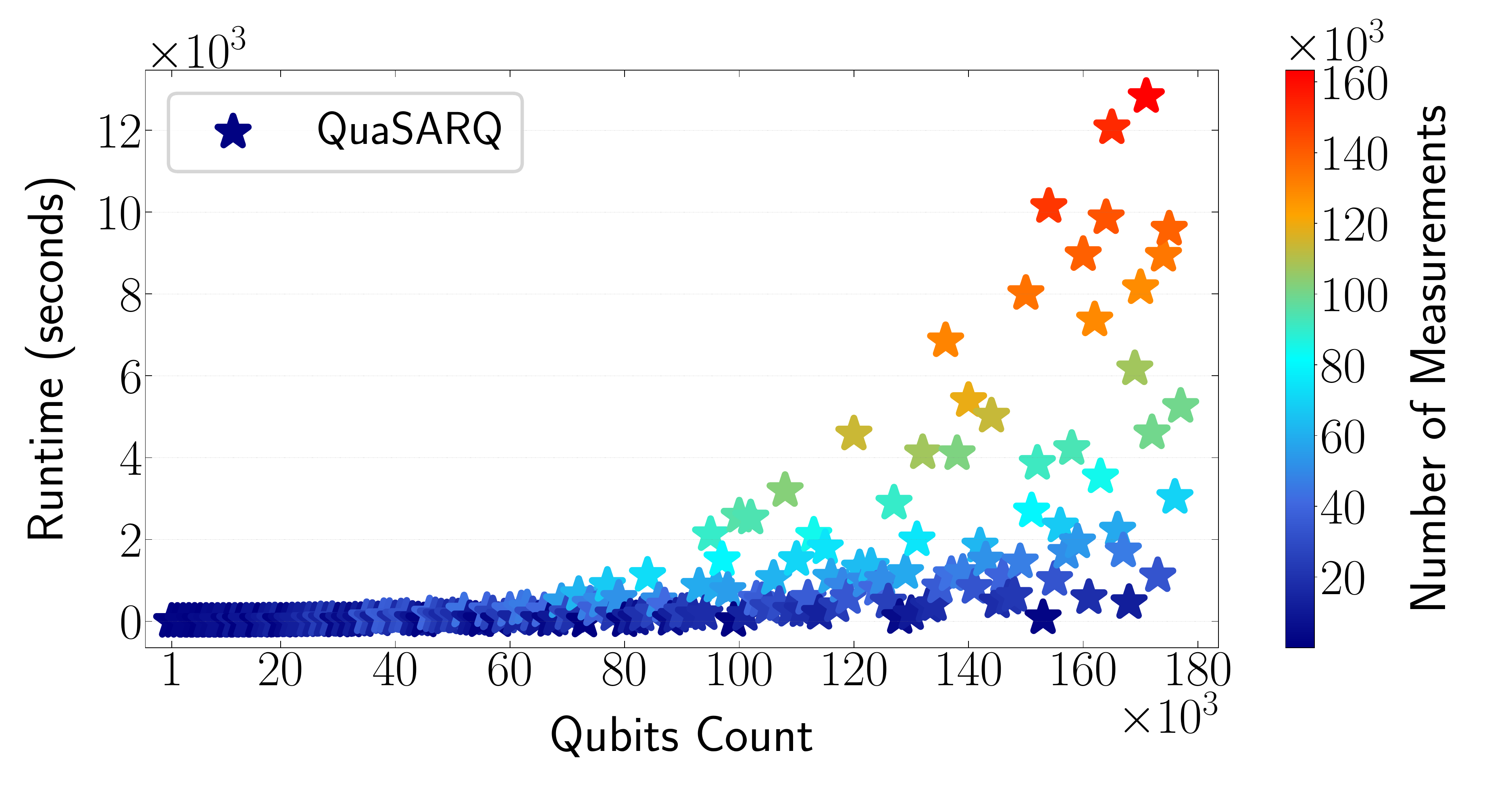}
		}
	}
	
	\caption{Distribution of successful measurements during simulation in \ourTool\ against \stim\ across 180 circuits (\(\text{qubits} \in [1K, 180K] , \text{depth}\  (\depth) \in \{100, 500, 1000\}\)).}
	\label{fig:measurements}
\end{figure*}

Figure~\ref{fig:measurements} plots runtime versus qubit count (both in thousands), with color indicating the number of nondeterministic measurements per bin: darker (blue) regions correspond to fewer measurements, while warmer (red) regions indicate higher counts.

In \stim’s plots, the nearly uniform dark coloring shows that most successful runs involve very few nondeterministic measurements. Even at large qubit counts, \stim rarely reaches warmer regions, indicating that it collapses only a small number of qubits non-deterministically before timing out.

In contrast, \ourTool exhibits a clear shift from dark to warmer colors as runtime increases. Even at 100K qubits, it successfully handles thousands of nondeterministic measurements, with the highest counts appearing in the reddest regions. These results show that \ourTool not only matches \stim on low-measurement instances but also scales effectively to circuits with substantially heavier nondeterminism, validating the efficiency of our pivot compaction and three-pass prefix-XOR elimination algorithms.

\Cref{tab:speedup} reports detailed per-algorithm runtimes (in milliseconds) for 30 representative circuits (the ten fastest at each depth), comparing \stim with our GPU-based simulator. For each circuit, we decompose the runtime into tableau evolution (TO), tableau transpose (T), and Gaussian elimination (GE) in \stim, and into TO, T, pivot compaction (CMP), and GE in \ourTool. The final three columns report speedup ratios (\stim over \ourTool) for TO, T, and the combined CMP+GE step, ensuring a fair comparison by folding compaction into elimination.

\begin{table}[htp]
  \centering
  \small
  \renewcommand{\arraystretch}{1.2}
  \setlength{\tabcolsep}{3.5pt}
  \caption{Runtime of our GPU-based simulator (\ourTool) against \stim on a best-performing sample of 30 circuits. All times are reported in milliseconds. Runtimes are decomposed into tableau evolution (TO), transpose (T), pivot compaction (CMP), and Gaussian elimination (GE); for a fair comparison, CMP is included in our total GE time. Bold entries indicate circuits where \ourTool outperforms \stim\ by a significant margin. The \emph{Speedup} columns report the ratio of \stim’s runtime to \ourTool’s runtime.}
  \label{tab:speedup}
  \resizebox{0.62\textheight}{!}{\input{stats_table}}
\end{table}

Across this sample, \ourTool applies gates about 105$\speedup$ faster at 12K qubits and depth 100 (8 ms versus 843 ms in \stim), showing that our Algorithm \ref{alg:buildTableau} scales linearly with qubit count while {\stim}'s routines incur super-linear overhead.  The transpose kernel maintains 20-49$\speedup$ speedups across all instances, thanks to our tiled shared-memory bit‐shuffling (\cref{alg:transpose}). In the critical collapse step, the combined CMP+GE accelerations range from roughly 30$\speedup$ on smaller circuits to about 12$\speedup$ on the largest or deepest circuits.  For example, at 10K qubits and depth 1,000, we reduce {\stim}'s 17 seconds GE into under 1 second, achieving 30$\speedup$ improvement. These consistent gains derived from our efficient pivots compaction (\cref{alg:compactPivots}) and three-pass prefix-XOR elimination (\cref{alg:paraGE}) designs.

%% file: stats_table.tex
 \begin{tabular}{rrrr|rrr|rrrr|ccc}
 	\hline\hline
 	\multirow{2}{*}{\textbf{\textsc{Qubits}}} 
 	& \multirow{2}{*}{\textbf{\textsc{Depth}}} 
 	& \multirow{2}{*}{\textbf{\textsc{Gates}}} 
 	& \multirow{2}{*}{\textbf{\textsc{Measures}}} 
 	& \multicolumn{3}{c|}{\textbf{\stim}} 
 	& \multicolumn{4}{c|}{\textbf{\ourTool}} 
 	& \multicolumn{3}{c}{\textbf{\textsc{Speedup}}} \\
    \cline{5-14}
 	& & &
 	& \textbf{TO}
 	& \textbf{T}
 	& \textbf{GE}
 	& \textbf{TO}
 	& \textbf{T}
 	& \textbf{CMP}
 	& \textbf{GE}
 	& \textbf{TO} 
 	& \textbf{T} 
 	& \textbf{CMP \code{+} GE} \\
 	\hline
	   6,000   & 100   & 418,285    &  5,646    
	 &   94   &   11   &   5,875    
	 & \textbf{3.7}   & \textbf{0.4}   & \textbf{83}    & \textbf{225}    
	 & \textbf{26}\speedup & \textbf{30}\speedup & \textbf{20}\speedup \\
	 7,000   & 100   & 486,959    &  5,422    
	 &  267   &   18   &   7,684    
	 & \textbf{4.4}   & \textbf{0.5}   & \textbf{80}    & \textbf{256}    
	 & \textbf{62}\speedup & \textbf{37}\speedup & \textbf{23}\speedup \\
	 8,000   & 100   & 554,796    &  4,225    
	 &  291   &   24   &   7,762    
	 & \textbf{5.1}   & \textbf{0.6}   & \textbf{62}    & \textbf{228}    
	 & \textbf{58}\speedup & \textbf{40}\speedup & \textbf{27}\speedup \\
	 9,000   & 100   & 624,149    &  5,087    
	 &  300   &   27   &  11,366    
	 & \textbf{5.8}   & \textbf{0.8}   & \textbf{75}    & \textbf{332}    
	 & \textbf{52}\speedup & \textbf{35}\speedup & \textbf{28}\speedup \\
	 11,000   & 100   & 764,886    &  7,848    
	 &  451   &   57   &  23,981    
	 & \textbf{6.8}   & \textbf{1.2}   & \textbf{117}   & \textbf{842}    
	 & \textbf{67}\speedup & \textbf{49}\speedup & \textbf{26}\speedup \\
	 12,000   & 100   & 828,593    &  3,831    
	 &  843   &   49   &  13,545    
	 & \textbf{8.1}   & \textbf{1.5}   & \textbf{57}    & \textbf{597}    
	 & \textbf{105}\speedup & \textbf{34}\speedup & \textbf{21}\speedup \\
	 13,000   & 100   & 903,631    &  9,242    
	 &  996   &   61   &  37,033    
	 & \textbf{14.0}  & \textbf{2.0}   & \textbf{138}   & \textbf{1,766}  
	 & \textbf{74}\speedup & \textbf{32}\speedup & \textbf{20}\speedup \\
	 14,000   & 100   & 975,315    & 12,705    
	 &1,172   &   72   &  57,466    
	 & \textbf{20.0}  & \textbf{2.5}   & \textbf{189}   & \textbf{2,914}  
	 & \textbf{58}\speedup & \textbf{29}\speedup & \textbf{19}\speedup \\
	 86,000   & 100   &5,928,542   & 15,596    
	 &26,544  & 2,876  &1,849,630  
	 & \textbf{641.0} & \textbf{106}   & \textbf{441}   & \textbf{111,136}
	 & \textbf{42}\speedup & \textbf{28}\speedup & \textbf{17}\speedup \\
	 116,000  & 100   &7,975,271   &    70    
	 &48,092  & 6,799  & 14,479    
	 & \textbf{1,199} & \textbf{204}   & \textbf{2.5}   & \textbf{1,034}  
	 & \textbf{41}\speedup & \textbf{34}\speedup & \textbf{14}\speedup \\
	 8,000   & 500   &2,755,985   &  6,339    
	 &1,199   &   15   &  11,408    
	 & \textbf{25.0}  & \textbf{0.6}   & \textbf{93.3}  & \textbf{342}    
	 & \textbf{49}\speedup & \textbf{25}\speedup & \textbf{27}\speedup \\
	 10,000   & 500   &3,440,969   &  2,961    
	 &1,510   &   24   &   7,731    
	 & \textbf{34.0}  & \textbf{1.0}   & \textbf{43.5}  & \textbf{227}    
	 & \textbf{46}\speedup & \textbf{25}\speedup & \textbf{29}\speedup \\
	 12,000   & 500   &4,133,639   &  8,561    
	 &3,053   &   33   &  30,041    
	 & \textbf{39.0}  & \textbf{1.5}   & \textbf{127.6} & \textbf{1,336}  
	 & \textbf{78}\speedup & \textbf{23}\speedup & \textbf{21}\speedup \\
	 13,000   & 500   &4,481,534   & 11,941    
	 &3,528   &   48   &  47,996    
	 & \textbf{68.0}  & \textbf{2.0}   & \textbf{178.4} & \textbf{2,286}  
	 & \textbf{53}\speedup & \textbf{25}\speedup & \textbf{20}\speedup \\
	 14,000   & 500   &4,824,016   & 11,332    
	 &4,035   &   53   &  51,287    
	 & \textbf{101.0} & \textbf{2.5}   & \textbf{169.6} & \textbf{2,602}  
	 & \textbf{40}\speedup & \textbf{21}\speedup & \textbf{19}\speedup \\
	 83,000   & 500   &28,535,171  &  4,403    
	 &124,012 & 3,475  & 486,459    
	 & \textbf{3,117} & \textbf{101}   & \textbf{123}   & \textbf{33,018} 
	 & \textbf{40}\speedup & \textbf{35}\speedup & \textbf{15}\speedup \\
	 86,000   & 500   &29,585,153  & 20,358    
	 &132,558 & 2,866  &2,409,790    
	 & \textbf{3,201} & \textbf{106}   & \textbf{575}   & \textbf{148,758}
	 & \textbf{42}\speedup & \textbf{27}\speedup & \textbf{17}\speedup \\
	 105,000  & 500   &36,098,518  &  3,506    
	 &198,221 & 4,294  & 613,795    
	 & \textbf{4,889} & \textbf{165}   & \textbf{102}   & \textbf{42,590} 
	 & \textbf{41}\speedup & \textbf{26}\speedup & \textbf{15}\speedup \\
	 107,000  & 500   &36,783,060  &  5,166    
	 &203,882 & 4,446  & 911,298    
	 & \textbf{4,981} & \textbf{165}   & \textbf{149}   & \textbf{61,264} 
	 & \textbf{41}\speedup & \textbf{27}\speedup & \textbf{15}\speedup \\
	 140,000  & 500   &48,126,478  &  1,672    
	 &348,276 & 7,762  & 482,996    
	 & \textbf{8,539} & \textbf{296}   & \textbf{63}    & \textbf{38,240} 
	 & \textbf{41}\speedup & \textbf{27}\speedup & \textbf{13}\speedup \\
	 10,000   &1000   & 6,881,923  &  6,733    
	 & 2,564  &   24   & 17,875     
	 & \textbf{67.0}  & \textbf{1.0}   & \textbf{99}    & \textbf{508}    
	 & \textbf{39}\speedup & \textbf{25}\speedup & \textbf{30}\speedup \\
	 58,000   &1000   &39,878,531  &  3,065    
	 &123,420 &   933  & 167,891    
	 & \textbf{3,028} & \textbf{49}    & \textbf{59}    & \textbf{11,435}
	 & \textbf{41}\speedup & \textbf{20}\speedup & \textbf{15}\speedup \\
	 65,000   &1000   &44,694,333  &  9,390    
	 &153,727 & 1,718  & 672,873    
	 & \textbf{3,663} & \textbf{58}    & \textbf{225}   & \textbf{39,768}
	 & \textbf{42}\speedup & \textbf{30}\speedup & \textbf{17}\speedup \\
	 67,000   &1000   &46,063,395  &  3,030    
	 &162,051 & 2,113  & 230,627    
	 & \textbf{4,054} & \textbf{64}    & \textbf{68}    & \textbf{15,046} 
	 & \textbf{40}\speedup & \textbf{34}\speedup & \textbf{16}\speedup \\
	 73,000   &1000   &50,186,126  &     35    
	 &191,483 & 2,069  &   3,077    
	 & \textbf{4,801} & \textbf{78}    & \textbf{1.1}  & \textbf{204}   
	 & \textbf{40}\speedup & \textbf{27}\speedup & \textbf{16}\speedup \\
	 80,000   &1000   &55,000,355  &    871    
	 &229,682 & 2,709  &  90,947    
	 & \textbf{5,689} & \textbf{93}    & \textbf{25}    & \textbf{6,220}  
	 & \textbf{41}\speedup & \textbf{30}\speedup & \textbf{15}\speedup \\
	 88,000   &1000   &60,501,128  &  4,722    
	 &277,716 & 3,812  & 591,435    
	 & \textbf{6,887} & \textbf{114}   & \textbf{133}   & \textbf{39,784}
	 & \textbf{41}\speedup & \textbf{34}\speedup & \textbf{15}\speedup \\
	 99,000   &1000   &68,063,957  &    742    
	 &350,677 & 3,830  & 113,640    
	 & \textbf{8,710} & \textbf{147}   & \textbf{21}    & \textbf{7,928}  
	 & \textbf{41}\speedup & \textbf{27}\speedup & \textbf{15}\speedup \\
	 128,000  &1000   &88,000,855  &  4,433    
	 &583,717 & 6,382  &1,089,120  
	 & \textbf{14,136}& \textbf{240}   & \textbf{166}   & \textbf{72,111}
	 & \textbf{42}\speedup & \textbf{27}\speedup & \textbf{16}\speedup \\
	 153,000  &1000   &105,195,653 &  2,023    
	 &827,808 & 9,259  & 703,303    
	 & \textbf{20,626}& \textbf{372}   & \textbf{78}    & \textbf{58,590} 
	 & \textbf{41}\speedup & \textbf{25}\speedup & \textbf{12}\speedup \\
	\hline
\end{tabular}%

%% file: contributions.tex
\section*{Author Contributions}

MO designed the parallel algorithms, implemented them in the \ourTool simulator, performed numerical analysis, and wrote the initial draft.
DT co-authored the manuscript, contributed to the stabilizer-formalism foundations, presented the measurement algorithms, and verified that the technical innovations are supported by theory.
AL conceived the project and supervised the research.
All authors discussed the results, reviewed the manuscript, and approved the 
final version, taking full responsibility for the work.

AI tools were used only to generate unit tests and to check the spelling of this document.

\section*{Acknowledgements}
This work was supported by the Dutch National Growth Fund (NGF), as part of the Quantum Delta NL program.
